\documentclass[twocolumn]{aa}
\usepackage{graphicx}
\usepackage{times}
\usepackage{longtable}
\usepackage{natbib}
\usepackage{psfig}

\bibpunct{(}{)}{;}{a}{}{,} 

\def\5{\footnotesize V\normalsize}
\def\4{\footnotesize IV\normalsize}
\def\3{\footnotesize III\normalsize}
\def\2{\footnotesize II\normalsize}
\def\1{\footnotesize I\normalsize}
\def\lam{$\lambda$}
\def\kms{$\mbox{km s}^{-1}$}
\def\p{$\phantom{:}$}
\def\a{$\phantom{^\ast}$}

\def\pp{$\phantom{-}$}
\def\o{$\phantom{1}$}

\begin{document}

\title{The VLT-FLAMES Survey of Massive Stars: Observations centered on the Magellanic
Cloud clusters NGC\,330, NGC\,346, NGC\,2004, and the N11 region
\footnote{Based on observations 
at the European Southern Observatory Very Large Telescope in programmes 68.D-0369 and 171.D-0237.}
\footnote{Tables \ref{346}, \ref{330}, \ref{lh910}, \ref{2004}, \& \ref{cftypes} are only available in electronic form at http://www.edpsciences.org}}

\author{C.~J.~Evans\inst{1}, D.~J.~Lennon\inst{2}, 
S.~J.~Smartt\inst{3}, and C.~Trundle\inst{3}
}

\offprints{C.~J.~Evans at cje@roe.ac.uk}

\authorrunning{C.~J.~Evans et al.}
   
\titlerunning{FLAMES Survey of Massive Stars: LMC \& SMC clusters}
   
\institute{UK Astronomy Technology Centre, 
           Royal Observatory Edinburgh, 
           Blackford Hill, Edinburgh, EH9 3HJ, UK
             \and
           The Isaac Newton Group of Telescopes,
           Apartado de Correos 321, E-38700,
           Santa Cruz de La Palma, Canary Islands, Spain
             \and 
           Department of Physics \& Astronomy,
           Queen's University Belfast,
           Belfast BT7 1NN, Northern Ireland, UK
}
\date{}

\abstract{
We present new observations of 470 stars using the Fibre Large Array
Multi-Element Spectrograph (FLAMES) instrument in fields centered on
the clusters NGC\,330 and NGC\,346 in the Small Magellanic Cloud
(SMC), and NGC\,2004 and the N11 region in the Large Magellanic Cloud
(LMC).  A further 14 stars were observed in the N11 and NGC\,330
fields using the Ultraviolet and Visual Echelle Spectrograph (UVES)
for a separate programme.  Spectral classifications and stellar radial
velocities are given for each target, with careful attention to checks
for binarity.  In particular, we have investigated previously
unexplored regions around the central LH9/LH10 complex of N11, finding
$\sim$25 new O-type stars from our spectroscopy.
We have observed a relatively large number of Be-type stars that display
permitted Fe~{\scriptsize II} emission lines.  These are primarily not
in the cluster cores and appear to be associated with classical
Be-type stars, rather than pre main-sequence objects.  The presence of
the Fe~{\scriptsize II} emission, as compared to the equivalent width
of H$\alpha$, is not obviously dependent on metallicity.  We have also
explored the relative fraction of Be- to normal B-type stars in the
field-regions near to NGC\,330 and NGC\,2004, finding no strong evidence of
a trend with metallicity when compared to Galactic results.
A consequence of service observations is that we have reasonable
time-sampling in three of our FLAMES fields.  We find lower limits to
the binary fraction of O- and early B-type stars of 23 to 36\%.  One of
our targets (NGC\,346-013) is especially interesting with a massive, 
apparently hotter, less luminous secondary component. }
\maketitle 

\section{Introduction}\label{intro}
As part of a European Southern Observatory (ESO) Large Programme we
have completed a new spectroscopic survey of massive stars in fields
centered on open clusters in the Large and Small Magellanic Clouds
(LMC and SMC respectively) and the Galaxy.  The survey has employed
the Fibre Large Array Multi-Element Spectrograph (FLAMES) instrument
at the Very Large Telescope (VLT), that provides high-resolution
($R\sim$20,000) multi-object spectroscopy over a 25' diameter
field-of-view.  The scientific motivations for the survey, and the
observational information for the three Galactic clusters (NGC\,3293,
NGC\,4755, and NGC\,6611), were presented by \citet[][hereafter
Paper~I]{mwpaper}.

In this paper we present the FLAMES observations in the Magellanic
Clouds.  The material presented is largely a discussion of the
spectral classifications and radial velocities of each star, and
provides a consistent and thorough overview of what is a particularly
large dataset.  In parallel to this catalogue, subsets of the sample
are now being analysed by different groups.  The sources of photometry and 
astrometry used for target selection are given in Section~\ref{targets}, 
followed by details of the observations in Section ~\ref{obs}, and then
a discussion of the observed sample.

Two FLAMES pointings were observed in each of the Clouds, centered on
NGC\,346 and NGC\,330 in the SMC, and on NGC\,2004 and the N11 region
\citep{hen56} in the LMC.  NGC\,346 is a young cluster with an age in
the range of 1 to 3$\times$10$^6$ yrs \citep{k89,wal00,jc03,mfast2}, that has
clearly undergone prodigious star formation.  It is also the largest H~\2
region in the SMC.  The best source of spectroscopic information in
NGC\,346 is the study by \citet[][ hereafter MPG]{mpg}, who found
as many O-type stars in the cluster as were known in the rest of the
SMC at that time.  High-resolution optical spectra of five of the
O-type stars from MPG were presented by \citet[][together with
AzV~220 that is also within the FLAMES field-of-view]{wal00}.  These
were analysed by \citet{jc03}, in conjunction with ultraviolet data,
to derive physical parameters.

N11 is also a relatively young region and includes the OB associations
LH9, LH10 and LH13 \citep{lh}, that are of interest in the context
of sequential star-formation, see \citet{w92}, \citet[][ hereafter
P92]{p92}, and
\citet{b03}.  P92 illustrated how rich the region is in terms of
massive stars, presenting observations of 43 O-type stars in LH9 and
LH10, with three O3-type stars found in LH10.  These O3 stars were
considered by \citet{w02} in their extension of the MK classification
scheme to include the new O2 subtype, with one of the stars from P92
reclassified as O2-type.  The FLAMES observations in N11 presented an opportunity to
obtain good-quality spectroscopy of a large number of known
O-type stars, whilst also exploring the spectral content of this
highly-structured and dynamic region.

NGC\,330 and NGC\,2004 are older, more centrally condensed clusters.
NGC\,330 in particular has been the focus of much attention in recent
years.  \citet{f72} presented H$\alpha$ spectroscopy of 18 stars in
the cluster, and noted: `It is also an object of considerable
importance in discussion of possible differences between stars of the
same age in the SMC and in the Galaxy'.  The community has clearly
taken his words to heart -- in the past 15 years there have been
numerous studies in the cluster, most of which were concerned with the
large population of Be-type stars therein, namely \citet{grb92},
\citet{l93}, \citet{grb96}, \citet{maz96}, \citet{sk98}, \citet{sk99}, 
\citet{m99}, and \citet{l03}.   The paper from these with the widest implications 
is that from \citet{m99}, who compared the fraction of Be stars
(relative to all B-type stars) in a total of 21 clusters in the SMC,
LMC and the Galaxy.  The fraction of Be stars appears to increase with
decreasing metallicity, although their study was limited to only one
cluster (NGC\,330) in the SMC.  This trend led \citeauthor{m99} to
advance the possibility of faster rotation rates at lower
metallicities -- one of the key scientific motivations that prompted
this FLAMES survey.

In contrast to NGC\,330, with the exception of abundance analyses of a
few B-type stars by \citet{korn02,korn06}, relatively little was known
about the spectroscopic content of NGC\,2004 until recently.  A new
survey by \citet{mhf}, also with FLAMES, has observed part of the
field population near NGC\,2004 using the lower-resolution mode of the
Giraffe spectrograph (and with a different field-centre to ours).
\citeauthor{mhf} concluded that the Be-fraction in their LMC field is
not significantly different to that seen in the Galaxy.

The FLAMES observations for the current survey were obtained
in service mode and so span a wide range of observational epochs,
giving reasonable time-sampling for the detection of binaries.  There
are surprisingly few multi-epoch, multi-object spectroscopic studies
of stellar clusters in the literature in this respect, with one such
study in 30 Doradus summarized by \citet{bsm98}.  Placing a lower
limit on the binary fraction in dense star-forming regions such as
NGC\,346 and N11, combined with stellar rotation rates, will help to provide
useful constraints in the context of star formation and the initial
mass function.

\section{Target selection}\label{targets}

\subsection{SMC photometry}
The adopted photometry and astrometry for the targets in NGC\,346 and
NGC\,330 is that from the initial ESO Imaging Survey (EIS) pre-FLAMES
release by \citet{eis}.  

\subsection{N11 photometry}

The N11 region was not covered by the EIS pre-FLAMES Survey and so
we obtained 60s $B$ and $V$ images with the Wide Field
Imager (WFI) at the 2.2-m Max Planck Gesellschaft (MPG)/ESO telescope,
on 2003 April 04.  These were processed by Dr.~M. Irwin using a
modified version of the Isaac Newton Telescope--Wide Field Camera
(INT--WFC) data reduction pipeline \citep{mji_wfc}, we then
calibrated the photometry to the Johnson-Cousins system using results
from P92.

\begin{figure}
\begin{center}
\includegraphics[width=9cm]{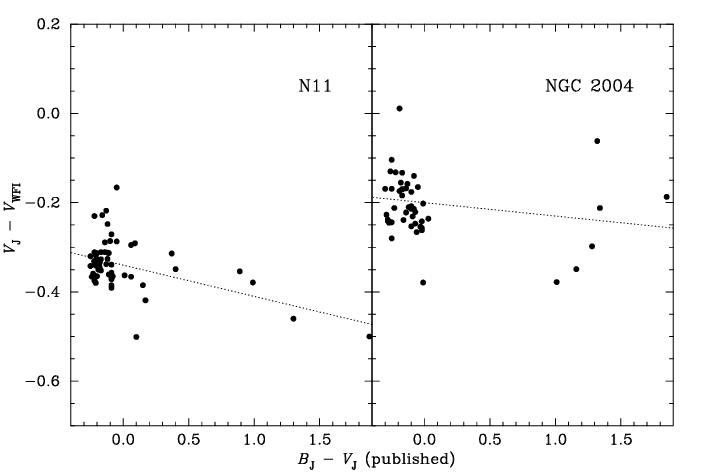}
\caption{Comparison of $V_{\rm J} - V_{\rm WFI}$ with published colours in N11
from \citet{p92}, and primarily from \citet{r74} for NGC\,2004.}
\label{v}
\end{center}
\end{figure}

\begin{figure}
\begin{center}
\includegraphics[width=9cm]{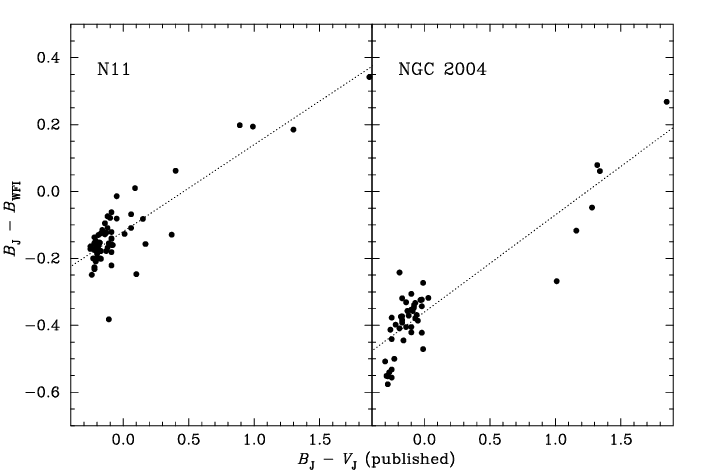}
\caption{Comparison of $B_{\rm J} - B_{\rm WFI}$ with published colours in N11
from \citet{p92}, and primarily from \citet{r74} for NGC\,2004.}
\label{b}
\end{center}
\end{figure}

As in Paper~I, for photometric calibration we prefer to visually match
stars using published finding charts.  In crowded regions such as those
in our FLAMES fields this ensures accurate cross-identification.
Cross-referencing the finding charts of P92 with the WFI
images yielded matches of 41 stars (for which both $B$ and $V$ WFI
photometry were available), with relatively sparse sampling in terms
of $(B-V)$. To increase the number of cross-matched stars an
astrometric search was performed between the WFI sources and the full
P92 catalogue (available on-line at the Centre de
Donn$\acute{\rm e}$es astronomiques de Strasbourg).  The mean (absolute)
offsets found between the WFI and P92 astrometry
(for the 41 stars with visual matches), were $<|{\Delta\delta}|>
\sim0.4''$ and $<|{\Delta\alpha}|> \sim0.06^{\rm s}$; these were used 
to then expand the sample to 60 stars, with $V <$~17.0.  The $V$ and
$B$ colour terms found for the WFI data in N11 are shown in Figures
\ref{v} and \ref{b}, and the transformation equations found were:
\begin{equation}
V_{\rm J} = V_{\rm WFI} - 0.07\times(B - V)_{\rm J} - 0.34, 
\end{equation}
\begin{equation}
B_{\rm J} = B_{\rm WFI} + 0.26\times(B - V)_{\rm J} - 0.12.
\end{equation}

After transformation we find mean (absolute) differences of
$\sim$0.05$^{\rm m}$, with $\sigma \sim$0.05$^{\rm m}$ for both $V$
and $(B-V)$.  In the centre of LH\,10, in which there is significant
nebular emission, the INT--WFC pipeline detected some of the sources
as `noise-like' and did not yield sensible results; for these 11 stars
(marked with a $\ast$ in Table \ref{lh910}) the photometry here is
that from P92.  Their photometry is also included in the Table for N11-105
which was in the gap between CCDs in the WFI $V$-band image.

\subsection{NGC\,2004 photometry}
NGC\,2004 was observed in two of the EIS pre-FLAMES fields, LMC 33 and
34.  At the time of the FLAMES observations a full EIS
data release was not available for these fields, so the raw images were
acquired from the ESO archive and reduced using the INT--WFC pipeline
(also by Dr.~M. Irwin).

CCD photometry in NGC\,2004 has been published by \citet{bbbc} and
\citet{bal93}.  The former study is calibrated to the 
photographic work of \citet{r74}, albeit with consideration of
two of the photoelectric standards from \citet[][]{mh84}.  The
study by \citeauthor{bal93} is independent of previously published
photometry in the cluster, however the `finding charts' and format of
the catalogue (in terms of pixel positions) are less than ideal for
successfully recovering matched stars; without access to the raw
frames, accurate cross-matching between their photometry and our WFI
images cannot be ensured.  Instead we employ visual
matches of 48 stars from the identification charts of \citet{r74} to
calibrate the WFI data, taking photoelectric results from
\citet[][2 stars]{mh84} and \citet[][7 stars]{w87}, with the data for 
the remaining 39 stars taken from \citeauthor{r74}.  The $V$ and $B$ colour
terms found for the NGC\,2004 WFI data are also shown in Figures
\ref{v} and \ref{b}, and the transformation equations found were:
\begin{equation}
V_{\rm J} = V_{\rm WFI} - 0.03\times(B - V)_{\rm J} - 0.20, 
\end{equation}
\begin{equation}
B_{\rm J} = B_{\rm WFI} + 0.29\times(B - V)_{\rm J} - 0.36.
\end{equation}

Whilst there is a relatively large scatter in $(V_{\rm J} - V_{\rm
WFI})$, the colour term is very robust; the same result is found if
only the photoelectic values are used or, similarly, if only the
photographic results are used.  After transformation we find mean
(absolute) differences of $\sim$0.06$^{\rm m}$, with $\sigma
\sim$0.04$^{\rm m}$ for both $V$ and $(B-V)$.

\subsection{Selection effects}\label{seffects}
Following similar methods to those in Paper~I for target selection,
the photometric data were used to create input catalogues for the FLAMES Fibre
Positioner Observation Support Software (FPOSS) that allocates the
Medusa fibres in a given field.  

In Paper~I we were perhaps too hasty with our description of avoiding
known Be-type stars.  To recap, the observed samples in the three
Galactic clusters from Paper~I were relatively complete for blue stars
in the FLAMES pointing, down to $\sim V$=13.  In terms of the final
samples no external selection effects were present with regard to
Be-type stars, with only 12 stars classified as Be and one as Ae-type,
from a combined total of 319 FLAMES and FEROS targets.  It is worth
noting that NGC\,3293 and NGC\,4755 are two of the three clusters
included by \citet{m99} in their `inner Galaxy' sample (with the
lowest Be-fraction compared to normal B-type stars), i.e. both
NGC\,3293 and NGC\,4755 have genuinely small numbers of
Be-type stars.

As mentioned in Paper~I, one of our primary objectives
concerns the process of nitrogen enrichment in OB-type stars and its
correlation (or not) with rotational velocities.  Spectroscopic
analysis of Be stars is more involved (i.e. difficult!) than for
`normal' B-type stars.  Thus in our FPOSS input catalogue, in an
attempt to ensure that we didn't observe a preponderance of Be stars
in NGC\,330, we excluded 15 stars classified as Be-type from
previous spectroscopy \citep{l93,sk98}.  However, this isn't such a strong selection
effect as it may sound - all of the excluded stars are from the
\citet{r74} survey and therefore are in (or near to) the main body of
the cluster.  Obviously the FLAMES survey cannot be used to comment on
the incidence of the Be-phenomenon in the cluster itself, but it
could in principle offer constraints on the field population around
the cluster.  With regard to the non-cluster population, we also
excluded 15 stars from \citet{eh04} that have emission in their
Balmer lines.  However, these are not necessarily Be-type stars as
the narrow emission in many of the stars from \citeauthor{eh04} is
likely attributable to nebular origins (see discussion in their Section 7).  
Given the high density of potential targets in the field, the exclusion of the 
stars from \citeauthor{eh04} is very unlikely to bias the final sample.

In NGC\,2004, \citet{sk99} reported 42 Be-type stars from their
photometric study.  However, due to the relative dearth of published
spectroscopy in this field at the time of our observations, no
potential targets were excluded when using FPOSS for
fibre configuration.  Lastly, the only weighting in NGC\,346 and N11
was to give higher priority of fibre-assignment to known OB-type stars
from MPG and P92.

In summary, the only strong external bias in our observed targets was in the
main body of NGC\,330.  A much stronger selection effect in both
NGC\,330 and NGC\,2004 is that the sheer density of the cluster cores
prevents Medusa fibres being placed so close together, with many
stellar targets blended -- in any one FLAMES pointing only a few stars
near the centre could be observed with the Medusa fibres. Follow-up of
these two clusters with integral field spectroscopy, such as that
offered by FLAMES-ARGUS at the centre of the FLAMES field plate, is an
obvious project to allow a comprehensive exploration of the cluster
populations.  However, the field populations of both NGC\,330 and
NGC\,2004 should be a relatively unbiased sample of the true population, subject
to cluster membership issues and the faint magnitude cut-off of the
FLAMES survey.  These issues are discussed further in
Section~\ref{discussbe}.

\section{Observations and data reduction}\label{obs}

\subsection{FLAMES-Giraffe spectroscopy}

All of the FLAMES observations were obtained in service mode, with the
majority acquired over a 6 month period from 2003 July to 2004
January.  The same high-resolution settings were used as for the
Galactic observations, i.e. HR02 (with a central wavelength of
\lam3958\,\AA), HR03 (\lam4124), HR04 (\lam4297), HR05 (\lam4471),
HR06 (\lam4656), and HR14 (\lam6515).  The bulk of the observations in
NGC\,330 were obtained prior to installation of a new grating in 2003
October, so the spectral coverage and resolution of these data are
identical to that of our Galactic observations (see Paper~I).
However, the observations in NGC\,346, NGC\,2004 and N11
were taken with the new grating in place.  The characteristics of the
NGC\,346 observations are given in Table \ref{arc_fwhm}.  In comparison
to the older grating the effective resolving power is decreased in
some setups (in particular in the HR14 setting, yielding a
correspondingly wider wavelength coverage), however the new grating is
more efficient in terms of throughput.

In addition to observations at 6 central wavelengths, each field was
observed at each wavelength for 6 exposures of 2275s.  The repeat
observations at each central wavelength were taken in batches of 3
exposures, each triplet forming an observing block for the ESO service
programme.  The observational constraints on our programme were not
particularly demanding (the required seeing was 1.2'' or better), but
in some cases these conditions were not fully satisfied.  Therefore,
at some settings we have more than 6 observations, for instance the
HR14 setting was observed 9 times in our NGC\,346 field.  Although the
conditions may not have been ideal for every observing block, $all$
the completed observations were reduced -- in general the seeing was
not greater than 1.5'' and the data are of reasonable quality.  This
is of particular relevance in terms of the radial velocity information
contained in the spectra.  The majority of the observations in
NGC\,330 were spread over a relatively short time ($\sim$10 days).
Otherwise we have reasonable time coverage, e.g. the NGC\,346
observations spanned almost 3~months so we are in an excellent
position to detect both single and double-lined binaries.  In Appendix
\ref{mjd} we list the modified Julian dates (MJD) of each of the
observations.

The data were reduced using the Giraffe Base-Line Reduction Software
(girBLDRS), full details of which are given by \citet{girbldrs}.  For
consistency with our performance tests in Paper~I, v1.10 of girBLDRS
was used.  More recent releases include the option to perform sky
subtraction within the pipeline, but we prefer to employ the methods
discussed in Paper~I.

\begin{table*}
\caption[]{Summary of the wavelength coverage, mean FWHM of the arc lines 
and effective resolving power, $R$, at each Giraffe central wavelength setting,
\lam$_c$, for the observations in NGC\,346.}
\label{arc_fwhm}
\begin{center}
\begin{tabular}{lccccc}
\hline\hline
Setting & \lam$_c$ & Wavelength coverage & \multicolumn{2}{c}{FWHM}& $R$\\
        & (\AA)    & (\AA)               & (\AA) & (pixels) & \\
\hline 
HR02    & 3958     & 3854-4051           & 0.18  & 3.7    & 22,000\\
HR03    & 4124     & 4032-4203           & 0.15  & 3.4    & 27,500\\
HR04    & 4297     & 4187-4394           & 0.19  & 3.5    & 22,600\\
HR05    & 4471     & 4340-4587           & 0.23  & 3.7    & 19,450\\
HR06    & 4656     & 4537-4760           & 0.20  & 3.5    & 23,300\\
HR14    & 6515     & 6308-6701           & 0.39  & 3.9    & 16,700\\
\hline
\end{tabular}
\end{center}
\end{table*}

The final signal-to-noise is, of course, slightly variable between the
different wavelength regions for a given target depending on the exact
conditions when the observations were taken.  As a guide, the
signal-to-noise of the reduced, coadded spectra is $\sim$200 for the
brightest stars in each field, decreasing to 110 in N11 (with the
brightest faint cut-off), 75 in NGC\,2004, and 60 in NGC\,346.
Inspection of the reduced spectra in the NGC\,330 field revealed
particularly low signal-to-noise in the six HR03 observations.  The
Si~\2 and Si~\4 lines that are important for quantitative analysis are
included in this region, so the NGC\,330 field was reobserved with the
HR03 setup on 2005 July 20 \& 24.  These more recent observations are
included in Table~\ref{330_obs} as HR03\#07-12.  The spectra were
reduced using the same routines and software as the rest of the
survey.

The NGC\,330 field features the faintest stars in the entire survey.
Besides the problems with the initial HR03 observations, the
signal-to-noise ratio of the fainter stars in this field (even in the
co-added spectra) decreases to $\sim$35-40 per unbinned pixel (at
$V~=$~16) and down to 25-30 for the very faintest stars.  Binning of
the data helps to some degree, but it is clear that detailed analysis
of individual stars (such as those presented by Hunter et al.,
submitted) will not be feasible for some of the faintest targets.
Nevertheless, the data presented here serve as a novel investigation
of the spectral content of the region.

\subsection{FLAMES-UVES spectroscopy}
\label{uves}

In addition to observations with the Giraffe Spectrograph, one can
observe up to an additional 6 objects simultaneously with the red-arm
of UVES (with a central wavelength setting of 5200\,\AA).  In the SMC
and LMC fields the UVES fibres were used to observe a small number of
`red' targets.  The FLAMES-UVES \lam5200 set-up gives spectral
coverage from approximately \lam4200 to 5150 and 5250 to 6200\,\AA~(with the
break in coverage arising from the gap between the two CCDs).  These
data were reduced by Dr.~A. Kaufer using the standard pipeline
\citep[][]{uvespipe}, which runs under MIDAS and performs the 
extraction, wavelength calibration, background subtraction and
subsequently merges the individual orders to form two continuous
spectra, i.e., from \lam4200 to \lam5150 and \lam5250 to \lam6200\,\AA.

Note that these data were observed over multiple epochs, and the
pipeline does not correct the spectra to the heliocentric frame.  For
the purposes of approximate classification the very large number of
individual UVES spectra were simply co-added, without correction to
the heliocentric frame.  Therefore we do not quote radial velocities
for these stars.

\subsection{Additional UVES spectroscopy in N11 and NGC\,330}
Normal UVES spectroscopy of a further 14 stars was obtained in 2001
November 1-3, as part of programme 68.D-0369(A); these data are
included here to supplement the FLAMES sample.  The spectra cover
\lam3750 to \lam5000, \lam5900 to \lam7700, and \lam7750 to \lam9600\,\AA, 
at a resolving power of $R\sim$20,000.  These data were also reduced
using the UVES pipeline, contemporaneously to those presented by
\citet{tl04}.

The 9 targets in N11 complement the FLAMES programme well, providing a
number of sharp-lined, luminous early B-type spectra.  The spectra of
the 5 targets in NGC\,330 were only examined after collation of the
FLAMES data hence their inclusion (out of sequence) at the end of
Table \ref{330}.  Note that the 14 stars observed with UVES in the
traditional (non-fibre) mode are marked in Tables \ref{330} and
\ref{lh910} as simply `UVES target' (cf. `FLAMES-UVES target' for
those observed in the more limited, fibre-fed mode).

\section{Spectral classifications and stellar radial velocities}\label{data}

The FLAMES spectra were classified by visual inspection, largely
following the precepts detailed in Paper~I, with additional
consideration of the lower metallicity of stars in the LMC and SMC.
The principal reference for O- and early B-type spectra is the digital
atlas from \citet{wf90}, with the effects of metallicity explored 
by \citet[][]{wal95,wal00}.  Later-type stars in the SMC were classified
using the critieria outlined by \citet{l97}, \citet{eh03}, and
\citet{eh04}.  In the LMC, A-type supergiants were classified from
interpolation between the Galactic and SMC spectra given by
\citeauthor{eh03}, and B-type supergiants were classified with
reference to \citet{f88,f91}.  Intermediate types of B0.2 and B0.7
have been used for some spectra -- these are interpolated types that,
in an effort to avoid metallicity effects, are largely guided by the
intensity of the weak He~\2 \lam4686 line.

Assignment of luminosity classes in early B-type stars, with due
attention to metallicity effects and stellar rotation rates, can be
difficult.  Very subtle changes in the observed spectra can yield
different classes.  Indeed, the reason that we do not regularly employ
the class IV notation for stars in the Magellanic Clouds, arises from
these issues -- in a large dataset such as the FLAMES survey, one
should be careful not to over-emphasize groupings as `dwarfs' and
`giants', remembering that luminosity and gravity are continuous
quantities.  Also, as in our previous studies, we are cautious with
regard to employing the Be notation -- many of our targets have
H$\alpha$ profiles displaying narrow core emission, accompanied by
[N~\2] emission lines which are clearly nebular in origin. Where
possible we give precise classifications for the Be spectra.  In those
spectra with Fe~\2 emission (discussed further in
Section~\ref{bestars}) contamination of lines such as Si~\3 \lam4552
makes classification more difficult.  However, as for normal B-type
stars, other absorption features are also of use as temperature
diagnostics, such as O~\2 \lam\lam4415-17 and the O~\2/C~\3 blend at
\lam4650~\AA\/ (remembering to account for metallicity effects).  
Of course, our Be classifications do not consider the likely
contribution to the continuum by the emission region, and thus may not
correlate to the same temperatures as for normal B stars.

In Tables~\ref{346}, \ref{330}, \ref{lh910}, and \ref{2004} we present
the observational properties of each of the stars observed in the four
FLAMES fields in the Magellanic Clouds.  Spectral classifications and
stellar radial velocities are given, together with cross-references to
existing catalogues and comments regarding the appearance of the
H$\alpha$ profiles and binarity.  For NGC\,346, NGC\,330 and NGC\,2004
we also include the radial distance (r$_{\rm d}$, in arcmin) of each
star from the centre of the cluster core.\footnote{For completeness,
the coordinates (J2000.0) used in calculating the radial distances
were:

NGC\,346:$\phantom{4}$ $\alpha$ = 00$^{\rm h}$59$^{\rm m}$18.0$^{\rm s}$, $\delta$ = $-$72$^{\circ}$10$'$48.0$''$,

NGC\,330:$\phantom{4}$ $\alpha$ = 00$^{\rm h}$56$^{\rm m}$18.8$^{\rm s}$, $\delta$ = $-$72$^{\circ}$27$'$47.2$''$,

NGC\,2004: $\alpha$ = 05$^{\rm h}$30$^{\rm m}$40.2$^{\rm s}$, $\delta$ = $-$67$^{\circ}$17$'$14.3$''$.}

Finding charts are included for each of the
fields in Figures~\ref{fchart_346}, \ref{fchart_330}, \ref{fchart_n11},
and \ref{fchart_2004}.  The finding charts employ images from the
Digitized Sky Survey (DSS) with our targets overlaid.  The WFI
pre-imaging used for target selection is significantly superior to the
DSS images, however our intent here is solely to give a clear overview
of the locations of each of the stars in our sample.  The DSS images
provide such clarity, in particular they are free of the chip-gaps
arising from the WFI CCD array.  One clear point evident from the
finding charts is that NGC\,330 and NGC\,2004 are very compact
clusters and that, whilst the FLAMES observations sample some peripheral
cluster members, the majority of the targets sample the local
field population.  Furthermore, the radius of the ionized region in
NGC\,346 is given by \citet{rpb} as $\sim$3.5 arcmin; many of the
FLAMES targets are beyond this radius highlighting that, even in this
pointing, the observations sample both the cluster and field
population.

The N11 region is more complex, with both LH10 (the denser region
above the centre in Figure~\ref{fchart_n11}) and LH9 (the relatively
`open' cluster just below centre).  The $V$-band WFI image is shown in
Figure~\ref{n11_mess}, in false-colour to better highlight the
nebulosity.  The full complexity of the region is revealed in the
original photographic plate from \citet{hen56}.  Star N11-004 is the
principal object in N11G, N11-063 and N11-099 are both in N11C, and some of our
targets south of LH9 are in N11F.  Further insight into the structure of
this region is given by the H$\alpha +$ [N~\2] photograph from Malcolm
G. Smith published by \citet{w92}, and from a near-IR image from
Dr.~R. Barb$\acute{\rm a}$ (private communication), both of which show
a wide variety of arcs and filaments of nebular gas.

Radial velocities ($v_{\rm r}$ in \kms) are given for each star,
excepting those that appear to be multiple systems.  The measurements here are
the means of manual estimates of a number of line centres, as
indicated in parentheses in the tables.  The primary lines used are
generally those of He~\1, He~\2 and Si~\3, with a typical standard
deviation of the individual measurements around 5-10~\kms.  We prefer
manual measurements such as these because of the highly variable
nebular contamination in a region such as NGC\,346 (which may led to
spurious results if more automated methods are employed).  In a small
number of stars with apparently large rotational velocities (or lower
signal-to-noise), precise determination of the line centres is more
difficult thereby yielding a less certain value, as indicated here using the
usual `:' identifier in the tabulated results.  The precision of
manual measurements may also be biased by the infilling of helium
lines, either by nebular contamination in morphologically normal
spectra, or by infilling of the lines in Be-type stars; where these
effects are obvious we have avoided the relevant lines but naturally
there may be cases in which the effects are not so prominent yet
affect the (apparent) line centre.  

In tables~\ref{346}, \ref{330}, \ref{lh910}, and \ref{2004} we present
stellar radial velocities for each of our targets.  In the process of
measuring the velocities we find a relatively large number of spectra
displaying evidence of binarity.  These are shown in the tables as
`Binary', with SB1 and SB2 added to indicate single-lined and
double-lined binaries where it was possible to identify the nature of
the system.  Finally, in the comments column of each table we note a
small number of stars as `variable $v_{\rm r}$?'.  In these one or two
lines have anomalous velocities, with it unclear whether they arise
from binarity.

\subsection{Cross-references with x-ray observations}

We have cross-referenced our FLAMES targets in the N11 and NGC\,346
fields with the x-ray sources from \citet{n03,n04,n04b}.  We find that
NGC\,346-067 is $\sim$1.9$''$ from source \#6 in the
\citet{n03} study.  In fact, it is the same counterpart suggested by
them, i.e. MA93\#1038 \citep{ma93}.  The tabulated separation between
optical and x-ray positions is 0.5$''$, suggesting a small offset
between their astrometric solution and that from the EIS data.
Widening our search radius from 3$''$ to 5$''$ we find one further
potential cross-match: NGC\,346-078 (an early B-type binary) is 3.8$''$
from their source \#17; although in the opposite sense to NGC\,346-067
and \#6 and it seems unlikely that this is a valid match.

Similar searches with the {\it XMM-Newton} sources reported by
\citet{n04} in N11, and additional sources in NGC\,346 from \citet{n04b}, 
revealed no further cross-matches with our FLAMES sample, suggesting
that none of our observed stars (excepting NGC\,346-067) are
particularly strong x-ray sources.

\section{Comments on individual stars}\label{stars}

There is a tremendous amount of new spectroscopic information
available in our LMC and SMC fields.  In nearly all cases the FLAMES
spectra offer significantly better-quality data than previously.  Also
of note is that many stars, particularly in the NGC\,330 and NGC\,2004
fields, have not been observed before spectroscopically.

In the following sections we discuss spectra with interesting or
peculiar morpholgies.  In general we do not address the specific
details of detected binaries, restricting ourselves to simply
indicating their binarity in Tables~\ref{346}, \ref{330}, \ref{lh910},
and \ref{2004}.  In many cases the FLAMES spectra are sufficient to
determine orbital periods and, for some systems, offer an insight into
the physical nature of the individual components.  A comprehensive
treatment of the binary spectra will be presented elsewhere.  Emission-line
(i.e. Oe and Be) stars are discussed separately in Section \ref{bestars}.

\subsection{NGC\,346-001}
NGC\,346-001 is the well-studied star Sk~80 \citep{sk68}, also
identified as AzV~232 \citep{av75, av82} and MPG~789.  It was
classified as O7~Iaf$+$ by \citet{w77}, with the f$+$ suffix employed
to denote the strong Si~\4 \lam4116, N~\3 \lam4634-40-42, and He~\2
\lam4686 emission features; indeed, Sk~80 serves as the principal O7
supergiant in the \citet{wf90} spectral atlas.  A detailed atmospheric
analysis of a VLT-UVES spectrum of this star was given by \citet{paul}.

\begin{figure}
\begin{center}
\includegraphics[width=9cm]{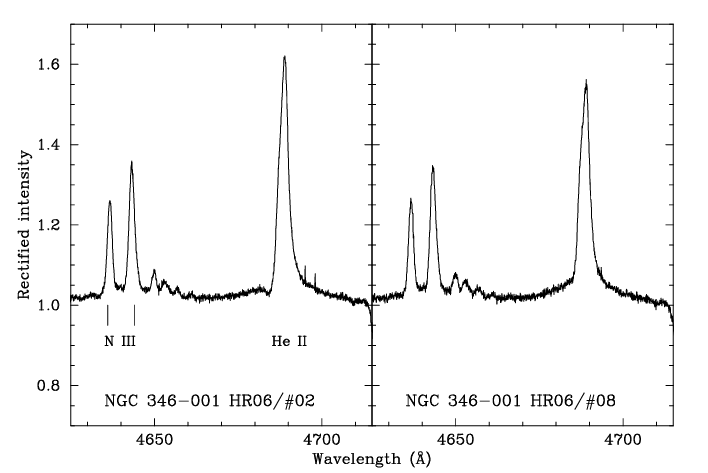}
\caption{Variations in the He~{\scriptsize II} \lam4686 line seen in NGC\,346-001.
The left panel shows exposure HR06/\#02, with the right panel showing
HR06/\#08, note that the intensity of the N~{\scriptsize III}  \lam4634, 4640-41 
emission is identical whereas the He~{\scriptsize II} \lam4686 morphology and intensity
differs.}
\label{346_001}
\end{center}
\end{figure}

Close inspection of the individual FLAMES-Giraffe spectra reveals
several features that suggest this object to be a binary.  Perhaps the
most distinctive of these is shown in Figure \ref{346_001}.  In the
first five exposures there is a weak absorption feature in the
blueward wing of the He~\2 \lam4686 emission line (left-hand panel of
figure).  Compare this to the HR06/\#06, \#07 and \#08 frames in which
it is absent, with slightly weaker \lam4686 emission (right-hand
panel).  From the same observations, a small shift ($\sim$10 \kms) is
seen in the He~\1 \lam4713 line.  A similar, though less significant
effect to that seen in the \lam4686 line, is also seen in the blueward
wing of the H$\alpha$ profile (we do not consider the core intensity
given the problems of nebular subtraction).  Inspection of the spectra
from the HR05 observations (which also include He~\2 \lam4542) reveals
a shift of $\sim$10 \kms between the two epochs, which is also
mirrored in the He~\1 \lam4471 line.  That these shifts are seen
in the mainly photospheric He~\1 features suggests there
is a companion and that we are not seeing evidence of wind
variability.

Compared to the VLT-UVES data from \citeauthor{paul}, the resolving
power from the high-resolution mode of FLAMES-Giraffe is roughly a
factor of two lower, but the signal-to-noise of the new spectrum is
$\sim$400.  Aside from consideration of binarity, we note the presence
of two weak emission lines that we attribute to N~\3, with rest
wavelengths of \lam3938.5 and \lam4379.0 (Dr.~F. Najarro, private
communication).  Upon reinspection of the UVES spectrum from
\citeauthor{paul} these features can also be seen, but are less
obvious due to the inherent problems associated with blaze removal
etc.  The \lam4379.0 line can also be seen in the high-resolution
spectrum presented by \citet{wal95}, although there are some
comparable artifacts nearby; the line is also seen in emission in
AzV~83 \citep{wal00}.

\subsection{NGC\,346-007}
High-resolution optical spectra of NGC\,346-007 (MPG~324) have been
published by \citet[][NGC346\#6]{wal95} and \citet{wal00}.  In the
FLAMES spectrum the Si~\4 \lam4116 feature is very weakly in
emission (such that it was within the noise level of previous spectra)
and therefore the \citeauthor{wal00} classification of O4~V((f)) is
revised slightly to O4~V((f$+$)) to reflect this. 
Small radial velocity shifts are seen in some lines in the
spectra of NGC\,346-007 (of order 20-30~\kms) suggesting it as a
single-lined binary.

\subsection{NGC\,346-013}
\begin{figure}
\begin{center}
\includegraphics[width=9cm]{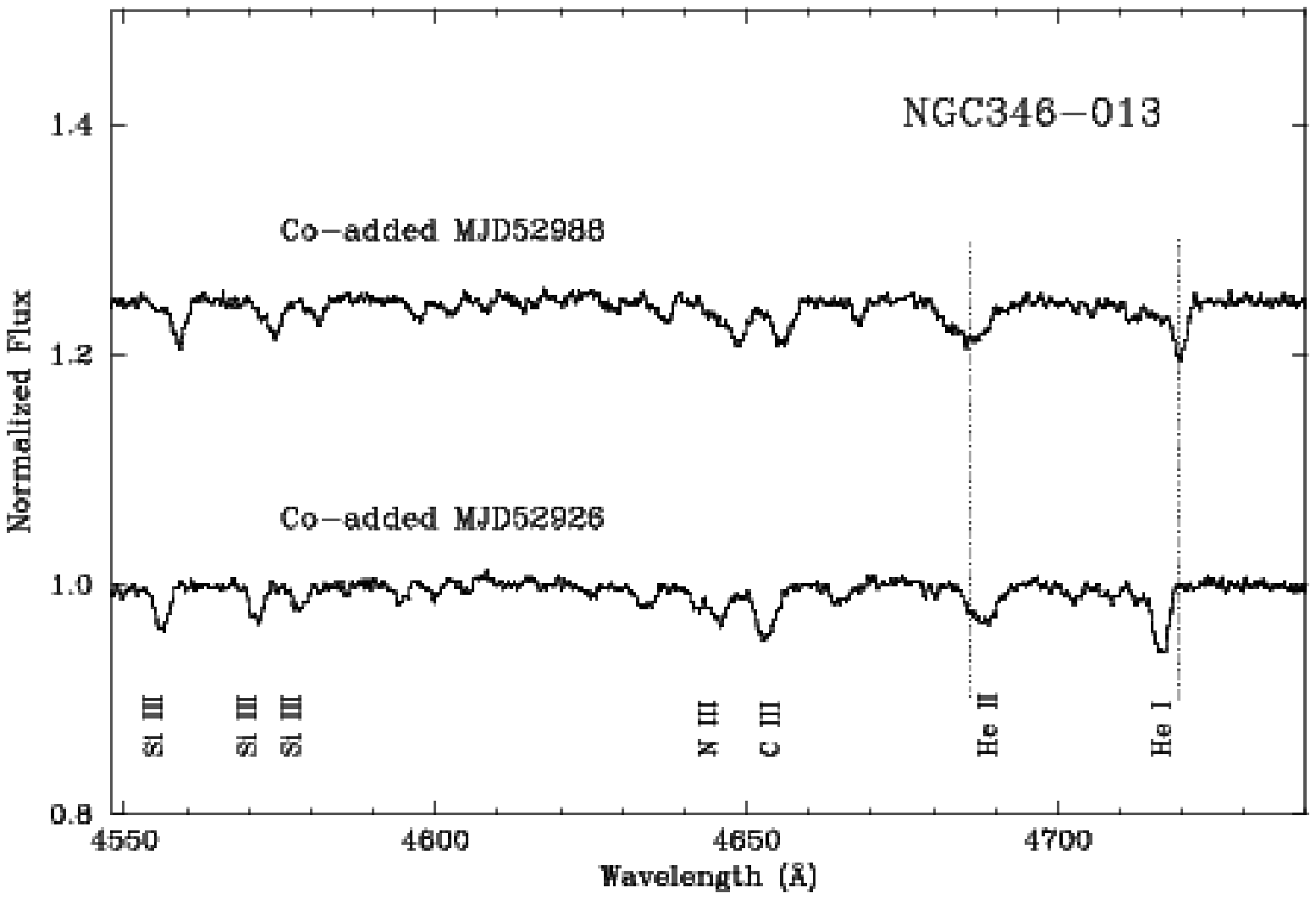}
\caption[]{The HR06 observations of NGC\,346-013.  The spectra from the 
two epochs have been co-added, and then 11-pixel median filtered to
aid clarity.  Identified lines are Si~{\scriptsize III} \lam\lam4552-68-75; 
N~{\scriptsize III} \lam\lam4634, 4640-42; C~{\scriptsize III} \lam4650; 
He~{\scriptsize II} \lam4686; He~{\scriptsize I} \lam4713.
The vertical dotted lines highlight the wavelengths of the He~{\scriptsize I} and
{\scriptsize II} lines in the second batch of observations; these are seen to move
in an opposite sense between epochs.  }\label{013}
\end{center}
\end{figure}

This object is one of the most intriguing in the FLAMES survey.  At
first glance the spectrum is that of an early B-type star with a
spectral type of B1 or B1.5, but there is also anomalous
absorption from He~\2 \lam4686.  Inspection of the individual spectra
reveals significant velocity shifts in the He~\1 and Si~\3 absorption
lines, with a maximum amplitude over the time coverage of the FLAMES
observations of approximately 400~\kms.   Other lines such as the C~\3
and N~\3 blends move in the same sense.  Interestingly, the
He~\2 \lam4686 moves in the {\it opposite} sense to the other
features and with a smaller amplitude, suggesting that it is the more
massive object in the system (and presumably hotter by virtue of the
4686 line).  These features are illustrated in Figure~\ref{013}.

We lack other diagnostics of the companion associated with the He~\2
\lam4686 feature in the current data.  The He~\1 \lam4471 absorption
appears to be marginally double-lined in the first three HR05
observations, but it is within the noise level preventing accurate
characterization.  No strong absorption is seen from He~\2 \lam4542 in
the HR04 data suggesting a narrow temperature range for the companion.
We note that NGC\,346-013 does not appear in the x-ray catalogue of
\citet{n03}.  A more detailed treatment of its orbital properties will
be presented elsewhere.  Further spectroscopic monitoring of this
object is clearly a high priority.

\subsection{NGC\,346-026, \& NGC\,346-028}
High-resolution echelle spectra of NGC\,346-026 (MPG~012) and
NGC\,346-028 (MPG~113) were presented by \citet{wal00}, with a
detailed discussion of their morphologies.  For NGC\,346-028 we adopt
the \citeauthor{wal00} classification of OC6~Vz. In the case of
NGC\,346-026 we prefer the unique classification of B0 (N str), over
the O9.5-B0 (N str) from \citeauthor{wal00}  The line that is least
consistent with a B0-type is He~\2 \lam4200, which is marginally
stronger in NGC\,346-026 than in $\upsilon$~Ori, the B0~V standard
\citep{wf90} -- given the `N~str' remark, perhaps part of the
explanation for this lies with a stronger \lam4200 N~\3 line.
Interestingly, the Si~\4 lines in the spectrum are in reasonable
agreement with those in $\upsilon$~Ori, which is surprising given the
lower metallicity of the SMC.  This point strengthens the suggestion
by \citeauthor{wal00} that luminosity class IV might be more befitting
of the star, {\it a priori} of the luminosity and gravity results
found by \citet{jc03}.  Thus, we classify the star as B0 IV (N str).

\subsection{N11-020}
The spectrum of N11-020 displays broad, asymmetric H$\alpha$ emission,
combined with strong N~\3 \lam4634-41 and He~\2 \lam4686 emission.
Such morphology is similar to that seen in $\zeta$~Pup \citep{wf90}.
The FLAMES spectrum appears slightly cooler than $\zeta$~Pup, resulting
in the classification of O5~I(n)fp.  There are clear variations in
both the H$\alpha$ and He~\2 \lam4686 lines, which could be indicative
of wind variability.  However, close inspection of other lines such as
He~\2 \lam4542 and H$\delta$ reveal subtle morphological changes
indicative of a companion, suggesting that N11-020 is a
further binary.

\subsection{N11-026, N11-014 \& N11-091}
\begin{figure*}
\begin{center}
\includegraphics[width=12cm]{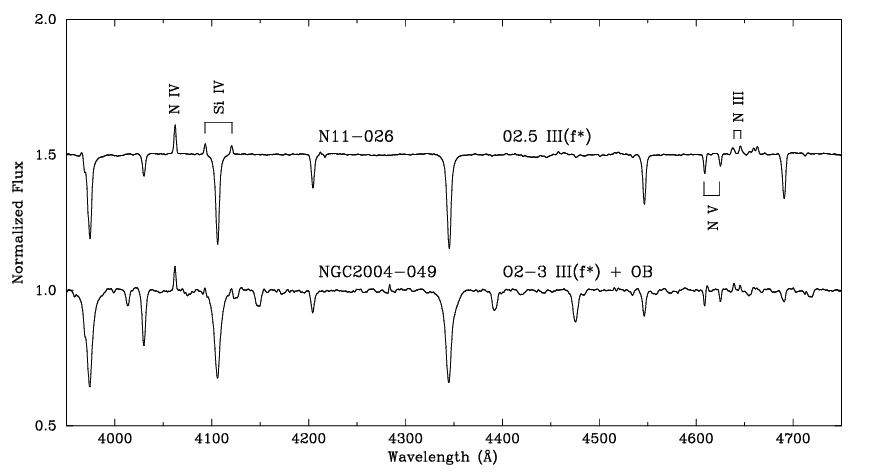}
\caption{Two early O-type stars: N11-026 and NGC\,2004-049.  The lines identified in 
N11-026 are, from left to right, N~{\scriptsize IV} \lam4058; Si~{\scriptsize IV} 
4089-4116; N~{\scriptsize V} \lam\lam4604-4620; N~{\scriptsize III} 
\lam\lam4634-4640-4642.  For clarity the spectra have been smoothed by a 
1.5 \AA~{\sc fwhm} filter.}
\label{o2stars}
\end{center}
\end{figure*}

The spectrum of N11-026 is shown in Figure~\ref{o2stars} and is
between the O2 and O3-type standards published by \citet{w02,w04}.
Thus we employ an intermediate type of O2.5~III(f$^\ast$).  N11-026 is
$\sim$4.5$'$ to the north of LH10 (see Figure~\ref{fchart_n11}) and
the photograph of N11 from \citet{w92} shows a likely
ionization front just to the east of the star.  In fact, the FLAMES
targets adjacent to \#026 offer a good illustration of the rich
star-formation history of the region.  Cross-matching our targets
again with the image from \citet{w92}, N11-014 (to the west of
\#026) is a B2 supergiant in a more rarefied region, and N11-091 is an O9-type
spectroscopic binary (with what appears to be a B-type secondary)
lying in a dense knot of gas.

N11-026 has the third largest velocity in the N11 targets, with $v_r
=$~330~\kms~(with standard deviation, $\sigma =$ 5).  The median
result for the N11 stars with measured radial velocities is 295~\kms
suggesting N11-026 as a run-away object,
cf. the usual critierion of $\Delta v_r$ $\sim$40~\kms \citep{b61}.
N11-026 is $\sim$1$'$ away from N11-091, which at the distance of
the LMC corresponds to $\sim$14~pc.  An ejected young star
could conceivably cover this distance in its short lifetime \citep[such arguments
are discussed in their definition of a field star by][]{m95}.  Indeed,
in 2~Myr the star may have even travelled the $\sim$70~pc from the
northern edge of LH10.

\subsection{N11-028}
N11-028 is the primary star of N11A, a well-studied nebular knot to
the north-east of LH10 (N11B).  \citet{hm85} suggested that the
principal source of ionization was a star with T$_{\rm eff}
\sim$44,000~K, i.e. an O-type star.  This was confirmed by P92, 
who found an embedded primary (their star 3264) with two faint
companions.  P92 show the spectrum of 3264 in their Figure~5, 
classifying it as O3-6~V.  The uncertainty in the classification
arises from the strong nebular emission in the spectrum, as one would
expect from the near-IR study by \citet{ik91} who concluded that N11A
was dominated by such emission.

More recently, impressive images of N11A from the Wide Field Planetary
Camera 2 (WFPC2) on the {\it Hubble Space Telescope} were published by
\citet{hm01}.  Their imaging (see their Figure~2) revealed at least
five objects in the core of N11A, which they labelled as numbers 5 to 9.
Their star \#7 was the brightest with Str$\ddot{\rm o}$mgren~$v =$
14.59, with three of the others (5, 6, and 9) fainter than 18th
magnitude.  

In the WFI imaging used to select our FLAMES targets, N11A appears as
an unresolved, dense blob, with the fibre placed on the bright core.
Our photometric methods here are relatively simple and were primarily
concerned with target selection.  The multiplicity in N11A, combined
with the effects of the nebulosity, highlight that a more tailored
photometric analysis is warranted for a star such as this, beyond the
scope of our programme (and imaging).  As such, the quoted magnitude
in Table~\ref{lh910} is brighter than that given by P92 and
\citet{hm01}.

The MEDUSA fibres used with FLAMES have an on-sky aperture of 1.2$''$.
From comparisons with the figure from \citeauthor{hm01}, it is likely
that the MEDUSA fibre was not centered perfectly on the principal star (owing to
the resolution of the WFI image).  It is also possible that the FLAMES
spectrum may include some contribution from their star \#8, which is
two magnitudes fainter than the primary.

\begin{figure*}
\begin{center}
\includegraphics[width=12cm]{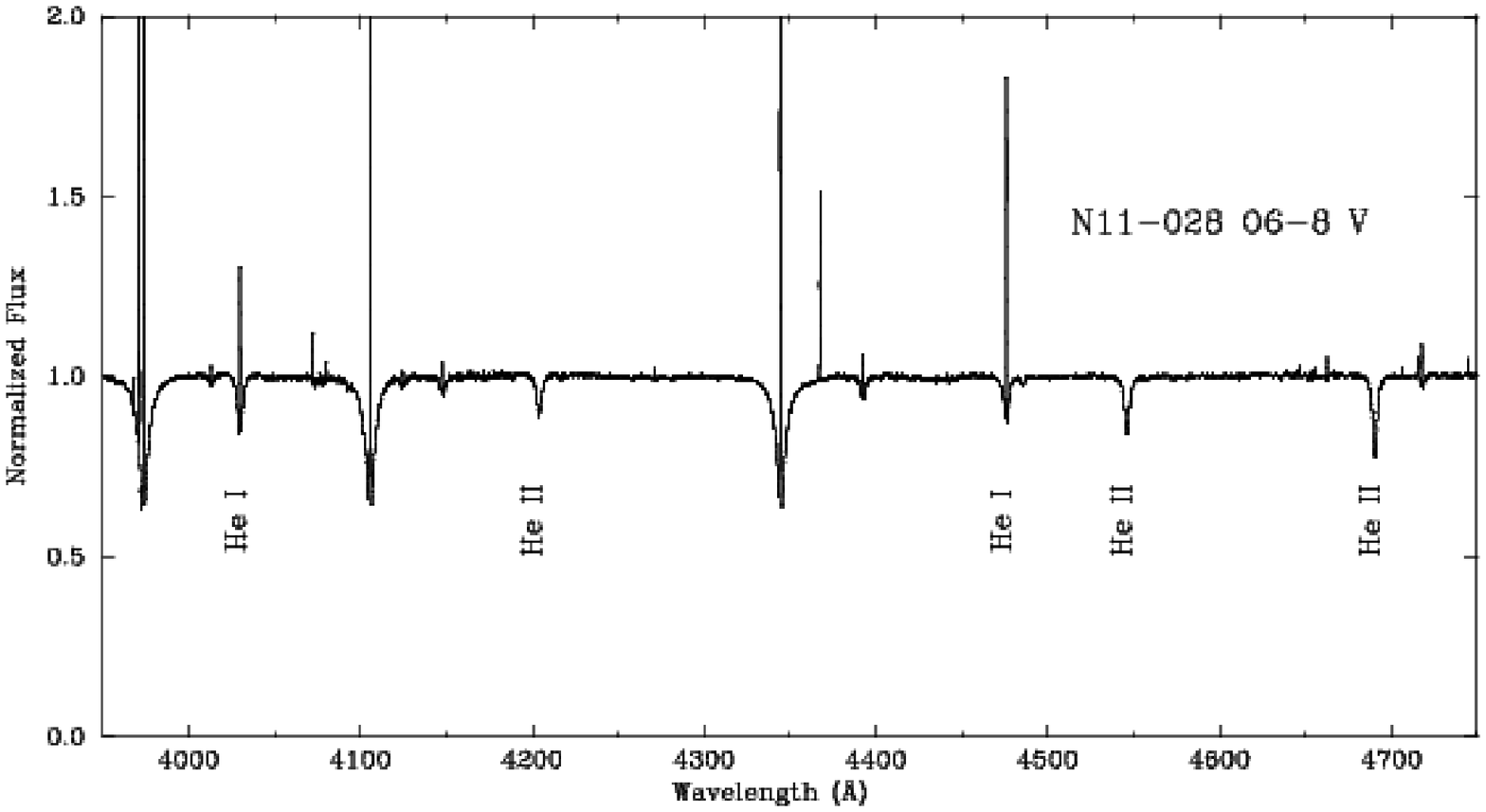}
\caption{The FLAMES spectrum of the principal star in N11A.  The 
He~{\scriptsize II} lines identified are, from left to right, \lam\lam4200, 4542, 4686;
the two strongest He~{\scriptsize I} lines at \lam\lam4026, 4471 are also labelled.
Although the nebular spectrum somewhat dominates, there is strong
absorption seen in both He~{\scriptsize I} lines, necessitating a later spectral
type than that adopted by \citet{p92}.  For clarity the spectrum has
been smoothed using a 7-pixel median filter.}
\label{n11a}
\end{center}
\end{figure*}

The FLAMES spectrum of N11-028/Parker 3264 is shown in
Figure~\ref{n11a}, and is classified here as O6-8~V, cf. O3-6 V from
P92.  We see no obvious evidence of binarity, or of other features
from a possible companion.  The absorption at He~\1 \lam4026 is larger
than that seen in He~\2 \lam4200, thereby requiring a spectral type of
O6 or later.  The uncertainty in the current classification arises
because of the He~\1 \lam4471 line, in which the degree of infilling
is unclear.  Obviously such a revision of spectral type will have a
significant effect on the perceived ionizing flux in the nebula.
Indeed, \citet{hm01} concluded that an O7.5 or O8 main-sequence star
could account for the observed ionization, commensurate with our
classification.  One interesting feature is the weak Mg~\2 \lam4481
absorption sometimes seen in mid O-type spectra \citep[e.g.][]{mwf91},
attributed by \citet{m72} to deviations from the assumptions of local
thermodynamic equilibrium, i.e. non-LTE effects.

\subsection{N11-038}
There is a striking similarity between the spectrum of this star and
that of AzV~75 in the SMC \citep{wal00}, with the spectrum here
classified as O5 III(f$+$).  Interestingly this star was observed by
P92 (their star 3100) and classified as `O6.5 V((f)):'.  The
FLAMES data shows strong He~\2 absorption at \lam4200, cf. the P92
spectrum in which it is relatively weak.  The relative ratio of He~\1
\lam4471 to He~\2 \lam4542 also appears different in the P92 spectrum
in comparison to the FLAMES data; although we see no evidence for
binarity in our spectra, variability of this object cannot be ruled
out.

\begin{figure*}
\begin{center}
\includegraphics[width=12cm]{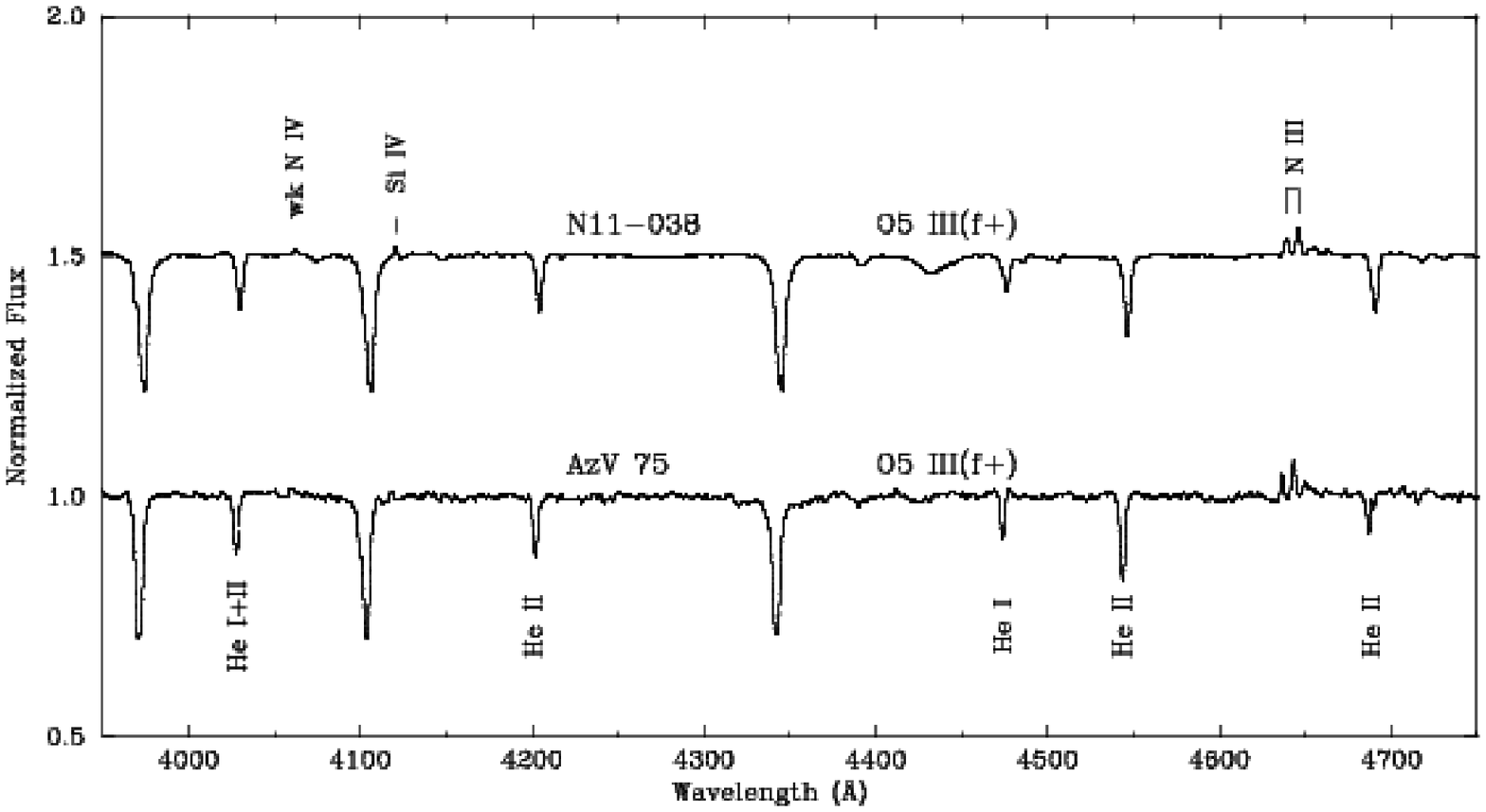}
\caption{The FLAMES spectrum of N11-038 compared with that of 
AzV~75 in the SMC, taken from \citet{wal00}.  The lines identified are
the same as in Figures~\ref{o2stars} and \ref{n11a}.  For clarity the
spectra have been smoothed by a 1.5 \AA~{\sc fwhm} filter.}
\label{n11_038}
\end{center}
\end{figure*}

N11-038 and AzV~75 are shown together in Figure~\ref{n11_038}.  Note
the very weak N~\4 emission feature in N11-038, also consistent with
an early O-type classification.  The stronger N~\3 emission and weaker
He~\2 \lam4686 absorption in AzV~75 suggest a more luminous star than
N11-038, and these are bourne out by the brighter $V$ magnitude
(taking into account the different distances and reddening).  However
we argue that both stars fall into the same morphological bin, cf. the
standards in \citet{wf90}.

\subsection{N11-040}
Classification of this spectrum was relatively difficult, resulting in
a final type of B0:~IIIn.  There is apparent emission at the edge of
both wings in the H$\alpha$ profile.  There are also indications of
binarity, with the He~\2 lines appearing to have a slightly larger
radial velocity than the other features, although the broadened lines
make precise measurements very difficult.

\subsection{N11-045}
Classified here as O9-9.5~III, the hydrogen and helium spectrum of
N11-045 is very similar to that of $\iota$~Ori (O9 III) from
\citet{wf90}.  However, we do not rule out a later O9.5 type; very weak
Si~\3 absorption is seen and the Si~\4 lines are slightly stronger
than one would expect at O9, especially when one considers the reduced
metallicity of the LMC.

\subsection{NGC\,2004-049}
The spectrum of NGC\,2004-049 is shown in Figure~\ref{o2stars} and is
classified here as O2-3 III(f$^\ast$)$+$OB; the spectrum is very
similar to that of LH10-3209 \citep{p92,w02}.  The N~\4 emission has a
greater intensity than the N~\3 lines, necessitating an early
type of O2-3 for the primary.  The secondary component in
NGC\,2004-049 appears cooler than in LH10-3209, with the He~\1 \lam4471
absorption greater than that from He~\2 \lam4542.  Given the
relatively weak He~\2 \lam4686 in the composite spectrum, one can
speculate there the secondary does not contribute significantly to
this line and that the companion is of early B-type.

\subsection{NGC\,2004-057}
Some 5$'$ east of the cluster, NGC\,2004-057 \citep[Brey 45;][]{brey} is a
Wolf-Rayet star classified as WN4b by \citet{ssm96}, in which the `b'
denotes broad emission at He~\2 \lam4686.  Given the broad nature of
the emission features, and the relatively short spectral range of each
Giraffe wavelength-setting, it is not possible to rectify the data for
NGC\,2004-057 reliably.  However, we retain the star in Table~\ref{2004} as
it highlights interesting objects found in the field.  The FLAMES data
appear to reveal small changes in the structure of emission lines such
as He~\lam4542, perhaps indicative of wind variability.

\section{Emission-line stars}\label{bestars}

\subsection{Be-type stars}

\begin{figure*}
\begin{center}
\includegraphics[width=12cm]{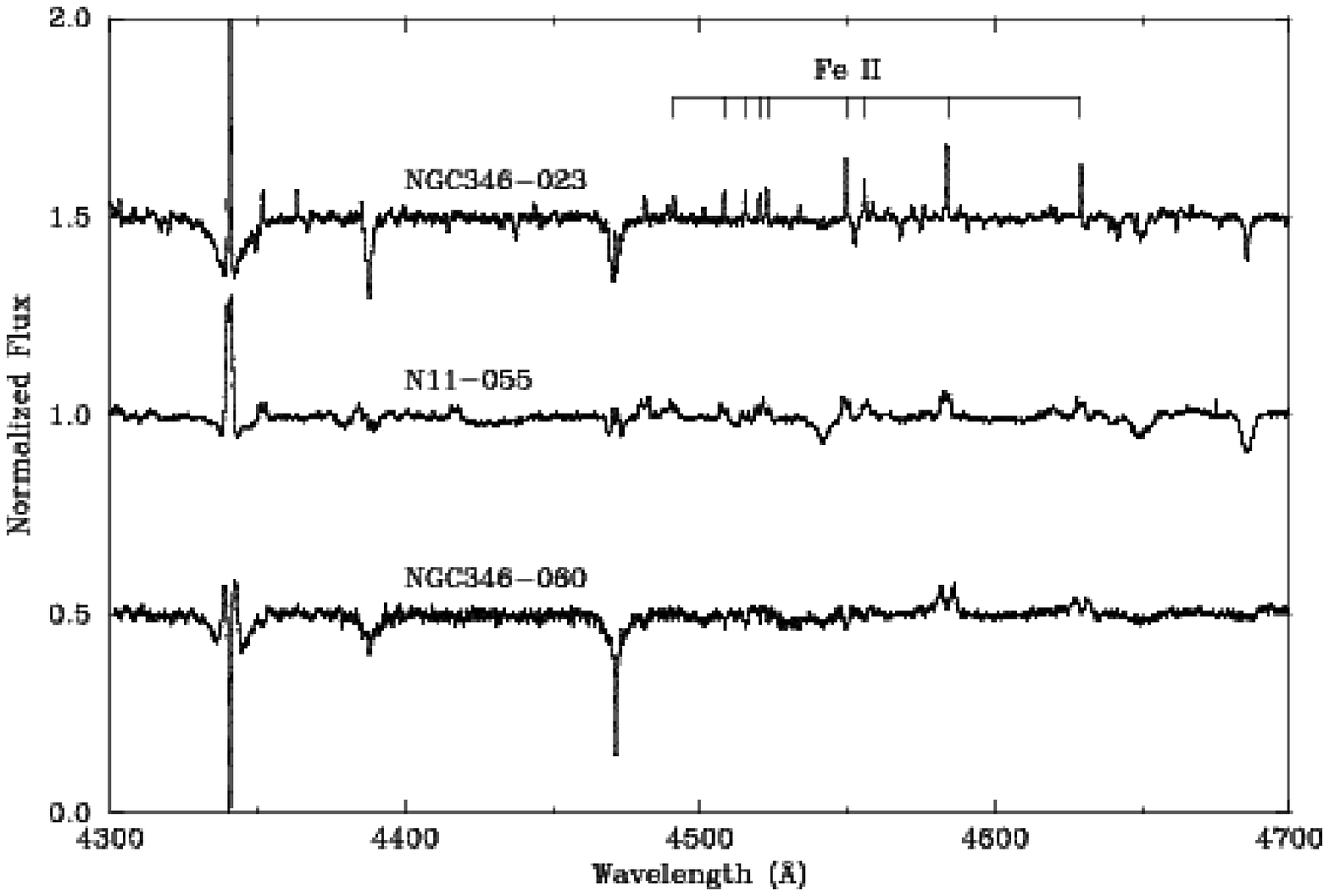}
\caption{Examples of Be/Oe-type spectra with Fe~{\scriptsize II} emission lines.  
The Fe~{\scriptsize II} lines identified in NGC\,346-023 are
\lam\lam4491, 4508, 4515, 4520, 4523, 4549, 4556, 4584, and 4629.  For
clarity the spectra have been filtered by a 7-pixel median filter and
to ease the comparison between SMC and LMC spectra they are corrected
for their tabulated radial velocities to rest wavelengths.}
\label{befe}
\end{center}
\end{figure*}

The H$\alpha$ profile of NGC\,346-023 has asymmetric, double-peaked
emission, with infilling in the other Balmer lines.  The absorption
spectrum suggests a fairly hot star and is classified as B0.2.  Of
particular note are numerous Fe~\2 emission lines, together with
emission from other metallic species, e.g. Mg~\2 (\lam4481) and Si~\2
(\lam\lam6347, 6371).  The \lam4300 to \lam4700~\AA\/ region of its
spectrum is shown in Figure~\ref{befe}.  Permitted emission lines from
Fe~\2 are a well-known phenomenon in some Be-type stars
\citep[e.g.][]{w52,s82}, with more than a dozen found in the
Magellanic Clouds by \citet{m95}.  The most striking example of these
stars in the FLAMES survey is NGC\,346-023.  Also shown in
Figure~\ref{befe} is another example, NGC\,346-060, in which
twin-peaked emission lines are seen, likely indicating a different
viewing angle.  Spectra with Fe~\2 in emission are identified in
Tables~\ref{346}, \ref{330}, \ref{lh910}, and \ref{2004} as `Be-Fe'.
The interesting Oe-type star, N11-055 (see Section~\ref{oestars}) is
also shown in Figure~\ref{befe}.

Prompted by the recent study in NGC\,2004 by \citet{mhf} we measured
the equivalent widths of H$\alpha$ emission, EW(H$\alpha$), for each
of the Be-type FLAMES spectra.  These values were then compared with
the presence or absence of Fe~\2 emission.  Given the problems of
nebular subtraction with multi-fibre data (which introduces a variable
uncertainty, proscribing precise measurements in Be-type stars), we do
not tabulate the EW(H$\alpha$) values, and limit ourselves to a
general discussion of the results.  \citet{h87} reported EW(H$\alpha$)
= 7~\AA\/ as the minimum for Fe~\2 to be seen in Galactic Be stars.
With the lower metallicity in the Clouds one might expect to find
that the EW(H$\alpha$) threshold for Fe~\2 emission increases.
\citet{mhf} find that Fe~\2 emission is seen in their LMC targets when
EW(H$\alpha$) $>$ 20~\AA, with the inference that this higher
threshold reflects the metallicity difference.  

In our NGC\,346 and
NGC\,330 fields we find the threshold for Fe~\2 emission is
EW(H$\alpha) >$15-20 \AA, above which Fe~\2 emission is seen in all
stars (except in NGC\,346-045).  Below this threshold Fe~\2 emission
lines are not seen except in NGC\,330-031 [EW(H$\alpha$) = 8.5
\AA~($\pm$1)] which has very weak Fe~\2 emission at \lam\lam4549,
4556, and 4584.
Five of the six Be stars in N11 have Fe~\2 emission lines (N11-056 is
the exception with very weak, double-peaked H$\alpha$ emission).  Four
of these five stars have EW(H$\alpha$) $>$ 20~\AA, with the fifth,
N11-078, having EW(H$\alpha$) $\sim$ 4~\AA\/ (although as in
NGC\,330-031, the Fe~\2 features are very weak).  Lastly, in NGC\,2004
both NGC\,2004-023 and NGC\,2004-035 have strong (i.e. EW $>$ 20~\AA) H$\alpha$
emission, accompanied by Fe~\2.  Again we find one star (NGC\,2004-025)
with EW(H$\alpha$) below 20~\AA\/ (in this case $\sim$12.5) with very
weak Fe~\2 emission.  

In summary, Fe~\2 is seen in emission in most stars for which
EW(H$\alpha$) $>$ 15~\AA; a strong H$\alpha$ emission profile does not
guarantee Fe~\2 emission, but it is very probable.  Also, the
threshold isn't completely exclusive, below it we see weak (but clear)
emission in the stronger Fe~\2 lines in three stars.  Following this
investigation of the LMC and SMC spectra, we revisited the 12 Be-type
stars from Paper~I to check for Fe~\2 emission.  In Paper~I we did not
pay particular attention to Fe~\2 emission, apart from in the
Herbig-type star 6611-022.  In fact, there are three stars from
Paper~I with EW(H$\alpha$) $>$ 20~\AA\/ (3293-011, 4755-014, and
4755-018) and Fe~\2 emission is seen in each of these.  There are two
further stars with moderate H$\alpha$ emission, 6611-010 and 6611-028,
for which EW(H$\alpha$) = 15.5 and 13.5 ($\pm$1) respectively.  There
appears to be very weak Fe~\2 emission in both of these spectra,
analogous to the stars with moderate H$\alpha$ emission in our LMC and
SMC sample.  The weak emission in 6611-010 \citep[W503,][]{w61} was
also noted by \citet{hmsm}, leading to their classification of the
spectrum as a Herbig-type object.

With the benefit of new instrumentation and detectors we can detect
Fe~\2 emission that may have gone un-noticed previously. It seems
clear that irrespective of environment, if EW(H$\alpha$) $>$ 20~\AA,
it is very probable that Fe~\2 emission will also be observed.
The statement from \citet{h87} was that `one should not expect to
find {\it notable} amounts of Fe~\2 emission for stars with
EW(H$\alpha$) $<$ 7~\AA.'   This remains true and, if anything, should be
revised upwards slightly, depending on one's definition of `notable.'
However, we dispute the claim by \citet{mhf} that their observations
in NGC\,2004 display the expected offset arising from the lower
metallicity - although we have not measured the intensity of the Fe~\2
emission, we see no obvious trend with metallicity for the {\it
presence} of Fe~\2 emission.

Given that both NGC\,346 and N11 are relatively young regions it does suggest
the question of whether these stars are classical Be-type stars, or are they
associated with younger (pre-main-sequence) objects.  Indeed, the cores of 
both regions are younger than the 10~Myr threshold given by \citet{ft00} for
classical Be-type stars to form.

In NGC\,346 the three Be-Fe stars closest to the centre of the cluster
are NGC\,346-023, 036 and 065, all of which are on the periphery of
the central ionized region (see Figure~\ref{fchart_346}).  Similarly
in N11, the Be-Fe stars are not in the obviously dense star-forming
regions.  Given that also we find such stars in the NGC\,2004 and
NGC\,330 pointings (that sample the more general field population) it
seems most plausible that the Be-Fe stars are associated with
classical Be-types, rather than younger objects.  However, it should
be noted that \citet{rmsmc} find evidence for coeval
O-type stars beyond the ionized region in NGC\,346 -- detailed
analysis of the Be-Fe stars is required to test whether those in the
NGC\,346 field are actually coeval with the young O-stars, or whether
they are older.

\subsection{Shell stars}
\begin{figure*}
\begin{center}
\includegraphics[width=12cm]{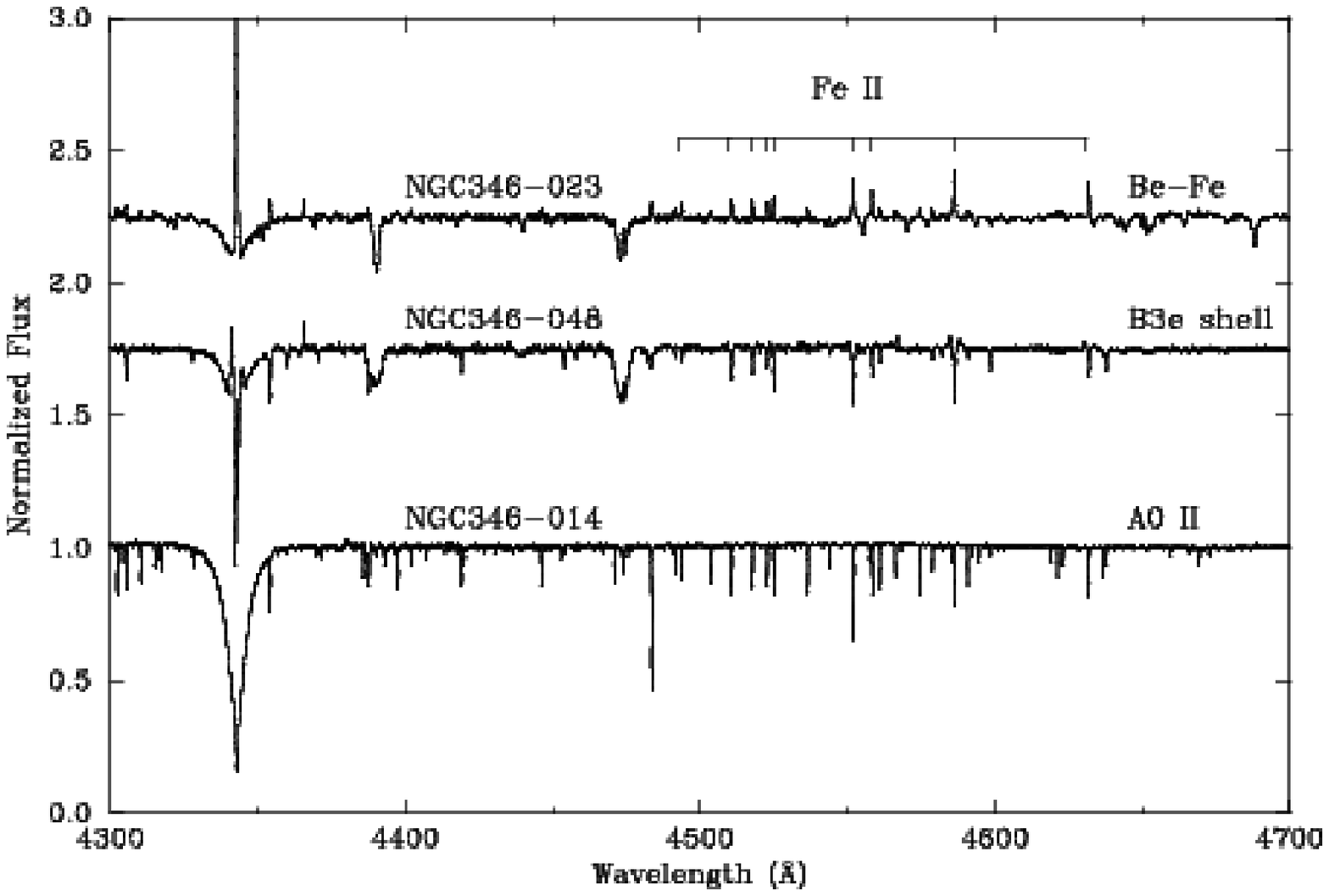}
\caption{A comparison of Be-type stars with Fe~{\scriptsize II} emission and absorption lines.
The Fe~{\scriptsize II} lines identified are the same as in Figure~\ref{befe}, and the spectra
have also been filtered by a 7-pixel median filter.  The spectrum of 
NGC\,346-014 is included to highlight that the Fe~{\scriptsize II} lines are indicative
of a much cooler temperature, comparable to that seen in early A-type stars.}
\label{befe2}
\end{center}
\end{figure*}

In Figure \ref{befe2} we show the spectrum of NGC\,346-048, which permits
an elegant comparison with NGC\,346-023.  The H$\alpha$ profile of NGC\,346-048
is double-peaked, leading to moderate `shell' effects in the H$\gamma$
line shown in Figure~\ref{befe2}.  Such a spectrum is thought to
originate from absorption by a cooler disk in front of the star
\citep[e.g.][]{h95,hv00} and also leads to the Fe~\2 absorption
seen in the figure, uncharacteristic of a star classified as
B3-type and more comparable to those seen in the sharp-lined, A0~II
spectrum of NGC\,346-014.  Note the weak emission in the wings of the
\lam4584 Fe~\2 line in NGC\,346-048.  Similar absorption from Fe~\2 is also
seen in NGC\,330-093, although the signal-to-noise of the data is lower
given the fainter magnitude of the star.

\subsection{Oe-type stars}\label{oestars}

\subsubsection{N11-055: An `Oe-Fe' star}

This star has a very complex spectrum, a section of which is shown in
Figure~\ref{befe}.  Emission is seen in all of the observed Balmer
lines, and most of the He~\1 lines are infilled by a twin-peaked
profile (e.g. \lam4026), with the \lam4713 line seen solely in
(twin-peaked) emission.  A large number of Fe~\2 emission lines are
also seen.  We find no strong evidence for binarity, with all the
features apparently consistent with one radial velocity; the mean
radial velocity from the 3 He~\2 lines is 293~\kms.  We classify the
star here as O7-9~IIIne, with the uncertainty in spectral type arising
due to the significant infilling in the He~\1 lines.  Assuming that
there is no significant companion, the intensity of the \lam4200 and
\lam4542 He~\2 lines suggests that the later-type is most likely the best
description of the spectrum.  Also, the He~\2 \lam4686 line appears
too weak for the star to be classified as a dwarf, hence the adoption
of class III.

In a sense, this star appears as a much less extreme version of iron
stars such as HD~87643 \citep{wf00}.  However, the iron stars tend to
show strong P Cygni features in some of their lines, which are not
seen here.  The H$\alpha$ emission in N11-055 is strong and symmetric,
suggesting this star might be more closely related to Be-type stars, with
HD~155806 a likely Galactic counterpart \citep{w80}.

\subsubsection{NGC\,346-018}

NGC\,346-018 (MPG~217) is classified as O9.5 IIIe.  The spectrum
displays very strong and broad H$\alpha$ emission, with the
higher-order lines of the Balmer series somewhat in-filled.  The He~\1
\lam6678 line is strongly in emission and there is evidence of
significant infilling in He~\1 \lam4471 (and perhaps other He~\1 lines
in the blue region); we have taken this into account when classifying
the spectrum, as discussed by \citet{nsb05}.

\subsubsection{NGC\,330-023}\label{330_023}
\begin{figure*}
\begin{center}
\includegraphics[width=12cm]{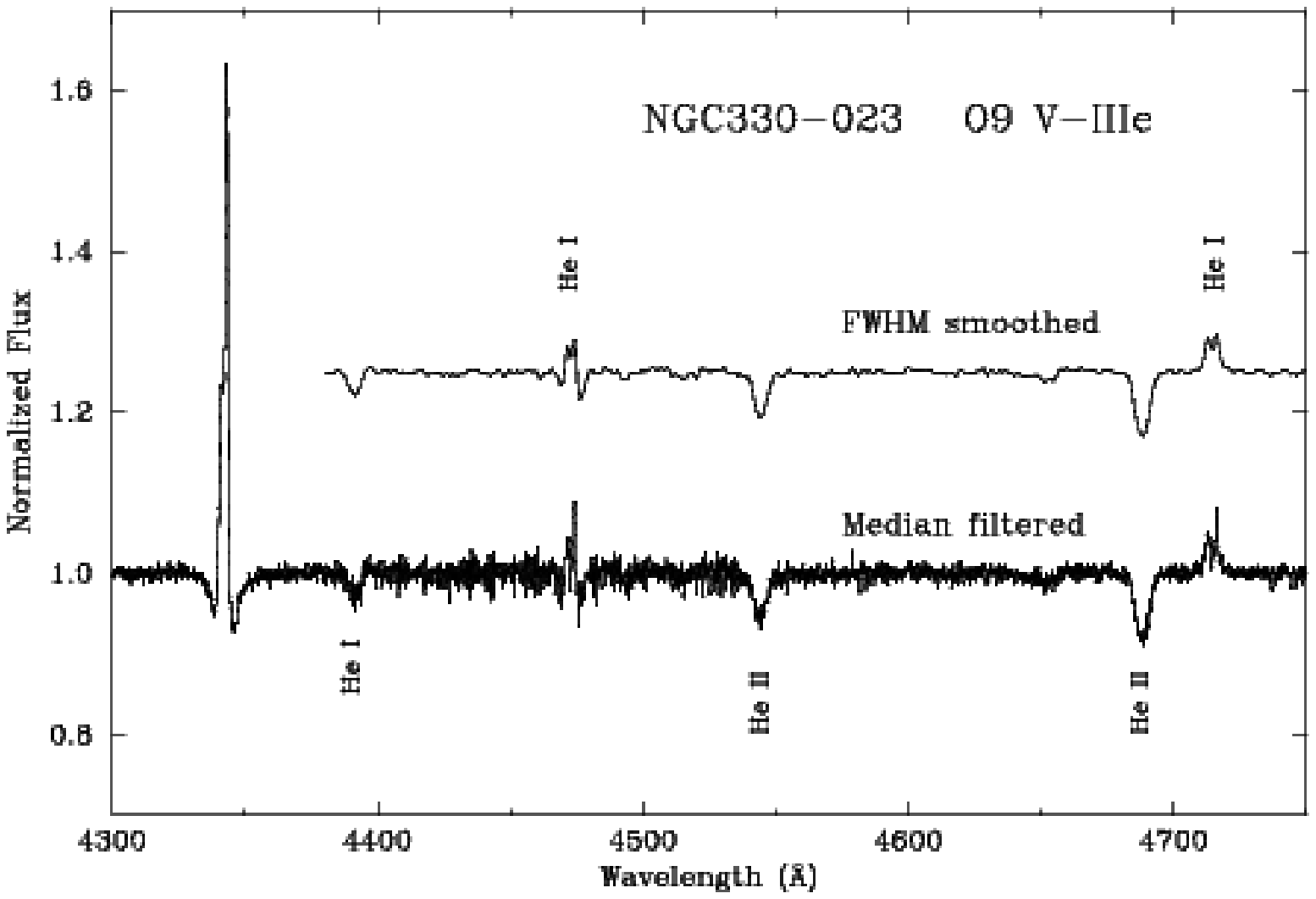}
\caption{A section of the blue-region spectrum of NGC\,330-023, highlighting
the double-peaked emission in the He~{\scriptsize I} lines at \lam4471 and
\lam4713.  The lower spectrum is at full-resolution, but has been
filtered to remove cosmic rays using a 7-pixel median filter.  The
upper spectrum has been smoothed with a 1.5 \AA~{\sc fwhm} filter.
The lines marked in the lower spectrum are He~{\scriptsize I} \lam4388, and 
He~{\scriptsize II} \lam\lam4542, 4686.}\label{023}

\includegraphics[width=12cm]{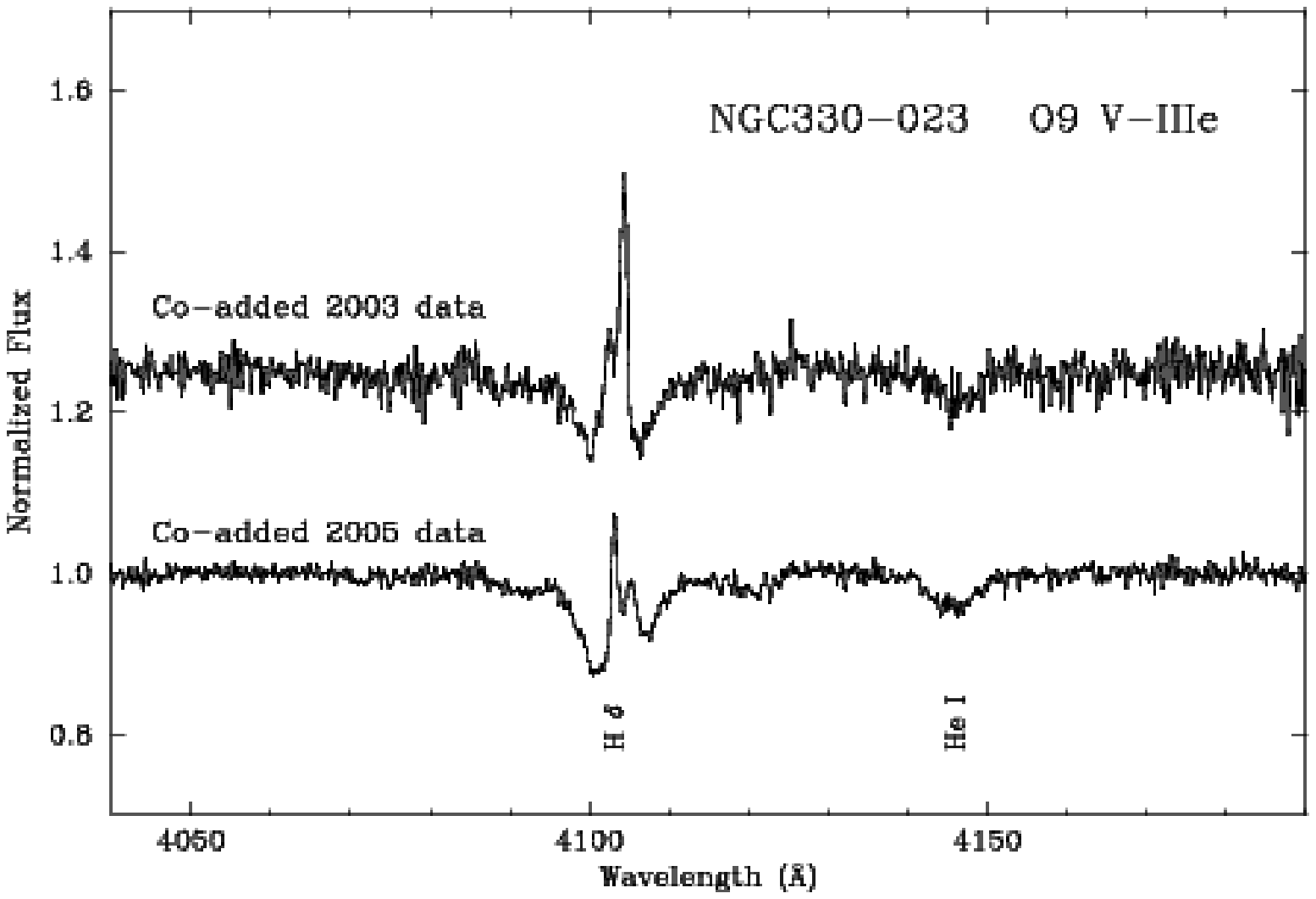}
\caption{Co-added spectra of NGC\,330-023 from the two epochs of HR03 
observations.  Both spectra have been 7-pixel median filtered for 
clarity.  Note the change in structure of the emission at H$\delta$, 
and the apparent wavelength invariance of the (albeit broadened or infilled)
He~{\scriptsize I}  \lam4143 line.}\label{023b}
\end{center}
\end{figure*}

The spectrum of NGC\,330-023 has strong, broad H$\alpha$ emission,
with an equivalent width of 28 $\pm$3~\AA\/ (with the uncertainty
arising from the rectification in the wings of the emission).  It's
conceivable that part of this emission may be nebular in origin, but
the absence of [N~\2] emission suggests that the majority is intrinsic
to the star.  The blue-region spectrum is particularly interesting,
with He~\2 absorption lines consistent with a late O-type star.  It is
difficult to assign a precise classification due to significant
infilling of the He~\1 lines; however from intensity of the He~\2
lines we classify the star as approximately O9-type.  Of note (see
Figure~\ref{023}) is the twin-peaked emission seen in both He~\1
\lam4471 and 4713.  The luminosity class is harder to ascertain, with
a final classification of O9~V-IIIe adopted.

Prior to the new HR03 spectra obtained in 2005, radial velocity
measurements were limited to the three He~\2 lines, with a mean value
of 159~\kms (with standard deviation, $\sigma =$10), and with two of
the lines included in the same wavelength setting.  Inspection of the
spectra acquired in 2005 reveal significant variations in the
asymmetric H$\delta$ emission feature, as shown in Figure~\ref{023b}.
With no obvious radial velocity shifts seen in other lines, such
differences resemble the V/R (violet / red) variability seen in
Be-type stars, suggested by \citet{hhds} as arising from oscillations
in the disk.  \citet{hum01} studied Be-type stars in NGC\,330 to see
if this phenomenon was metallicity dependent as theory would suggest,
but found that the fraction of stars undergoing disk oscillations was
comparable to that in Galactic targets.  We note that NGC\,330-023 is
the photometric variable \#95V reported by \citet{sw94}, with the
variability presumably arising from changes such as those reported
here.

\section{Compilation of previous spectroscopy}

For completeness, in Table~\ref{cftypes} the FLAMES classifications
presented here are compared with previous types.  Published
classifications are included from: W77 \citep{w77}; FB80 \citep{fb80};
CJF85 \citep{cjf85}; NMC \citep{nmc}; G87 \citep{gar87}; MPG
\citep{mpg}; F91 \citep{f91}; P92 \citep{p92}; L93 \citep{l93}; M95 \citep{m95}; G96 \citep{grb96}; 
HM00 \citep{hm00}; W00 \citep{wal00}; L03 \citep{l03}; EH04
\citep{eh04}.  Note that we do not include published types from
objective-prism spectroscopy, preferring to limit our comparisons to
long-slit/multi-object observations.

With the benefit of the high-quality FLAMES spectra, many of the
classifications from \citet{eh04} are now refined.  There are few
significant differences between the FLAMES classifications and those
extant in the literature.  Two stars (N11-028 and N11-038) have
already been discussed in Section~\ref{stars} -- one further target
worthy of comment is N11-080 which appears to be comprised of two
late-type O stars from the FLAMES data, compared to the O4-6~V
classification from P92.  This is not surprising given that P92 had
noted the spectrum as a possible composite object.

\section{Discussion}

In Table~\ref{overview} we give an overview of the entire sample in the FLAMES
survey, incorporating the LMC and SMC observations reported here, with the
Galactic data from Paper~I.  Analysis of many of these data is now well advanced
by different groups.  Here, with the benefit of a broad view of the whole sample, 
we discuss some of the more general features of the survey.

\subsection{H-R diagrams for each of the fields}
\begin{figure*}
\begin{center}
\includegraphics[angle=180,width=12cm]{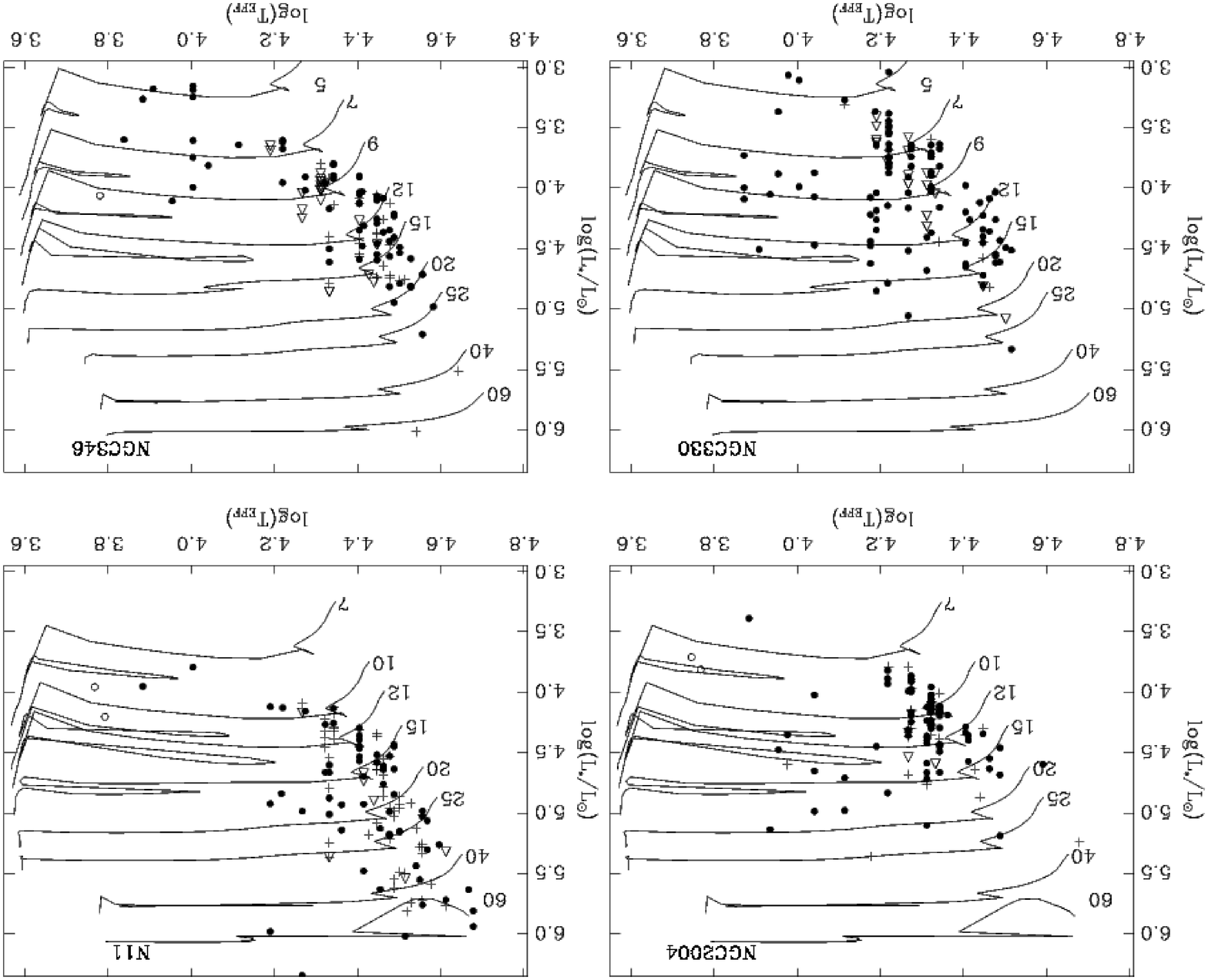}
\vspace*{0.5cm}\caption{H-R diagrams for the four FLAMES fields in the Magallanic Clouds.
Open circles denote foreground stars, open triangles are emission-line 
objects, and likely binaries are indicated with crosses.  The evolutionary
tracks are taken from \citet{s93} for N11 and NGC\,2004, and from 
\citet{char93} for NGC\,330 and NGC\,346.}\label{hrds}
\end{center}
\end{figure*}

In Figure~\ref{hrds} we show Hertzsprung-Russell (H-R diagrams) for
each of our LMC and SMC fields.  These have been compiled by employing
various published calibrations -- clearly the detailed studies of
different subsets of the survey will yield precise determinations of
temperatures and luminosities, but we take the opportunity now to show
the full extent of the LMC/SMC stars in the H-R diagram.  Objects
listed as likely foreground objects are plotted as open circles,
likely binaries as '+`, and emission-line objects as triangles.  For
illustrative purposes the evolutionary tracks shown are from
\citet{s93} for the LMC targets and from \citet{char93} for the SMC.

Temperatures were adopted on the basis of spectral type, luminosity and
metallicity from \citet[][O-type stars]{mfast2}, \citet[][B-type supergiants]{clw06}, 
\citet[][A-type superigants]{eh03}, and \citet[][for other types]{sk82}.  Objects
that luminosity classes of III or V were assigned temperatures from
calibrations for dwarfs; more luminous stars (i.e. II, Ib, Iab and Ia)
adopted temperatures from calibrations of supergiants.  With no
metallicity-dependent temperature scale in the literature for early
B-type supergiants, temperatures were taken from the Galactic results
\citep{clw06}, which were found to be in good agreement with those
from analysis of individual stars in the SMC \citep{tl04,tl05}.

Luminosities were calculated using intrinsic colours from \citet{f70},
extrapolating or interpolating where required; bolometric corrections
were calculated using the relations from \citet{vacca} for the
earliest types, and from
\citet{balbc} for the cooler stars; the ratio of total to
selective extinction ($R$) in the LMC was taken as 3.1
\citep[e.g.][]{h83} and as 2.7 in the SMC \citep{b85};
distance moduli were taken as 18.5 to the LMC \citep[e.g.][]{gibson}
and 18.9 to the SMC \citep{hhh}.

Figure~\ref{hrds} highlights the predominantly less-massive
populations observed in the two older clusters (i.e. NGC\,2004 and
NGC\,330), particularly when compared to N11.  The N11
observations also sample a more luminous, more massive population
than those in NGC\,346.  This is, in part, influenced by the fact that
some of the O-type stars observed by \citet{wal00} were explicitly
avoided in the FLAMES survey as state-of-the-art analyses have already
been presented in the literature \citep[e.g.][]{jc03}.  

\begin{table*}
\caption[]{Overview of the distribution of spectral types of the 
FLAMES survey, including the Galactic clusters from Paper~I.}\label{overview}
\begin{center}
\begin{tabular}{lp{1.5cm}p{1.5cm}p{1.5cm}p{1.5cm}p{1.5cm}}
\hline
Field & O & Early-B & Late-B & AFG & Total \\
      &   & (B0-3)  & (B5-9) &     &       \\
\hline
MW:\o NGC\,3293  & $-$ & 48 & 51 & 27 & 126 \\
MW:\o NGC\,4755  & $-$ & 54 & 44 & 10 & 108 \\
MW:\o NGC\,6611  & 13  & 28 & 12 & 32 & 85 \\
SMC: NGC\,330     & 6 & 98 & 11 & 10 & 125 \\
SMC: NGC\,346     & 19 & 84 & 2 & 11 & 116 \\
LMC: NGC\,2004   & 4 ($+$ 1 WR) & 101 & 6 & 7 & 119 \\
LMC: N11              & 44 & 76 & $-$ & 4 & 124 \\ 
\hline
Total            & 87 & 489 & 126 & 101 & 803 \\
\hline
\end{tabular}
\end{center}
\end{table*}

\subsection{Exploring new regions of N11}
As mentioned earlier, part of our intention in the FLAMES observations
was to explore some of the less dense, apparently star-forming regions
in N11.  In addition to the discovery of the O2.5-type star to the
north of LH10, the survey has revealed a large number of O-type stars
in the regions surrounding LH9 and LH10.  In Figure~\ref{n11_mess} we
show the $V$-band WFI image used for target selection in N11, with the
O- and B-type stars marked in different colours; LH9, LH10 and LH13
are also identified in the image.  Note the O-type stars to the south
of LH9, newly discovered by this survey.  These include N11-020
[classified as O5 I(n)fp] on the northern edge of the N11F region
(cf. Figure~\ref{fchart_n11}).

Furthermore, N11-004 [O9.7 Ib] is the bright star in N11G, and N11-058
[O5.5 V((f))] is one of two visually bright stars in N11I, the other
being N11-102 [B0.2~V].  Both N11G and N11I appear as small `bubbles'
in near-IR images, presumably driven by the ionization and/or the stellar
winds from these stars.  We have also observed N11-029 [OC9.7 Ib] in
N11H, which appears radially smaller than N11G and N11I.

The densest regions in N11, i.e. N11B (LH10) and N11C (LH13) have been 
demonstrated by other authors \citep{p92,w92,hm00,b03} to have rich, 
young populations of early-type stars.  The general consensus is that
this region is a two-stage starburst, with the evolution of LH9
triggering star-formation around it.  For the first time the FLAMES
survey has observed the regions to the south and west of LH9 -- whilst
much smaller in size and stellar content, we find newly discovered O-type stars
as the likely source of ionization and dynamic energy for the observed
nebulae.  

To place this region in context, the `bigger picture' is dramatically
illustrated by the cover image of Edition \#80 of the National Optical
Astronomy Observatory/National Solar Observatory (NOAO/NSO)
Newsletter.  This features an image of N11 from the Magellanic Clouds
Emission Line Survey (credited to Drs.~S. Points, C. Smith, and M.
Hanna).  In addition to N12, N13, and N14, the image includes the
significantly extended shell of gas that constitutes N10 from
\citet{hen56}.

\subsection{`Blue stragglers' in NGC\,330}

Blue stragglers were reported in NGC\,330 by \citet{l93}, with 
\citet{grb96} arguing that they might be a product of binary evolution.  
Blue stragglers are thought to arise from either mass-exchange or
stellar mergers \citep[e.g.][]{pm94}, with recent studies of globular
clusters suggesting that both channels play a role \citep{d04}.
Six of our targets in the NGC\,330 observations are late O-type stars,
three of which are less than 5$'$ from the centre of the cluster:
NGC\,330-023 (see Section \ref{330_023}), NGC\,330-049, and NGC\,330-123.  

In particular, NGC\,330-123 (R74-B18) is classified here as O9.5 V,
cf.  B0~Ve from \citet{l93} and O9~III/Ve from \citet[][who were
quoting unpublished types from Lennon]{grb96}.  Narrow (and weak)
emission is seen in the core of the H$\alpha$ Balmer line in the UVES
spectrum of NGC\,330-123, but is accompanied by [N~\2] emission so the
origins are likely nebular.  Indeed, \citet{sk98} note that there is
an elliptical region of diffuse nebular emission centred on this star,
it would appear that NGC\,330-123 is the ionization source of this
nebulosity.  Interestingly, the radial velocity of NGC\,330-123 is 177
km/s (with standard deviation, $\sigma$ = 5), i.e. different from the
systemic velocity of the cluster which is $\sim$155 km/s
\citep[e.g.][]{l03}.  Binarity is a possible explanation of this
difference, but additional spectra at other epochs (October 1992 and
October 1995), though of lower resolution, are in good agreement with
the FLAMES data. If this star is coeval with the cluster then its
current radial velocity might indicate a previous, though relatively
recent ejection event.  Its radial velocity however is similar to the
three more distant O-type stars in the NGC\,330 field: NGC\,330-013,
NGC\,330-046 and NGC\,330-052, which have $v_r =$ 176, 177 and
166~\kms\/ respectively.  Two of these, NGC\,330-013 and NGC\,330-052,
are found to be helium rich \citep{rmsmc}, whilst
NGC\,330-123 has also been reported as helium rich by \citet[][note
that this paper erroneously refers to R74-A01 as helium rich, when it
should actually refer to R74-B18, i.e. NGC\,330-123]{l93}.  All four
stars could perhaps be considered as members of the general field
population which, from the H-R diagram (Figure~\ref{hrds}), can be
seen to extend well beyond the notional cluster turn-off -- as
represented by the $\sim$20 stars with spectral types earlier than B1.
We note that the majority of these stars have radial velocities rather
different from the cluster radial velocity, suggesting that the
probable number of true blue stragglers in this field belonging to
NGC\,330 is small.

Both NGC\,330-023 and NGC\,330-049 are late O-type stars with velocities
consistent with the cluster \citep[cf.][]{l03}, although the
pecularities of NGC\,330-023 have already been discussed in
Section~\ref{330_023}.  At respective radial distances of 2.3$'$ and
4.5$'$ they are not immediately proximate to the cluster core, but are
perhaps outer members of the cluster and therefore could be true blue stragglers.

The situation for NGC\,2004 is qualitatively similar to that of
NGC\,330.  The Wolf-Rayet and O2-3 stars stand out as potential blue
stragglers but are more likely field stars, though the presence of an
O2-3 star in the field is unusual in itself.  One further candidate
blue straggler is NGC\,2004-019, which is close to the core, with
$r_{\rm d} =$ 1.5$'$.

\subsection{Incidence of Be-stars}\label{discussbe}

The numbers of Be-type stars in each FLAMES pointing are summarised in
Table \ref{overview2}.  The observed field-population of B- and Be-type
stars in our NGC\,330 and NGC\,2004 pointings should be relatively
unaffected by specific selection effects (Section \ref{seffects}),
aside from the slightly different ($\sim$0.5$^{\rm m}$) faint cut-off
of the observations.

The Be-stars in Table~\ref{overview2} include some in the outer
regions of the clusters themselves.  As an experiment, we considered
the B-type stars with $r_d > {\rm 2}'$ in NGC\,330 and NGC\,2004 to
sample the field population away from the main clusters.  All stars in
the range B0-3 were counted and then the relative percentage of those
that are seen to be Be-type stars was found.  In NGC\,2004 this ratio
for the field population (i.e. Be / [Be$+$B]) was 16\% (13/81),
compared to 23\% (18/77) in NGC\,330.  It is difficult to quantify the
uncertainties in these results.  Is there any effect in terms of
absolute magnitude?  As a further experiment (and still with the $r_d
> {\rm 2}'$ condition) we considered the early B-type stars in our
NGC\,330 data with $V \leq$~16.03.  Taking the difference of the
distance moduli of the Clouds as 0.4$^{\rm m}$, and allowing for the
fact that the `typical' extinction toward the LMC is $\sim$0.1$^{\rm
m}$ greater than the SMC \citep[][]{m95}, this notionally imposes a
similar cut-off as that in the NGC\,2004 data.  The Be-fraction for
the NGC\,330 field population remains robust at 24\% (11 Be/45 Be+B).

Our results are in reasonable agreement with those from
\citet{sk99} and, for NGC\,2004, match those of \citet{mhf}.
Both \citeauthor{sk99} and \citeauthor{mhf} advanced their
results as a lack of evidence for a strong dependence of the
Be-fraction with metallicity when compared to the general
result for the Milky Way of $\sim$17-20\% \citep{zb}.  This 
is in contrast to the results for clusters from \citet{m99}.
The statistics in our NGC\,346 and N11 data do not provide a
meaningful test of the relative numbers of Be stars at different
metallicities.  Aside from a likely mix of field and cluster
populations in the NGC\,346 pointing, the N11 region is even less
suited given its complex star-formation history.

\subsection{Binaries and the binary fraction}

With the time-sampling provided by the service-mode observations we
are reasonably sensitive to detection of binarity in our SMC and LMC
targets.  In Appendix~\ref{mjd} we give full details of the
observational epochs of the FLAMES spectra.  The observations in both
N11 and NGC\,2004 spanned a total of 57 days, and those in NGC\,346
covered 84 days.  The time-coverage of the majority of the
NGC\,330 data is not as extensive as for the other fields, covering 10
days.  The new HR03 (\lam4124) data offer some extra information in
this regard (e.g. Section~\ref{330_023}).  However, given the
relatively poor signal-to-noise in the 2003 data it is difficult to
compare measured velocities meaningfully -- from these comparisons
a number of stars are flagged as having potentially variable velocities.

The FLAMES data are sufficient to derive periods for some of the newly
discovered systems, the most interesting of which will receive a
detailed treatment in a future study.  The incidence of binaries in
summarised in Table~\ref{overview2}.  We stress that stars considered
as multiple in this discussion are those listed in Tables~\ref{346},
\ref{330}, \ref{lh910}, and \ref{2004} as `Binary' (be they SB1, SB2
or not specified).  Stars suggested to perhaps demonstrate $v_r$ variations
are not considered further, pending follow-up.  As such, the
percentages in Table~\ref{overview} serve as strong, {\it lower}
limits on the binary fraction in our fields.

In young, dense clusters multiplicity seems a common, almost
ubiquitous feature.  \citet{gm01} found a significant binary fraction
(79\%) in the very young Galactic cluster NGC\,6231, and
\citet{wbp99} and \citet{p99} highlighted the high degree of
multiplicity found in the massive members of the Orion Nebula,
suggesting a different mode of star formation to that at lower masses
(in which the binary fraction is lower).  

The formation mechanism of massive stars is still a point of
significant debate.  In their discussion of the competing scenarios of
massive-star formation from accretion versus stellar mergers,
\citet{bz05} suggest that the multiple star fraction will be larger if
merging dominates.  \citeauthor{bz05} also suggest that mergers may be
the dominant process in ultradense regions such as 30 Doradus; perhaps
the high binary fraction of O- and early B-type stars in N11 is
indicative of this mode of star-formation, remembering that 36\% is a
solid, lower limit obtained from a programme that was not optimised
for detection of binaries.

By comparison, the low binary fraction in the NGC\,330 targets is
somewhat puzzling.  Although fewer binaries are generally found in the
field population \citep[e.g.][]{mgh98}, the NGC\,330 fraction is
significantly lower than that for NGC\,2004, which similarly samples
the field (see Figures~\ref{fchart_330} and \ref{fchart_2004}).  We
speculate that, whilst the new HR03 data added an additional epoch to
the time-sampling for the NGC\,330 targets, it is still not as
thorough as for the other fields.

\begin{table*}
\caption[]{Overview of the number of Be-type stars and binaries in the 
LMC and SMC sample}\label{overview2}
\begin{center}
\begin{tabular}{lp{1.2cm}p{1.2cm}p{1.2cm}p{1.2cm}p{1.2cm}}
\hline
Field            & O ($+$ Oe) & Early-B & Be & Total & Binary \\
                 &   & (B0-3)  &    & (O$+$Early-B)   & Fraction \\
\hline
SMC: NGC\,330    & 6 & 76 & 22 & 104 &  \\
{\it Binary}     & $-$ & {\it 3} & {\it 1} & {\it 4} & {\it 4\%} \\
&&&&&\\
SMC: NGC\,346    & 19 & 59 & 25 & 103 & \\
{\it Binary}     & {\it 4} & {\it 19} & {\it 4} & {\it 27} & {\it 26\%} \\
&&&&&\\
LMC: NGC\,2004   & 4 & 83 & 18 & 105 & \\
{\it Binary}     & {\it 1} & {\it 21} & {\it 2} & 24 & {\it 23\%} \\
&&&&&\\
LMC: N11         & 44 & 68 & 8 & 120 & \\
{\it Binary}     & {\it 19} & {\it 21} & {\it 3} & {\it 43} & {\it 36\%} \\ 
\hline
\end{tabular}
\end{center}
\end{table*}

\section{Summary}

Quantitative analyses of different subsets of the FLAMES survey are now
underway by various groups, e.g. Dufton et al. (submitted); Hunter et
al. (submitted); \citet{rmsmc}. Here we have presented a
significant dataset of high-resolution observations of early-type
stars in the Magellanic Clouds, giving stellar classifications and
radial velocities, and noting evidence of binarity from the
multi-epoch observations.  Spectral peculiarities were
discussed, in particular with regard to emission-line spectra.  We
have found a relatively large number of Be-type stars that display
permitted Fe~\2 emission lines.  We find that in nearly all the spectra 
with EW(H$\alpha$) $>$ 20~\AA, Fe~\2 is seen in emission.  We
do not find evidence for a metallicity-dependent scaling of the
required H$\alpha$ equivalent width for Fe~\2 emission to be present,
contrary to the suggestion by \citet{mhf}.  We have tentatively explored
the relative fraction of Be- to normal B-type stars in the
field-regions of NGC\,330 and NGC\,2004, finding no compelling evidence of
a strong trend with metallicity {\it for field stars}.

We have investigated previously unexplored regions around the central
LH9/LH10 complex of N11, finding $\sim$25 new O-type stars from our
spectroscopy.  Furthermore, to the north of LH10 we have discovered a
very hot, potential runaway star (N11-026) that we classify as O2.5 III(f$^\ast$).

For three of our fields we find lower limits to the binary fraction of
O- and early B-type stars of 23 to 36\%.  Following identification of a
relatively large number of binaries, more sophisticated methods will
be used in a subsequent paper to characterise as many of these
systems as possible.  NGC\,346-013 is particularly interesting with
an apparently hotter, more massive, but less luminous secondary
component.

The overall distribution of the targets observed in the VLT-FLAMES
survey of massive stars is summarised in Table~\ref{overview}.  The
large number of O-type stars observed will provide allow for better
sampling of this domain, building on the significant studies by
\citet{mfast,mfast2}.  In the early B-type domain the FLAMES survey is
truly unique, allowing precise abundance determinations of a hitherto
impossible number of stars (Hunter et al., submitted), enabling us to
really address some of the unsolved questions regarding the dependence of
stellar evolution on metallicity.

\vspace{.5in}
{\it Acknowledgements:}
We are very grateful to Francesca Primas and the staff at Paranal
for their invaluable assistance with the observational programme, and to
Nolan Walborn for his comments on the manuscript.  We also thank Mike
Irwin for processing the N11 and NGC\,2004 images, and Andreas Kaufer for
reducing the FLAMES-UVES data.  Finally we thank the referee, Dr.~Anne-Marie
Hubert, for her useful comments \& suggestions.  CJE acknowledges financial support
from the UK Particle Physics and Astronomy Research Council (PPARC)
under grant PPA/G/S/2001/00131, and the continued support of the other
members of the FLAMES massive star consortium.  SJS acknowledges support
from the EURYI scheme.

\bibliographystyle{aa}
\bibliography{4988}

\newpage

\onecolumn

{\scriptsize
\begin{center}
\begin{longtable}{llccccclll}
\caption[]{NGC\,346: Observational parameters of target stars.  Cross-references
to identifications by \citet[][Sk]{sk68}, \citet[][AzV]{av75,av82}, \citet[][NMC]{nmc}, \citet[][MPG]{mpg},
\citet[][MA93]{ma93}, \citet[][KWB]{sk99}, and \citet[][2dFS]{eh04} are given in the final
column.  The radial velocities ($v{\rm r}$, in \kms) and radial
distances to the centre of the cluster ($r_{\rm d}$, in arcmin) are
given for each star.
\label{346}} \\
\hline
ID & EIS\# & $\alpha$(2000) & $\delta$(2000) & r$_{\rm d}$ & $V$ & $B-V$ & Sp. Type & $v_{\rm r}$  & Comments \\
\hline 
\endfirsthead
\caption[]{\it{continued}} \\
\hline
ID & EIS\# & $\alpha$(2000) & $\delta$(2000) & r$_{\rm d}$ & $V$ & $B-V$ & Sp. Type & $v_{\rm r}$  & Comments \\
\hline 
\endhead
\hline 
\multicolumn{10}{r}{\it{continued on next page}} \\
\endfoot
\hline 
\\
\multicolumn{10}{l}{$\dagger$: Further notes on individual stars:}\\
\multicolumn{10}{l}{{\it NGC\,346-001:} saturated in images, photometry is from MPG}\\
\multicolumn{10}{l}{{\it NGC\,346-011:} member of Lindsay 56/LHA 115-S26}\\
\multicolumn{10}{l}{{\it NGC\,346-024:} is also KWB346\#72, blended with SMC5\_078073}\\
\multicolumn{10}{l}{{\it NGC\,346-026:} faint companions in WFI images}\\
\multicolumn{10}{l}{{\it NGC\,346-033:} blended in images with SMC5\_080234}\\
\multicolumn{10}{l}{{\it NGC\,346-036:} bad column in $V$ image, photometry is from MPG}\\
\multicolumn{10}{l}{{\it NGC\,346-048:} is also MA93\#1101}\\
\multicolumn{10}{l}{{\it NGC\,346-057:} blended in images with SMC5\_032946}\\
\multicolumn{10}{l}{{\it NGC\,346-061:} is also MA93\#1048}\\
\multicolumn{10}{l}{{\it NGC\,346-065:} is also MA93\#1100}\\
\multicolumn{10}{l}{{\it NGC\,346-072:} correlated with MA93\#1067, which is resolved into two bright components in WFI image}\\
\multicolumn{10}{l}{{\it NGC\,346-076:} is also MA93\#1126}\\
\multicolumn{10}{l}{{\it NGC\,346-104:} part blended with SMC5\_082742/MPG\,635}\\
\multicolumn{10}{l}{{\it NGC\,346-115:} part blended with SMC5\_246984/MPG\,426}\\
\endlastfoot
\hline
NGC\,346-001$\dagger$& SMC5\_083194 &  00 59 31.94 & $-$72 10 46.05 &\o1.07 &  12.31 & $-$0.19   & O7 Iaf$+$  & Binary & Sk 80, AzV 232, MPG\,789, H$\alpha=$ broad em.\\
NGC\,346-002 & SMC5\_082844 &  00 58 06.39 & $-$72 07 05.47 &\o6.62 &  13.22 & $-$0.04   & A2: Iab        & $-$        & FLAMES-UVES target\\
NGC\,346-003 & SMC5\_082834 &  00 58 47.62 & $-$72 13 31.01 &\o3.58 &  13.56 & \pp0.62   & G0:            & $-$        & FLAMES-UVES target; foreground\\
NGC\,346-004 & SMC5\_082667 &  00 57 37.22 & $-$72 13 09.12 &\o8.06 &  13.69 & $-$0.16   & Be (B1:)       & 106:~(8)   & MA93\#1021, AzV 191; H$\alpha=$ twin em.\\
NGC\,346-005 & SMC5\_072019 &  00 57 21.47 & $-$72 13 33.70 &\o9.33 &  13.88 & $-$0.02   & A0 II          & 162\p~(5)  & \\
NGC\,346-006 & SMC5\_000969 &  00 59 51.39 & $-$72 14 49.23 &\o4.76 &  14.02 & \pp0.15   & F2:            & $-$        & FLAMES-UVES target\\
NGC\,346-007 & SMC5\_079542 &  00 58 57.40 & $-$72 10 33.53 &\o1.59 &  14.07 & $-$0.28   & O4 V((f$+$))   & Binary (SB1)& MPG\,324, NMC\,32\\
NGC\,346-008 & SMC5\_081043 &  00 59 15.99 & $-$72 04 44.46 &\o6.06 &  14.20 & $-$0.20   & B1e            & 158\p~(6)  & AzV 224; H$\alpha=$ twin em.\\
NGC\,346-009 & SMC5\_028909 &  00 59 51.32 & $-$72 11 28.57 &\o2.64 &  14.36 & $-$0.25   & B0e            & 154\p~(7)  & MPG\,845, MA93\#1167, H$\alpha=$ broad em.\\
NGC\,346-010 & SMC5\_000836 &  00 59 20.70 & $-$72 17 10.52 &\o6.38 &  14.37 & $-$0.22   & O7 IIIn((f))   & 208:~(8)   & AzV 226 \\
NGC\,346-011$\dagger$ & SMC5\_075100 &  00 57 29.49 & $-$72 16 00.40 &\o9.80 &  14.39 & $-$0.12   & B9 II          &160\p~(10)& \\
NGC\,346-012 & SMC5\_001361 &  00 58 14.46 & $-$72 07 29.51 &\o5.88 &  14.39 & $-$0.23   & B1 Ib          &181\p~(13)& AzV 202 \\
NGC\,346-013 & SMC5\_072124 &  00 59 30.36 & $-$72 09 09.59 &\o1.89 &  14.46 & \pp0.06   & B1:            & Binary (SB2)& MPG\,782\\
NGC\,346-014 & SMC5\_026814 &  00 59 50.36 & $-$72 13 57.16 &\o4.01 &  14.58 & \pp0.01   & A0 II          & 164\p~(5)  & 2dFS\#1425\\
NGC\,346-015 & SMC5\_022635 &  00 59 01.95 & $-$72 18 52.91 &\o8.17 &  14.62 & $-$0.30   & B1 V           & Binary (SB2)& AzV 217, 2dFS\#1357 \\
NGC\,346-016 & SMC5\_089653 &  00 59 36.19 & $-$72 15 56.74 &\o5.33 &  14.65 & $-$0.23   & B0.5 Vn        & Binary (SB2)& 2dFS\#5100\\
NGC\,346-017 & SMC5\_007202 &  00 58 11.94 & $-$72 04 02.06 &\o8.44 &  14.67 & $-$0.22   & Be (B1)        & Binary     & H$\alpha=$ twin em.\\
NGC\,346-018 & SMC5\_038701 &  00 58 47.10 & $-$72 13 01.57 &\o3.25 &  14.78 & $-$0.12   & O9.5 IIIe      & 164\p~(3)  & MPG\,217, H$\alpha=$ em.\\
NGC\,346-019 & SMC5\_031737 &  00 59 05.56 & $-$72 08 02.38 &\o2.92 &  14.82 & $-$0.03   & A0 II          & 162\p~(3)  & \\
NGC\,346-020 & SMC5\_005500 &  00 57 46.46 & $-$72 12 45.01 &\o7.27 &  14.89 & $-$0.22   & B1 V$+$early-B & Binary (SB2)& 2dFS\#1259\\
NGC\,346-021 & SMC5\_000965 &  00 59 19.58 & $-$72 14 50.47 &\o4.04 &  14.90 & $-$0.18   & B1 III         &166\p~(11)& 2dFS\#5099 \\
NGC\,346-022 & SMC5\_005834 &  00 59 18.59 & $-$72 11 09.89 &\o0.37 &  14.91 & $-$0.26   & O9 V           & 156\p~(6)  & MPG\,682, NMC\,40\\
NGC\,346-023 & SMC5\_029130 &  00 58 41.86 & $-$72 11 17.55 &\o2.81 &  14.92 & $-$0.07   & B0.2: (Be-Fe)  & 161\p~(6)& MPG\,178, MA93\#1089, NMC\,45, 2dFS\#5097\\
NGC\,346-024$\dagger$ & SMC5\_078074 &  00 59 06.38 & $-$72 07 44.97 &\o3.18 &  14.92 & $-$0.17   & B2: shell (Be-Fe) & 151\p~(5) & MA93\#1118, KWB346\#205,H$\alpha=$ twin\\
NGC\,346-025 & SMC5\_056190 &  00 59 52.93 & $-$72 10 49.07 &\o2.67 &  14.95 & $-$0.27   & O9 V           & Binary (SB1)  & MPG\,848\\
NGC\,346-026$\dagger$ & SMC5\_056277 &  00 58 14.10 & $-$72 10 44.18 &\o4.89 &  14.98 & $-$0.14   & B0 IV (Nstr)   &222\p~(17) & MPG\,12, 2dFS\#1299\\
NGC\,346-027 & SMC5\_032353 &  00 59 00.92 & $-$72 07 18.16 &\o3.73 &  15.00 & $-$0.24   & B0.5 V         & 166:~(6)   & \\
NGC\,346-028 & SMC5\_069506 &  00 58 31.77 & $-$72 10 57.92 &\o3.54 &  15.01 & $-$0.26   & OC6 Vz         & 178\p~(8)  & MPG\,113 \\
NGC\,346-029 & SMC5\_055586 &  00 59 14.53 & $-$72 11 59.77 &\o1.23 &  15.02 & $-$0.19   & B0 V           & Binary (SB1) & MPG\,637, NMC\,50 \\
NGC\,346-030 & SMC5\_000991 &  00 58 55.97 & $-$72 14 37.59 &\o4.18 &  15.02 & $-$0.21   & B0 V           & Binary     & 2dFS\#5098 \\
NGC\,346-031 & SMC5\_034478 &  00 59 54.04 & $-$72 04 31.28 &\o6.86 &  15.02 & $-$0.26   & O8 Vz          &159\p~(11) & \\
NGC\,346-032 & SMC5\_034263 &  00 59 24.81 & $-$72 04 47.68 &\o6.03 &  15.06 & $-$0.26   & B0.5 V         &174\p~(11) & \\
NGC\,346-033$\dagger$ & SMC5\_089286 &  00 59 11.62 & $-$72 09 57.52 &\o0.97 &  15.07 & $-$0.25   & O8 V           & 189:~(3)   & MPG\,593 \\
NGC\,346-034 & SMC5\_029400 &  00 59 05.90 & $-$72 10 50.28 &\o0.93 &  15.08 & $-$0.26   & O8.5 V         & Binary (SB1) & MPG\,467, NMC\,18\\
NGC\,346-035 & SMC5\_001453 &  00 59 46.61 & $-$72 05 32.45 &\o5.70 &  15.09 & $-$0.23   & B1 V           & Binary (SB2) & 2dFS\#1418\\
NGC\,346-036$\dagger$ & SMC5\_233890 &  00 59 23.33 & $-$72 12 00.62 &\o1.28 &  15.11 & $-$0.07   & B0.5 V (Be-Fe)& Binary?   & MPG\,729, MA93\#1144\\
NGC\,346-037 & SMC5\_032995 &  00 59 18.84 & $-$72 06 29.27 &\o4.31 &  15.18 & $-$0.21   & B3 III         &153\p~(13) & \\ 
NGC\,346-038 & SMC5\_079698 &  00 58 51.33 & $-$72 05 10.25 &\o5.99 &  15.18 & $-$0.24   & B1 V           & 180\p~(7)  & variable $v_r$?\\
NGC\,346-039 & SMC5\_075002 &  00 57 47.42 & $-$72 16 40.44 &\o9.08 &  15.19 & $-$0.20   & B0.7 V         &156\p~(13) & 2dFS\#1262\\
NGC\,346-040 & SMC5\_005698 &  00 58 23.59 & $-$72 11 53.20 &\o4.30 &  15.23 & $-$0.16   & B0.2 V         & Binary (SB1) & MPG\,61\\
NGC\,346-041 & SMC5\_024945 &  00 57 44.32 & $-$72 16 10.72 &\o8.96 &  15.24 & \pp0.01   & B2 (Be-Fe)    & 176\p~(9)  & MA93\#1030, H$\alpha=$broad em.\\
NGC\,346-042 & SMC5\_067681 &  00 59 36.26 & $-$72 18 26.32 &\o7.77 &  15.28 & \pp0.09   & A7 II          & 122\p~(5)  & \\
NGC\,346-043 & SMC5\_075935 &  00 58 13.95 & $-$72 09 19.04 &\o5.12 &  15.36 & $-$0.23   & B0 V           &172\p~(16) & MPG\,11\\
NGC\,346-044 & SMC5\_081656 &  00 59 26.46 & $-$72 13 11.78 &\o2.48 &  15.38 & $-$0.22   & B1 II          &146\p~(11) & MPG\,753\\
NGC\,346-045 & SMC5\_069093 &  00 58 24.19 & $-$72 12 46.82 &\o4.57 &  15.43 & $-$0.20   & B0.5 Vne       & 155\p~(5)  & MPG\,64, MA93\#1070, H$\alpha=$ broad em.\\
NGC\,346-046 & SMC5\_027160 &  00 59 31.84 & $-$72 13 35.24 &\o2.98 &  15.44 & $-$0.28   & O7 Vn          & 250:~(8)   & \\
NGC\,346-047 & SMC5\_006207 &  00 57 03.23 & $-$72 09 15.03 & 10.43 &  15.45 & $-$0.19   & B2.5 III       &137\p~(13) & 2dFS\#1189\\
NGC\,346-048 & SMC5\_028019 &  00 58 47.50 & $-$72 12 36.57 &\o2.95 &  15.47 & $-$0.21   & Be (B3 shell)  & 165\p~(7) & MPG\,222, KWB346\#377, $\dagger$, H$\alpha=$ twin\\
NGC\,346-049 & SMC5\_026567 &  00 57 37.55 & $-$72 14 17.54 &\o8.44 &  15.50 & $-$0.05   & B8 II          &144\p~(11) & \\
NGC\,346-050 & SMC5\_075959 &  00 58 55.22 & $-$72 09 06.57 &\o2.43 &  15.50 & $-$0.30   & O8 Vn          & 160\p~(5)  & MPG\,299\\
NGC\,346-051 & SMC5\_038979 &  00 59 08.68 & $-$72 10 14.08 &\o0.91 &  15.51 & $-$0.26   & O7 Vz          & 169\p~(9)  & MPG\,523, NMC\,38\\
NGC\,346-052 & SMC5\_000921 &  00 58 07.49 & $-$72 15 47.98 &\o7.36 &  15.52 & $-$0.23   & B1.5 V         & Binary (SB1) &  \\
NGC\,346-053 & SMC5\_004887 &  00 57 57.68 & $-$72 16 02.24 &\o8.07 &  15.52 & $-$0.25   & B0.5 V         & Binary (SB1) &  \\
NGC\,346-054 & SMC5\_038674 &  00 58 56.51 & $-$72 13 13.43 &\o2.93 &  15.59 & $-$0.15   & B1 V           & 164\p~(8)  & MPG\,309\\
NGC\,346-055 & SMC5\_001131 &  00 58 27.59 & $-$72 11 54.55 &\o4.01 &  15.59 & $-$0.15   & B0.5 V         & 176\p~(6)  & MPG\,84, MA93\#1072\\
NGC\,346-056 & SMC5\_079574 &  00 58 56.12 & $-$72 09 33.84 &\o2.08 &  15.59 & $-$0.26   & B0 V           &178\p~(11) & MPG\,310, NMC\,48\\
NGC\,346-057 & SMC5\_032947 &  00 59 52.69 & $-$72 06 29.50 &\o5.06 &  15.62 & $-$0.23   & B2.5 III       & 143\p~(9)  &  MA93\#1169, KWB346\#526,$\dagger$,str. nebular lines\\
NGC\,346-058 & SMC5\_001067 &  00 59 31.53 & $-$72 13 06.46 &\o2.53 &  15.63 & $-$0.22   & B0.5 V         & Binary (SB1)&  MPG\,787\\
NGC\,346-059 & SMC5\_076160 &  00 58 04.83 & $-$72 07 17.70 &\o6.60 &  15.64 & \pp0.09   & A5 II          & 188\p~(5)  & \\
NGC\,346-060 & SMC5\_033513 &  00 59 19.28 & $-$72 05 47.80 &\o5.00 &  15.68 & $-$0.13   & B0.5e (shell)  & 142\p~(5)  & MA93\#1140, KWB346\#236, H$\alpha=$ twin\\
NGC\,346-061 & SMC5\_082792 &  00 57 57.82 & $-$72 07 55.43 &\o6.78 &  15.70 & $-$0.21   & B1-2 (Be-Fe)   & 152:~(4) & KWB346\#122,2dFS\#1277,$\dagger$, H$\alpha=$ broad em.\\
NGC\,346-062 & SMC5\_039483 &  00 59 56.61 & $-$72 05 08.50 &\o6.38 &  15.70 & $-$0.24   & B0.2 V         &136\p~(15) & \\
NGC\,346-063 & SMC5\_005036 &  00 59 44.87 & $-$72 15 11.43 &\o4.85 &  15.72 & $-$0.03   & A0 II          & 118\p~(5)  & 2dFS\#1413\\
NGC\,346-064 & SMC5\_006226 &  00 57 55.47 & $-$72 09 05.86 &\o6.54 &  15.72 & $-$0.13   & B1-2 (Be-Fe)   & 162\p~(6)  & MA93\#1043, KWB346\#289, H$\alpha=$broad em.\\
NGC\,346-065 & SMC5\_083005 &  00 58 47.53 & $-$72 09 02.70 &\o2.92 &  15.73 & $-$0.15   & B3 (Be-Fe)    & 132\p~(8)  & MPG\,228, KWB346\#379, $\dagger$, H$\alpha=$ twin\\
NGC\,346-066 & SMC5\_001152 &  00 58 45.98 & $-$72 11 36.78 &\o2.58 &  15.75 & $-$0.25   & O9.5 V         & 163\p~(5)  & MPG\,213; variable $v_r$?\\
NGC\,346-067 & SMC5\_001335 &  00 57 50.38 & $-$72 07 56.04 &\o7.29 &  15.76 & $-$0.04   & B1-2 (Be-Fe)   & 172:~(8)   & MA93\#1038, H$\alpha=$ twin\\
NGC\,346-068 & SMC5\_076404 &  00 59 04.14 & $-$72 04 48.72 &\o6.08 &  15.77 & $-$0.17   & B0 V (Be-Fe)  & Binary     & H$\alpha=$broad em.\\
NGC\,346-069 & SMC5\_005388 &  00 57 07.29 & $-$72 13 19.31 & 10.31 &  15.80 & $-$0.10   & B1-2 (Be-Fe)   & 138:~(6)   & MA93\#981, H$\alpha=$ broad em.\\
NGC\,346-070 & SMC5\_026922 &  00 59 15.49 & $-$72 13 52.21 &\o3.08 &  15.82 & $-$0.22   & B0.5 V         &165:~(13) & \\
NGC\,346-071 & SMC5\_052484 &  01 00 04.21 & $-$72 16 47.56 &\o6.96 &  15.84 & $-$0.07   & A0 II          & 178\p~(2)  & \\
NGC\,346-072$\dagger$ & SMC5\_055979 &  00 58 22.65 & $-$72 11 17.86 &\o4.26 &  15.84 & $-$0.10   & B1-2 (Be-Fe)   & 153\p~(6) & MPG\,54,MWB143,H$\alpha=$ broad em.\\
NGC\,346-073 & SMC5\_025100 &  01 00 05.24 & $-$72 15 55.54 &\o6.27 &  15.84 & $-$0.17   & B1-2 (Be-Fe)   & 143\p~(5)  & MA93\#1183, H$\alpha =$ broad em.\\
NGC\,346-074 & SMC5\_000748 &  00 58 55.81 & $-$72 18 41.29 &\o8.07 &  15.85 & $-$0.22   & B3 III         &142\p~(12) & \\
NGC\,346-075 & SMC5\_006828 &  00 59 25.96 & $-$72 06 01.45 &\o4.81 &  15.85 & $-$0.26   & B1 V           & Binary (SB1)& 2dFS\#1389\\
NGC\,346-076 & SMC5\_055468 &  00 59 12.12 & $-$72 12 11.69 &\o1.47 &  15.87 & \pp0.05   & B2 (Be-Fe)    & 172:~(4)   &MPG\,596,,KWB346\#445,$\dagger$, H$\alpha=$ broad em.\\
NGC\,346-077 & SMC5\_030292 &  00 58 48.97 & $-$72 09 51.92 &\o2.41 &  15.88 & $-$0.18   & O9 V           & 165\p~(6)  & MPG\,238, NMC\,39\\
NGC\,346-078 & SMC5\_026141 &  00 58 22.07 & $-$72 14 45.69 &\o5.83 &  15.88 & $-$0.21   & B2 III         & Binary     & \\
NGC\,346-079 & SMC5\_029544 &  00 59 03.05 & $-$72 10 44.07 &\o1.15 &  15.88 & $-$0.23   & B0.5 Vn        & 168\p~(4) & MPG\,400\\
NGC\,346-080 & SMC5\_029906 &  00 58 49.61 & $-$72 10 19.52 &\o2.22 &  15.89 & $-$0.11   & B1 V           & 160:~(6)   & MPG\,243, NMC\,47, H$\alpha=$ broad em.,neb?\\
NGC\,346-081 & SMC5\_027589 &  00 58 00.38 & $-$72 13 07.86 &\o6.38 &  15.91 & $-$0.20   & B2 IIIn        &148\p~(10) & \\
NGC\,346-082 & SMC5\_026587 &  00 58 08.32 & $-$72 14 16.49 &\o6.36 &  15.91 & $-$0.21   & B2 III         & Binary (SB1) & \\
NGC\,346-083 & SMC5\_005921 &  00 59 28.07 & $-$72 10 42.23 &\o0.78 &  15.91 & $-$0.24   & B1 V           & 172:~(5)   & MPG\,767\\
NGC\,346-084 & SMC5\_001425 &  00 58 13.04 & $-$72 06 02.88 &\o6.88 &  15.92 & $-$0.23   & B1 V           & 147\p~(8)  & 2dFS\#1296\\
NGC\,346-085 & SMC5\_030007 &  00 57 37.95 & $-$72 10 13.78 &\o7.68 &  15.93 & $-$0.22   & B2 III         & Binary (SB1) & \\
NGC\,346-086 & SMC5\_056528 &  00 59 01.92 & $-$72 10 21.33 &\o1.31 &  15.94 & $-$0.25   & B0.2 V         & Binary (SB1) & MPG\,371\\
NGC\,346-087 & SMC5\_052697 &  00 59 21.91 & $-$72 16 27.37 &\o5.66 &  15.95 & \pp0.00   & A0 II          & 115\p~(6)  & \\
NGC\,346-088 & SMC5\_024131 &  00 59 07.82 & $-$72 17 06.01 &\o6.35 &  15.95 & $-$0.22   & B1 V           &174\p~(11) & \\
NGC\,346-089 & SMC5\_023010 &  00 59 10.87 & $-$72 18 27.92 &\o7.68 &  15.96 & $-$0.17   & B1-2 (Be-Fe)   & 185\p~(6) & MA93\#1123, H$\alpha=$ twin em.\\
NGC\,346-090 & SMC5\_026342 &  00 58 25.73 & $-$72 14 33.19 &\o5.48 &  15.96 & $-$0.18   & O9.5 V         & 172\p~(7)  & \\
NGC\,346-091 & SMC5\_022851 &  00 58 29.64 & $-$72 18 40.58 &\o8.70 &  15.98 & $-$0.18   & B1e            & Binary (SB1) & MA93\#1075, H$\alpha=$broad em. \\ 
NGC\,346-092 & SMC5\_032179 &  00 59 42.36 & $-$72 07 27.12 &\o3.83 &  16.00 & $-$0.27   & B1 Vn          & 157:~(6)   & \\
NGC\,346-093 & SMC5\_069704 &  00 58 55.60 & $-$72 10 07.19 &\o1.84 &  16.01 & $-$0.19   & B0 V           & 157\p~(6)  & MPG\,304\\
NGC\,346-094 & SMC5\_006635 &  00 59 57.22 & $-$72 07 00.28 &\o4.84 &  16.01 & $-$0.31   & B0.7 V         &138\p~(13) & \\
NGC\,346-095 & SMC5\_004916 &  00 57 40.34 & $-$72 15 51.42 &\o9.02 &  16.04 & $-$0.05   & B1-2 (Be-Fe)   & 162:~(6)  & MA93\#1027, H$\alpha=$broad em.\\
NGC\,346-096 & SMC5\_033169 &  00 57 28.87 & $-$72 06 19.25 &\o9.47 &  16.04 & $-$0.08   & B1-2 (Be-Fe)   & 185:~(6)   & MA93\#1009, H$\alpha=$broad em.\\
NGC\,346-097 & SMC5\_005824 &  00 59 08.53 & $-$72 11 12.56 &\o0.83 &  16.06 & $-$0.07   & O9 V           & 159\p~(8)  & MPG\,519, NMC\,6\\
NGC\,346-098 & SMC5\_005427 &  00 59 11.01 & $-$72 13 04.04 &\o2.33 &  16.09 & $-$0.22   & B1.5 V         & 158\p~(9)  & MPG\,568\\
NGC\,346-099 & SMC5\_060385 &  00 58 02.32 & $-$72 03 12.37 &\o9.55 &  16.10 & $-$0.18   & B3 III         &157\p~(10) & \\
NGC\,346-100 & SMC5\_007269 &  00 59 20.66 & $-$72 03 38.00 &\o7.17 &  16.10 & $-$0.24   & B1.5 V         & 152\p~(6)  & \\
NGC\,346-101 & SMC5\_056629 &  00 57 59.80 & $-$72 10 11.61 &\o6.01 &  16.15 & $-$0.23   & B1 V           &184\p~(13) & \\
NGC\,346-102 & SMC5\_058504 &  00 58 12.30 & $-$72 06 52.07 &\o6.38 &  16.16 & $-$0.17   & B3 III         &144\p~(11) & \\
NGC\,346-103 & SMC5\_007370 &  00 59 20.82 & $-$72 02 58.58 &\o7.83 &  16.16 & $-$0.30   & B0.5 V         &137\p~(16) & \\
NGC\,346-104$\dagger$ & SMC5\_089334 &  00 59 13.76 & $-$72 09 27.36 &\o1.38 &  16.16 & $-$0.34   & B0 V           & Binary (SB1) & MPG\,628,MA93\#1129, H$\alpha=$ str. em.\\
NGC\,346-105 & SMC5\_069135 &  00 58 28.48 & $-$72 12 34.27 &\o4.18 &  16.19 & $-$0.11   & B2 III         & Binary     & MPG\,88\\
NGC\,346-106 & SMC5\_004992 &  00 57 52.24 & $-$72 15 28.18 &\o8.05 &  16.19 & $-$0.17   & B1 V           & Binary     & \\
NGC\,346-107 & SMC5\_001206 &  00 59 10.37 & $-$72 10 28.45 &\o0.67 &  16.20 & $-$0.27   & O9.5 V         & 156\p~(7)  & MPG\,559, NMC\,22\\
NGC\,346-108 & SMC5\_006368 &  00 59 58.11 & $-$72 08 21.80 &\o3.92 &  16.21 & $-$0.28   & B1.5 V         & 160\p~(6)  & \\
NGC\,346-109 & SMC5\_089444 &  00 59 19.39 & $-$72 04 29.11 &\o6.32 &  16.21 & $-$0.29   & B1.5 V         & 169:~(6)   & \\
NGC\,346-110 & SMC5\_005120 &  00 58 11.49 & $-$72 14 42.70 &\o6.42 &  16.22 & $-$0.15   & B1-2 (Be-Fe)   & Binary    & KWB346\#313, H$\alpha=$ twin em.\\
NGC\,346-111 & SMC5\_082079 &  00 59 00.16 & $-$72 10 46.66 &\o1.37 &  16.24 & $-$0.24   & B0.5 V         & 163\p~(5)  & MPG\,344\\
NGC\,346-112 & SMC5\_005637 &  00 58 58.54 & $-$72 12 06.72 &\o1.98 &  16.24 & $-$0.24   & O9.5 V         & 162\p~(5)  & MPG\,327\\
NGC\,346-113 & SMC5\_006076 &  00 58 11.22 & $-$72 09 59.08 &\o5.17 &  16.27 & $-$0.27   & B0.5 V         &177:~(13) & variable $v_r$?\\
NGC\,346-114 & SMC5\_031881 &  01 00 02.66 & $-$72 07 48.68 &\o4.54 &  16.28 & $-$0.30   & B1 Vn          & 174:~(6)   & \\
NGC\,346-115$\dagger$ & SMC5\_075795 &  00 59 04.63 & $-$72 10 31.16 &\o1.06 &  16.30 & $-$0.28   & B0.2 V         & 182\p~(7)  & MPG\,443\\
NGC\,346-116 & SMC5\_076289 &  01 00 03.23 & $-$72 06 03.65 &\o5.87 &  16.30 & $-$0.30   & B1 V           &154\p~(13) &  \\
\end{longtable}
\end{center}
}

\newpage

{\scriptsize
\begin{center}
\begin{longtable}{llccccclll}
\caption[]{NGC\,330: Observational parameters of target stars.  
Cross-references to identifications given by \citet[][Sk]{sk68}, \citet{arp59}, \citet[][R74]{r74}, 
\citet[][AzV]{av75,av82}, \citet[][MA93]{ma93}, \citet[][KWB]{sk99} and \citet[][2dFS]{eh04} are given in the final column.  The 
radial velocities ($v{\rm r}$, in \kms) and radial distances to the 
centre of the cluster ($r_{\rm d}$, in arcmin) are given for each star.
\label{330}} \\
\hline
ID & EIS\# & $\alpha$(2000) & $\delta$(2000) & r$_{\rm d}$ & $V$ & $B-V$ & Sp. Type & $v_{\rm r}$  & Comments \\
\hline 
\endfirsthead
\caption[]{\it{continued}} \\
\hline
ID & EIS\# & $\alpha$(2000) & $\delta$(2000) & r$_{\rm d}$ & $V$ & $B-V$ & Sp. Type & $v_{\rm r}$  & Comments \\
\hline 
\endhead
\hline 
\multicolumn{10}{r}{\it{continued on next page}} \\
\endfoot
\hline
\\
\multicolumn{10}{l}{$\dagger$: Further notes on individual stars:}\\
\multicolumn{10}{l}{{\it NGC\,330-029:} is also 2dFS\#5093}\\
\multicolumn{10}{l}{{\it NGC\,330-053:} 2dFS\#1279 is likely a blend of 330-053 \& SMC5\_064320}\\
\multicolumn{10}{l}{{\it NGC\,330-065:} is also MA93\#920}\\
\multicolumn{10}{l}{{\it NGC\,330-073:} is also 2dFS\#1087}\\
\multicolumn{10}{l}{{\it NGC\,330-076:} is also MA93\#948}\\
\multicolumn{10}{l}{{\it NGC\,330-077:} is also MA93\#868}\\
\endlastfoot
\hline 
NGC\,330-001 & SMC5\_078698 &  00 55 48.34 & $-$72 30 52.90 &\o3.85 & 12.99 & \pp0.03 & B9-A0 Iab      & $-$        & FLAMES-UVES target \\
NGC\,330-002 & SMC5\_082820 &  00 56 24.85 & $-$72 27 41.79 &\o0.46 & 13.03 & $-$0.04 & B3 Ib          &154\p~(20)& R74-A02\\
NGC\,330-003 & SMC5\_081785 &  00 56 55.29 & $-$72 24 20.55 &\o4.41 & 13.17 & \pp0.00 & B2 Ib          &147\p~(17)& Sk 65, AzV 180, 2dFS\#1183; H$\alpha =$ infilled abs.\\
NGC\,330-004 & SMC5\_082763 &  00 56 20.80 & $-$72 28 33.98 &\o0.79 & 13.33 & $-$0.07 & B2.5 Ib        &157\p~(16)& R74-B37, 2dFS\#5090; H$\alpha =$ infilled abs.\\
NGC\,330-005 & SMC5\_082862 &  00 55 04.93 & $-$72 29 22.87 &\o5.79 & 13.35 & $-$0.06 & B5 Ib          &150\p~(15)& \\
NGC\,330-006 & SMC5\_082471 &  00 56 38.18 & $-$72 18 37.72 &\o9.27 & 13.41 & \pp0.02 & A3: II         & $-$        & FLAMES-UVES target \\
NGC\,330-007 & SMC5\_082854 &  00 56 58.26 & $-$72 31 17.34 &\o4.59 & 13.48 & \pp0.23 & A7-F0 II       & $-$        & FLAMES-UVES target \\
NGC\,330-008 & SMC5\_082532 &  00 56 09.63 & $-$72 32 28.56 &\o4.74 & 13.71 & \pp0.23 & A7-F0 II       & $-$        & FLAMES-UVES target \\
NGC\,330-009 & SMC5\_082706 &  00 53 53.75 & $-$72 26 49.99 & 10.97 & 13.72 & $-$0.06 & B5 Ib          &139\p~(11)& \\
NGC\,330-010 & SMC5\_082798 &  00 54 55.21 & $-$72 33 26.49 &\o8.46 & 13.89 & $-$0.03 & B5 Ib          &142\p~(10)& \\
NGC\,330-011 & SMC5\_051901 &  00 56 56.75 & $-$72 17 44.71 & 10.44 & 13.91 & $-$0.02 & B9 Ib          & 156\p~(6)  & AzV 181 \\
NGC\,330-012 & SMC5\_015870 &  00 56 07.95 & $-$72 26 03.74 &\o1.91 & 13.92 & $-$0.01 & A0 Ib          & 128\p~(6)  & Arp 211 \\
NGC\,330-013 & SMC5\_001959 &  00 57 26.97 & $-$72 33 13.30 &\o7.48 & 14.00 & $-$0.13 & O8.5 II-III ((f)) &176\p~(11)& AzV 186, 2dFS\#1230\\
NGC\,330-014 & SMC5\_081774 &  00 56 38.79 & $-$72 25 13.75 &\o2.97 & 14.07 & $-$0.12 & B1.5 Ib        &159\p~(15)& AzV 176 \\
NGC\,330-015 & SMC5\_003863 &  00 55 21.62 & $-$72 21 50.23 &\o7.35 & 14.27 & \pp0.15 & A7-F0 II       & $-$        & AzV 167; FLAMES-UVES target \\
NGC\,330-016 & SMC5\_073986 &  00 55 21.48 & $-$72 23 38.01 &\o5.99 & 14.28 & $-$0.14 & B5: II         &130\p~(12)& contaminated by arcs \\
NGC\,330-017 & SMC5\_064339 &  00 56 47.98 & $-$72 28 46.86 &\o2.41 & 14.35 & $-$0.11 & B2 II          &157\p~(15)& AzV 178, 2dFS\#1171\\
NGC\,330-018 & SMC5\_078858 &  00 56 09.41 & $-$72 27 58.71 &\o0.73 & 14.37 & $-$0.13 & B3 II          &153\p~(11)& R74-B30\\
NGC\,330-019 & SMC5\_037327 &  00 55 52.85 & $-$72 26 31.42 &\o2.33 & 14.38 & $-$0.03 & B9 Ib          & 127\p~(8)  & \\
NGC\,330-020 & SMC5\_002300 &  00 57 28.08 & $-$72 31 03.34 &\o6.16 & 14.43 & $-$0.03 & B3 II          &158\p~(14)& 2dFS\#1232\\
NGC\,330-021 & SMC5\_003855 &  00 57 37.34 & $-$72 21 53.94 &\o8.35 & 14.45 & $-$0.25 & B0.2 III       &173\p~(10)& AzV 192, 2dFS\#1242\\
NGC\,330-022 & SMC5\_002377 &  00 55 15.68 & $-$72 30 33.73 &\o5.51 & 14.56 & $-$0.13 & B3 II          &147\p~(16)& 2dFS\#1062\\
NGC\,330-023 & SMC5\_190576 &  00 56 44.31 & $-$72 29 06.33 &\o2.33 & 14.56 & $-$0.12 & O9 V-IIIe      & 159:~(3)   & KWB330\#111, H$\alpha =$ strong em.\\
NGC\,330-024 & SMC5\_065615 &  00 54 58.57 & $-$72 25 09.71 &\o6.59 & 14.72 & $-$0.06 & B5 Ib          &161\p~(11)& 2dFS\#1034\\
NGC\,330-025 & SMC5\_037578 &  00 57 23.98 & $-$72 23 56.49 &\o6.24 & 14.81 & $-$0.11 & B1.5e          & 110:~(7)   & 2dFS\#1224, H$\alpha =$ twin\\
NGC\,330-026 & SMC5\_019099 &  00 57 23.40 & $-$72 22 45.68 &\o7.00 & 14.82 & $-$0.15 & B2.5 II        &152\p~(11)& \\
NGC\,330-027 & SMC5\_003910 &  00 56 18.27 & $-$72 21 33.39 &\o6.23 & 14.86 & $-$0.16 & B1 V           &170\p~(11)& \\
NGC\,330-028 & SMC5\_002054 &  00 57 06.05 & $-$72 32 39.25 &\o6.03 & 14.94 & $-$0.11 & B1 V           & 172\p~(9)  & 2dFS\#1195\\
NGC\,330-029 & SMC5\_000205 &  00 57 00.18 & $-$72 30 09.78 &\o3.92 & 14.97 & $-$0.09 & B0.2 V (Be-Fe)& Binary     & MA93\#973, KWB330\#104, $\dagger$, H$\alpha =$ em.\\
NGC\,330-030 & SMC5\_020391 &  00 56 23.58 & $-$72 21 23.64 &\o6.40 & 14.99 & $-$0.26 & B0.5 V         & Binary     & \\
NGC\,330-031 & SMC5\_002476 &  00 56 13.84 & $-$72 30 00.77 &\o2.26 & 15.00 & $-$0.04 & B0.5 V (Be-Fe)& 146\p~(4)  & 2dFS\#5088, H$\alpha =$em.\\
NGC\,330-032 & SMC5\_076906 &  00 57 38.56 & $-$72 30 38.67 &\o6.65 & 15.07 & $-$0.10 & B0.5 V         &162\p~(19)& \\
NGC\,330-033 & SMC5\_079846 &  00 57 10.52 & $-$72 30 04.17 &\o4.52 & 15.08 & $-$0.11 & B1.5 V         & 177:~(9)   & 2dFS\#5094, variable $v_r$?, contaminated by arcs\\
NGC\,330-034 & SMC5\_000183 &  00 54 40.74 & $-$72 30 28.56 &\o7.86 & 15.09 & $-$0.09 & B1-2e      & 121:~(5)   & MA93\#794\\
NGC\,330-035 & SMC5\_009426 &  00 57 23.92 & $-$72 32 37.52 &\o6.89 & 15.09 & $-$0.07 & B3 II          &143\p~(12)& \\
NGC\,330-036 & SMC5\_044567 &  00 56 10.66 & $-$72 28 10.25 &\o0.72 & 15.10 & $-$0.16 & B2 II          &156\p~(16)& R74-B32\\
NGC\,330-037 & SMC5\_004102 &  00 55 09.80 & $-$72 20 24.65 &\o9.02 & 15.12 & $-$0.02 & A2 II          & 145\p~(5)  & 2dFS\#1058\\
NGC\,330-038 & SMC5\_014400 &  00 57 15.31 & $-$72 27 33.94 &\o4.26 & 15.21 & $-$0.19 & B1 V           & 133:~(9)   & 2dFS\#1206\\
NGC\,330-039 & SMC5\_003405 &  00 55 55.67 & $-$72 24 32.80 &\o3.68 & 15.26 & $-$0.22 & B0 V           & 173:~(9)   & 2dFS\#1109\\
NGC\,330-040 & SMC5\_007692 &  00 55 03.41 & $-$72 34 12.96 &\o8.58 & 15.29 & $-$0.05 & B2 III         & 152\p~(7)  & 2dFS\#1041\\
NGC\,330-041 & SMC5\_077354 &  00 57 37.17 & $-$72 23 55.97 &\o7.05 & 15.38 & $-$0.24 & B0 V           &184\p~(10)& 2dFS\#1241\\
NGC\,330-042 & SMC5\_016204 &  00 56 27.09 & $-$72 25 42.59 &\o2.17 & 15.41 & $-$0.16 & B2 II          &130\p~(12)& \\
NGC\,330-043 & SMC5\_046989 &  00 56 27.11 & $-$72 25 04.74 &\o2.78 & 15.47 & $-$0.19 & B0 V           & 161:~(6)   & \\
NGC\,330-044 & SMC5\_037013 &  00 55 13.61 & $-$72 29 13.69 &\o5.12 & 15.50 & $-$0.02 & B1-2 (Be-Fe)   & 158:~(5)   & MA93\#824\\
NGC\,330-045 & SMC5\_003118 &  00 56 05.59 & $-$72 26 21.48 &\o1.74 & 15.54 & $-$0.15 & B3 III         & 159\p~(3)  & 2dFS\#5087\\
NGC\,330-046 & SMC5\_088493 &  00 58 12.39 & $-$72 26 12.09 &\o8.70 & 15.56 & $-$0.14 & O9.5 V         & 177\p~(6)  & 2dFS\#1293\\
NGC\,330-047 & SMC5\_021183 &  00 57 03.45 & $-$72 20 35.41 &\o7.94 & 15.61 & $-$0.28 & B1 V           &176\p~(13)& 2dFS\#1190, variable $v_r$?\\
NGC\,330-048 & SMC5\_042483 &  00 55 35.03 & $-$72 30 51.53 &\o4.51 & 15.62 & $-$0.19 & B0.5 V         & 181\p~(6)  & \\
NGC\,330-049 & SMC5\_086635 &  00 55 59.58 & $-$72 23 31.30 &\o4.50 & 15.63 & $-$0.27 & O9 V           & 158\p~(8)  & \\
NGC\,330-050 & SMC5\_004159 &  00 56 56.44 & $-$72 20 07.40 &\o8.17 & 15.66 & $-$0.18 & B3e            & 113\p~(7)  & MA93\#969, H$\alpha =$ twin\\
NGC\,330-051 & SMC5\_082766 &  00 57 57.69 & $-$72 27 10.54 &\o7.47 & 15.66 & $-$0.16 & B1.5 V         & 174\p~(5)  & 2dFS\#1276\\
NGC\,330-052 & SMC5\_000744 &  00 56 31.00 & $-$72 18 52.99 &\o8.95 & 15.69 & $-$0.26 & O8.5 Vn        & 166:~(7)   & 2dFS\#1152\\
NGC\,330-053$\dagger$ & SMC5\_013120 &  00 57 58.81 & $-$72 28 50.68 &\o7.61 & 15.69 & $-$0.12 & B0.5 V         &167\p~(10)& variable $v_r$? \\
NGC\,330-054 & SMC5\_037029 &  00 55 28.20 & $-$72 29 06.50 &\o4.03 & 15.70 & $-$0.08 & B2 (Be-Fe)    & 127:~(6)   & MA93\#851, H$\alpha =$ em.\\
NGC\,330-055 & SMC5\_012510 &  00 56 49.24 & $-$72 29 28.96 &\o2.85 & 15.72 & $-$0.11 & B0.5 V         & 174:~(7)   & variable $v_r$? \\
NGC\,330-056 & SMC5\_002447 &  00 55 23.37 & $-$72 30 09.56 &\o4.80 & 15.76 & $-$0.19 & B2 III         & 122:~(9)   & \\
NGC\,330-057 & SMC5\_009833 &  00 54 54.63 & $-$72 32 09.58 &\o7.70 & 15.82 & $-$0.17 & B0.5 V         & 124:~(4)   & \\
NGC\,330-058 & SMC5\_036895 &  00 54 58.54 & $-$72 30 22.38 &\o6.58 & 15.84 & $-$0.11 & B3:            & 112:~(2)   & \\
NGC\,330-059 & SMC5\_002498 &  00 57 52.13 & $-$72 29 51.87 &\o7.33 & 15.85 & $-$0.08 & B3 III         &151:~(11) & \\
NGC\,330-060 & SMC5\_011393 &  00 55 59.43 & $-$72 30 34.33 &\o3.14 & 15.86 & $-$0.06 & B2.5 (Be-Fe)  & 156:~(7)   & MA93\#890, H$\alpha =$ em.\\
NGC\,330-061 & SMC5\_004133 &  00 56 00.52 & $-$72 20 15.94 &\o7.65 & 15.88 & $-$0.10 & A0 II          & 131\p~(3)  & 2dFS\#1118\\
NGC\,330-062 & SMC5\_045835 &  00 56 18.05 & $-$72 26 35.82 &\o1.19 & 15.90 & $-$0.10 & B3e            & 156:~(6)   & H$\alpha =$ twin \\
NGC\,330-063 & SMC5\_080598 &  00 56 06.81 & $-$72 30 28.62 &\o2.84 & 15.90 & $-$0.06 & B1-3           & 121:~(9)   & \\
NGC\,330-064 & SMC5\_061057 &  00 55 45.87 & $-$72 33 13.06 &\o5.97 & 15.96 & $-$0.09 & B3:e          & 135\p~(4)  & \\
NGC\,330-065 & SMC5\_002751 &  00 56 14.27 & $-$72 28 30.13 &\o0.79 & 15.99 & $-$0.04 & B1-3 (Be-Fe)   & 158:~(8)   & R74-B34, KWB330\#239, $\dagger$, H$\alpha =$ em.\\
NGC\,330-066 & SMC5\_010645 &  00 55 55.96 & $-$72 31 19.82 &\o3.94 & 15.99 & $-$0.10 & B3 III         & 152\p~(8)  & \\
NGC\,330-067 & SMC5\_079924 &  00 55 34.83 & $-$72 27 18.75 &\o3.35 & 16.00 & $-$0.17 & B2.5 III       & 141\p~(8)  & \\
NGC\,330-068 & SMC5\_009989 &  00 55 35.52 & $-$72 32 00.30 &\o5.33 & 16.02 & $-$0.06 & B1.5 (Be-Fe)  & 133\p~(9)  & MA93\#860, H$\alpha =$ twin\\
NGC\,330-069 & SMC5\_043375 &  00 56 48.62 & $-$72 29 40.11 &\o2.93 & 16.02 & $-$0.05 & B3 III         & 160:~(7)   & \\
NGC\,330-070 & SMC5\_077231 &  00 57 02.19 & $-$72 25 55.33 &\o3.76 & 16.02 & $-$0.05 & B0.5e          & 128\p~(4)  & H$\alpha =$ twin \\
NGC\,330-071 & SMC5\_014767 &  00 56 59.92 & $-$72 27 04.66 &\o3.18 & 16.03 & $-$0.09 & B3 III         &134\p~(10)& \\
NGC\,330-072 & SMC5\_017978 &  00 56 09.16 & $-$72 23 55.23 &\o3.93 & 16.03 & $-$0.20 & B0.5 V         & 136:~(6)   & \\
NGC\,330-073 & SMC5\_047763 &  00 55 42.62 & $-$72 23 58.22 &\o4.69 & 16.04 & $-$0.08 & B8 Ib          & 58\p~(6) & KWB330\#522, $\dagger$, H$\alpha =$ em., variable $v_r$?\\
NGC\,330-074 & SMC5\_002782 &  00 54 41.61 & $-$72 28 15.74 &\o7.34 & 16.09 & $-$0.24 & B0 V           & 156\p~(6)  & \\
NGC\,330-075 & SMC5\_004413 &  00 55 52.41 & $-$72 18 45.05 &\o9.25 & 16.10 & $-$0.17 & B8 II          & 131:~(3)   & \\
NGC\,330-076 & SMC5\_014864 &  00 56 33.11 & $-$72 27 04.99 &\o1.29 & 16.12 & $-$0.05 & B3 (Be-Fe)    &126:~(10) & R74-B09, KWB330\#419, $\dagger$, H$\alpha =$ em. \\
NGC\,330-077 & SMC5\_000166 &  00 55 43.22 & $-$72 30 49.00 &\o4.05 & 16.13 & $-$0.01 & B0-3 (Be-Fe)   & $-$        & $\dagger$, low S/N, strongly contaminated by arcs \\
NGC\,330-078 & SMC5\_003498 &  00 55 11.95 & $-$72 23 54.90 &\o6.35 & 16.15 & $-$0.03 & A0: III        & 112\p~(2)  & \\
NGC\,330-079 & SMC5\_037369 &  00 55 52.41 & $-$72 26 11.32 &\o2.55 & 16.16 & $-$0.13 & B3 III         & 155\p~(5)  & variable $v_r$?\\
NGC\,330-080 & SMC5\_037034 &  00 57 35.30 & $-$72 29 04.90 &\o5.91 & 16.17 & $-$0.07 & B1-3           & 141:~(5)   & MA93\#1019\\
NGC\,330-081 & SMC5\_044096 &  00 56 02.28 & $-$72 28 44.96 &\o1.57 & 16.17 & $-$0.16 & B1-3           & 130:~(8)   & \\
NGC\,330-082 & SMC5\_044447 &  00 57 24.60 & $-$72 28 17.06 &\o4.98 & 16.18 & $-$0.07 & B1-3           & 141:~(8)   & contaminated by arcs \\
NGC\,330-083 & SMC5\_087520 &  00 57 04.56 & $-$72 22 53.62 &\o5.99 & 16.19 & $-$0.13 & B3 III         & 140\p~(9)  & \\
NGC\,330-084 & SMC5\_044140 &  00 55 36.33 & $-$72 28 40.84 &\o3.32 & 16.20 & $-$0.08 & B3 III-V       & Binary (SB1) & \\
NGC\,330-085 & SMC5\_073581 &  00 56 26.60 & $-$72 26 23.02 &\o1.52 & 16.22 & $-$0.12 & B3:e      & 110:~(3)   & H$\alpha =$ twin \\
NGC\,330-086 & SMC5\_012975 &  00 56 33.74 & $-$72 29 01.55 &\o1.67 & 16.25 & $-$0.15 & B2.5 III       & 128:~(9)   & R74-B01\\
NGC\,330-087 & SMC5\_000135 &  00 55 43.50 & $-$72 31 32.61 &\o4.60 & 16.26 & $-$0.12 & Be-Fe          & $-$        & MA93\#869, H$\alpha =$ em.\\
NGC\,330-088 & SMC5\_009618 &  00 55 56.53 & $-$72 32 25.98 &\o4.94 & 16.30 & $-$0.14 & B1-3           & $-$        & \\
NGC\,330-089 & SMC5\_002393 &  00 55 37.99 & $-$72 30 27.19 &\o4.07 & 16.32 & $-$0.17 & B1-5           & 154:~(4)   & contaminated by arcs \\
NGC\,330-090 & SMC5\_073266 &  00 55 44.03 & $-$72 28 10.44 &\o2.65 & 16.34 & $-$0.15 & B3 III         & 117:~(5)   & \\
NGC\,330-091 & SMC5\_080521 &  00 57 05.74 & $-$72 33 41.81 &\o6.89 & 16.34 & $-$0.13 & B0e            & 162\p~(7)  & H$\alpha =$ twin$+$abs.\\
NGC\,330-092 & SMC5\_002411 &  00 57 48.75 & $-$72 30 19.28 &\o7.23 & 16.35 & $-$0.03 & B3:            & $-$        & \\
NGC\,330-093 & SMC5\_036967 &  00 55 40.11 & $-$72 29 44.78 &\o3.51 & 16.36 & $-$0.14 & Be-Fe          & $-$        & MA93\#863, KWB330\#528, H$\alpha =$ twin \\
NGC\,330-094 & SMC5\_082714 &  00 55 33.33 & $-$72 23 57.34 &\o5.14 & 16.36 & $-$0.21 & B1-5           & 179:~(2)   & contaminated by arcs \\
NGC\,330-095 & SMC5\_014467 &  00 56 15.39 & $-$72 27 29.32 &\o0.39 & 16.38 & $-$0.15 & B3 III         &155\p~(10)& \\
NGC\,330-096 & SMC5\_014644 &  00 54 39.14 & $-$72 27 16.74 &\o7.52 & 16.38 & $-$0.08 & B1-3 (Be-Fe)   & $-$        & MA93\#782, H$\alpha =$ em. \\
NGC\,330-097 & SMC5\_002965 &  00 56 17.13 & $-$72 27 17.95 &\o0.50 & 16.40 & $-$0.19 & B1 V           & 148:~(7)   & \\
NGC\,330-098 & SMC5\_065934 &  00 56 07.74 & $-$72 24 18.13 &\o3.58 & 16.40 & $-$0.24 & B0.2: V        & 175:~(2)   & contaminated by arcs \\
NGC\,330-099 & SMC5\_044674 &  00 55 01.81 & $-$72 28 01.35 &\o5.80 & 16.41 & $-$0.18 & B2-3           & $-$        & \\
NGC\,330-100 & SMC5\_002442 &  00 56 54.25 & $-$72 30 12.22 &\o3.60 & 16.43 & \pp0.00 & Be (B0-3)      & $-$        & H$\alpha =$twin \\
NGC\,330-101 & SMC5\_002817 &  00 56 35.34 & $-$72 28 06.57 &\o1.29 & 16.44 & $-$0.14 & B2.5 III       & 154\p~(8)  & \\
NGC\,330-102 & SMC5\_014776 &  00 57 12.14 & $-$72 27 10.37 &\o4.06 & 16.46 & $-$0.16 & B2-3 III       & 162:~(5)   & \\
NGC\,330-103 & SMC5\_002256 &  00 56 52.31 & $-$72 31 18.50 &\o4.33 & 16.47 & $-$0.07 & B1-3           & $-$        & contaminated by arcs \\
NGC\,330-104 & SMC5\_048561 &  00 55 01.81 & $-$72 22 49.67 &\o7.63 & 16.50 & $-$0.30 & B0: V          & 129:~(4)   & 2dFS\#1037 \\
NGC\,330-105 & SMC5\_020588 &  00 57 05.64 & $-$72 21 11.89 &\o7.47 & 16.53 & $-$0.17 & B1-3           & $-$        & \\
NGC\,330-106 & SMC5\_011915 &  00 56 35.13 & $-$72 29 58.90 &\o2.52 & 16.54 & $-$0.15 & B1-2           &135\p~(12)& \\
NGC\,330-107 & SMC5\_014046 &  00 55 57.78 & $-$72 27 51.05 &\o1.58 & 16.56 & $-$0.12 & B3: III-V      & 138\p~(5)  & \\
NGC\,330-108 & SMC5\_015203 &  00 55 14.50 & $-$72 26 43.73 &\o4.96 & 16.56 & $-$0.13 & B5 III         & 126:~(2)   & \\
NGC\,330-109 & SMC5\_065064 &  00 56 24.94 & $-$72 26 48.43 &\o1.08 & 16.56 & $-$0.17 & B3 III         & 144:~(4)   & \\
NGC\,330-110 & SMC5\_037341 &  00 56 20.67 & $-$72 26 25.49 &\o1.37 & 16.57 & $-$0.19 & B2 III         & 160:~(6)   & \\
NGC\,330-111 & SMC5\_013331 &  00 56 17.70 & $-$72 28 37.71 &\o0.85 & 16.58 & $-$0.17 & B1-3           & $-$        & \\
NGC\,330-112 & SMC5\_048045 &  00 55 56.33 & $-$72 23 33.34 &\o4.56 & 16.59 & $-$0.11 & B1-3 (Be-Fe)   & $-$        & MA93\#886, KWB330\#509, H$\alpha =$ em.\\
NGC\,330-113 & SMC5\_015068 &  00 55 48.73 & $-$72 26 51.42 &\o2.45 & 16.61 & $-$0.14 & B1-3           & $-$        & \\
NGC\,330-114 & SMC5\_078349 &  00 56 57.01 & $-$72 25 31.71 &\o3.66 & 16.62 & $-$0.14 & B2 III         &157\p~(11)& \\
NGC\,330-115 & SMC5\_061511 &  00 55 15.38 & $-$72 24 23.49 &\o5.86 & 16.65 & $-$0.26 & B1-3           & Binary     & \\
NGC\,330-116 & SMC5\_045013 &  00 55 26.53 & $-$72 27 33.66 &\o3.94 & 16.66 & $-$0.19 & B3 III         & 133\p~(6)  & \\
NGC\,330-117 & SMC5\_049053 &  00 56 39.75 & $-$72 22 06.43 &\o5.89 & 16.67 & $-$0.21 & B1-3           & 120:~(6)   & \\
NGC\,330-118 & SMC5\_037899 &  00 54 47.70 & $-$72 20 51.51 &\o9.75 & 16.69 & $-$0.26 & B1-2           & 170\p~(4)  & 2dFS\#1013\\
NGC\,330-119 & SMC5\_191730 &  00 56 45.39 & $-$72 27 55.47 &\o2.01 & 16.71 & $-$0.22 & B1-3           & $-$        & \\
NGC\,330-120 & SMC5\_191173 &  00 56 42.59 & $-$72 27 27.78 &\o1.82 & 16.89 & $-$0.42 & B3: III-V      & 147:~(5)   & \\
\hline
NGC\,330-121 & SMC5\_081980 &  00 56 22.57 & $-$72 28 35.68 &\o0.86 & 12.62 & \pp0.20 & A5 II          & 154\p~(5)  & R74-B38; UVES target \\
NGC\,330-122 & SMC5\_015195 &  00 56 18.56 & $-$72 26 45.21 &\o1.03 & 14.00 & \pp0.04 & A2 II          & 138\p~(5)  & R74-B16; UVES target \\
NGC\,330-123 & SMC5\_037226 &  00 56 03.92 & $-$72 27 13.01 &\o1.26 & 15.79 & $-$0.22 & O9.5 V         &177\p~(11)& R74-B18; UVES target \\
NGC\,330-124 & SMC5\_013293 &  00 56 06.81 & $-$72 28 35.08 &\o1.21 & 15.83 & $-$0.20 & B0.2 V         &155\p~(12)& R74-B28; UVES target \\
NGC\,330-125 & SMC5\_003014 &  00 56 20.18 & $-$72 27 02.02 &\o0.76 & 15.89 & $-$0.17 & B2 III         & 151\p~(8)  & R74-B13; UVES target \\
\end{longtable}
\end{center}
}

\newpage

{\scriptsize
\begin{center}
\begin{longtable}{lcccclll}
\caption[]{N11: Observational parameters of target stars.  The cross-references in the 
final column are to \citet[][Sk]{sk69} and \citet[][P]{p92}  The photometry for
stars flagged with an asterisk was taken from \citet{p92}.  Classifications for N11-031 and
N11-060 were taken from \citet{w02} and \citet{w04} respectively.
\label{lh910}} \\
\hline
ID & $\alpha$(2000) & $\delta$(2000) & $V$ & $B-V$ & Sp. Type & $v_{\rm r}$ & Comments \\
\hline 
\endfirsthead
\caption[]{\it{continued}} \\
\hline
ID & $\alpha$(2000) & $\delta$(2000) & $V$ & $B-V$ & Sp. Type & $v_{\rm r}$  & Comments \\
\hline 
\endhead
\hline 
\multicolumn{8}{r}{\it{continued on next page}} \\
\endfoot
\hline
\multicolumn{8}{l}{$\dagger$: Further notes on individual stars}\\
\multicolumn{8}{l}{{\it N11-023 \& N11-057:} These two stars are unresolved in the charts of \citet{sk69} and together were identified Sk$-$66$^\circ$24.}\\
\multicolumn{8}{l}{{\it N11-048:} Blended with Parker 3209, which together likely comprise Sk$-$66$^\circ$33.}\\
\multicolumn{8}{l}{{\it N11-092:} Blended with another star in the WFI image, which together comprise Sk$-$66$^\circ$19.}\\
\endlastfoot
\hline 
N11-001 & 04 57 08.85 & $-$66 23 25.1 &\a11.35$^\ast$ & \pp0.06 & B2 Ia             &299\p~(12) & Sk$-$66$^\circ$36, P3252; UVES spectrum, H$\alpha =$ P Cyg em. \\
N11-002 & 04 56 23.51 & $-$66 29 51.7 &\a11.90$^\ast$ & \pp0.37 & B3 Ia             &295\p~(12) & Sk$-$66$^\circ$27, P1062; UVES spectrum, H$\alpha =$ P Cyg em. \\
N11-003 & 04 56 50.59 & $-$66 24 34.9 &\a12.47$^\ast$ & $-$0.02 & B1 Ia             &287\p~(12) & P3157; UVES spectrum, H$\alpha =$ broad em. \\
N11-004 & 04 55 29.42 & $-$66 23 12.0 & 12.56         & $-$0.06 & O9.7 Ib           &304\p~(15) & Sk$-$66$^\circ$16 \\
N11-005 & 04 57 04.10 & $-$66 29 10.1 & 12.62         & \pp0.03 & B1 Ia             & Binary     & Sk$-$66$^\circ$34, H$\alpha =$ em. \\
N11-006 & 04 56 51.45 & $-$66 28 06.4 & 12.66         & \pp0.51 & F8:               & $-$        & FLAMES-UVES target, foreground \\
N11-007 & 04 56 17.29 & $-$66 31 03.6 & 12.74         & $-$0.03 & O8 Ib (f)         &302\p~(14) & Sk$-$66$^\circ$25, variable $v_r$ or wind var.? \\
N11-008 & 04 55 22.35 & $-$66 28 18.9 & 12.77         & $-$0.01 & B0.5 Ia           &288\p~(14) & Sk$-$66$^\circ$15 \\
N11-009 & 04 57 17.68 & $-$66 26 31.5 & 12.80         & \pp0.06 & B3 Iab            &288\p~(15) & \\
N11-010 & 04 56 40.94 & $-$66 27 40.1 & 12.89         & $-$0.14 & O9.5 III + B1-2:  & Binary (SB2)     & P1310\\
N11-011 & 04 55 55.50 & $-$66 28 20.3 & 12.89         & $-$0.08 & OC9.5 II          & Binary (SB1)    & Sk$-$66$^\circ$17 \\
N11-012 & 04 56 51.15 & $-$66 31 48.3 & 12.90         & \pp0.02 & B1 Ia             &295\p~(11) & varaible $v_r$? \\
N11-013 & 04 57 00.86 & $-$66 24 24.8 & 12.93         & $-$0.07 & O8 V              & Binary (SB2)    & P3223\\
N11-014 & 04 56 48.02 & $-$66 20 09.8 & 12.98         & $-$0.02 & B2 Iab            &298\p~(13) & Sk$-$66$^\circ$30 \\
N11-015 & 04 57 22.08 & $-$66 24 27.5 & 13.00         & $-$0.10 & B0.7 Ib           &307\p~(12) & Sk$-$66$^\circ$37, P3271; UVES spectrum \\
N11-016 & 04 56 20.59 & $-$66 27 14.0 & 13.05         & $-$0.12 & B1 Ib             &294\p~(12) & Sk$-$66$^\circ$26, P1036; UVES spectrum \\
N11-017 & 04 56 17.57 & $-$66 18 18.5 & 13.11         & \pp0.02 & B2.5 Iab          &291\p~(14) & Sk$-$66$^\circ$23 \\
N11-018 & 04 56 41.04 & $-$66 24 40.3 &\a13.13$^\ast$ & $-$0.09 & O6 II(f$^+$)  & 301\p~(6)  & P3053, variable $v_r$ or wind var.?\\
N11-019 & 04 56 11.72 & $-$66 31 59.1 & 13.14         & $-$0.17 & O8-9 III-V((f))   & Binary (SB2)     & Sk$-$66$^\circ$22 \\
N11-020 & 04 56 50.32 & $-$66 31 03.6 & 13.18         & $-$0.22 & O5 I(n)fp      & Binary     & H$\alpha =$ broad em.\\
N11-021 & 04 56 30.58 & $-$66 18 08.6 & 13.24         & \pp0.12 & A7 II             & 298\p~(4)  & \\
N11-022 & 04 56 00.90 & $-$66 26 16.4 & 13.33         & $-$0.15 & O6.5 II(f) var    & 289\p~(8)  & Sk$-$66$^\circ$20 \\
N11-023$\dagger$ & 04 56 15.44 & $-$66 27 34.9 & 13.40         & $-$0.14 & B0.7 Ib           &286\p~(12) & P1014; UVES spectrum \\
N11-024 & 04 55 32.93 & $-$66 25 27.7 & 13.45         & $-$0.10 & B1 Ib             &292\p~(12) & \\
N11-025 & 04 56 34.84 & $-$66 28 23.1 & 13.45         & \pp0.35 & O8.5 V            & $-$        & FLAMES-UVES target \\
N11-026 & 04 56 52.54 & $-$66 19 55.8 & 13.51         & $-$0.17 & O2.5 III(f$^\ast$)& 330\p~(8)  & \\
N11-027 & 04 56 53.19 & $-$66 19 01.9 & 13.56         & \pp0.68 & G2                & $-$        & FLAMES-UVES target, foreground \\
N11-028 & 04 57 16.24 & $-$66 23 20.8 & 13.63         & \pp0.10 & O6-8 V            & 304\p~(3)  & P3264; strong nebular contamination \\
N11-029 & 04 55 56.34 & $-$66 29 03.9 & 13.63         & $-$0.12 & OC9.7 Ib          &297\p~(15) & \\
N11-030 & 04 55 48.38 & $-$66 29 41.4 & 13.66         & $-$0.10 & B1e               & Binary     & H$\alpha =$ broad em.\\
N11-031 & 04 56 42.49 & $-$66 25 18.0 &\a13.68$^\ast$ & $-$0.01 & ON2 III (f$^\ast$)& 322\p~(6)  & P3061 \\
N11-032 & 04 56 54.46 & $-$66 24 15.6 &\a13.68$^\ast$ & $-$0.11 & O7 II(f)          & 305\p~(9)  & P3168; no observations in HR02/\lam3958 region \\
N11-033 & 04 56 11.04 & $-$66 28 23.9 & 13.68         & $-$0.16 & B0 IIIn           & 296:~(9)   & P1005 \\
N11-034 & 04 56 42.66 & $-$66 29 45.1 & 13.68         & $-$0.15 & B0.5 III          & 289:~(10)  & P1332; variable $v_r$? \\
N11-035 & 04 57 28.70 & $-$66 31 02.4 & 13.69         & \pp0.03 & O9 II(f)          & Binary (SB1)     & \\
N11-036 & 04 57 41.00 & $-$66 29 56.4 & 13.72         & $-$0.15 & B0.5 Ib           &292\p~(16) & \\
N11-037 & 04 56 22.33 & $-$66 28 04.6 & 13.77         & $-$0.10 & B0 III            & Binary (SB1)    & P1052 \\
N11-038 & 04 56 45.21 & $-$66 25 10.6 &\a13.81$^\ast$ & \pp0.00 & O5 III(f$+$)      & 318\p~(4)  & P3100 \\
N11-039 & 04 56 17.33 & $-$66 17 48.0 & 13.83         & $-$0.18 & B2 III            & 286\p~(9)  & \\
N11-040 & 04 57 16.72 & $-$66 28 22.7 & 13.84         & $-$0.10 & B0: IIIn          & Binary?  & H$\alpha =$ abs+twin em.\\
N11-041 & 04 56 52.32 & $-$66 32 52.5 & 13.87         & $-$0.19 & O6.5 Iaf          & Binary & H$\alpha =$ complex abs+broad em.\\
N11-042 & 04 56 15.57 & $-$66 27 21.2 & 13.93         & $-$0.23 & B0 III            &288\p~(15) & P1017 \\
N11-043 & 04 57 01.06 & $-$66 28 29.8 & 13.93         & $-$0.21 & O7 III $+$ B0:    & Binary (SB2)    & \\
N11-044 & 04 57 15.74 & $-$66 33 54.0 & 13.96         & $-$0.14 & O6-8 III-V((f))   & Binary (SB2)     & early B-type secondary \\
N11-045 & 04 56 58.32 & $-$66 31 32.9 & 13.97         & $-$0.15 & O9-9.5 III        &290\p~(13) & \\
N11-046 & 04 56 44.62 & $-$66 34 20.5 & 13.98         & $-$0.24 & O9.5 V            & Binary     & \\
N11-047 & 04 56 25.48 & $-$66 26 33.2 & 14.01         & $-$0.23 & B0 III            & Binary (SB1)     & \\
N11-048$\dagger$ & 04 56 58.79 & $-$66 24 40.7 &\a14.02$^\ast$ & $-$0.17 & O6.5 V((f))       & 299\p~(3)  & P3204 \\
N11-049 & 04 56 29.59 & $-$66 28 20.5 & 14.02         & $-$0.24 & O7.5 V            & Binary (SB1)   & P1110 \\
N11-050 & 04 57 00.88 & $-$66 23 57.1 & 14.03         & \pp0.10 & O4-5 $+$ O7:      & Binary  (SB2)   & P3224 \\
N11-051 & 04 56 29.72 & $-$66 21 38.5 & 14.03         & $-$0.26 & O5 Vn((f))        & 296:~(5)  & \\
N11-052 & 04 57 40.12 & $-$66 26 02.8 & 14.06         & $-$0.18 & O9.5 V            & Binary (SB2)    & \\
N11-053 & 04 56 54.48 & $-$66 27 32.0 & 14.09         & $-$0.16 & B1: (Be-Fe)       & Binary   & \\
N11-054 & 04 57 18.33 & $-$66 25 59.6 & 14.10         & $-$0.06 & B1 Ib             &294\p~(12) & \\
N11-055 & 04 57 10.57 & $-$66 18 06.8 & 14.11         & \pp0.18 & O8-9 IIIne        & 293\p~(3)  & H$\alpha =$ broad em.\\
N11-056 & 04 57 49.05 & $-$66 24 17.4 & 14.13         & $-$0.10 & B1e               & 304\p~(8)  & H$\alpha =$ weak twin em.\\
N11-057$\dagger$ & 04 56 15.48 & $-$66 27 41.5 & 14.13         & \pp0.03 & A0 II             & 284\p~(4) & P1015 \\
N11-058 & 04 55 52.35 & $-$66 34 13.4 & 14.16         & $-$0.23 & O5.5 V((f))       & 301\p~(3)  & \\
N11-059 & 04 56 31.02 & $-$66 28 40.8 & 14.23         & $-$0.25 & O9 V              & Binary (SB1)    & P1125 \\
N11-060 & 04 56 42.16 & $-$66 24 54.4 &\a14.24$^\ast$ & $-$0.06 & O3 V((f$^\ast$))  & 314\p~(3)  & P3058 \\
N11-061 & 04 57 25.73 & $-$66 21 05.4 & 14.24         & $-$0.06 & O9 V              &305\p~(10) & \\
N11-062 & 04 56 55.35 & $-$66 33 01.8 & 14.32         & $-$0.23 & B0.2 V            & Binary (SB1)    & \\
N11-063 & 04 57 40.53 & $-$66 27 25.1 & 14.35         & $-$0.18 & O9: Vn            & Binary (SB2)    & \\
N11-064 & 04 57 11.47 & $-$66 22 18.8 & 14.40         & $-$0.14 & B0.2: Vn          & Binary     & \\
N11-065 & 04 56 19.35 & $-$66 27 01.8 & 14.40         & $-$0.24 & O6.5 V((f))       & 303\p~(7)  & P1027 \\
N11-066 & 04 56 57.51 & $-$66 35 21.3 & 14.40         & $-$0.24 & O7 V((f))         & 282\p~(7)  & \\
N11-067 & 04 57 03.33 & $-$66 30 08.1 & 14.50         & $-$0.24 & B0.5: Vn          & Binary    & \\
N11-068 & 04 56 04.61 & $-$66 23 57.8 & 14.55         & $-$0.23 & O7 V((f))         &291\p~(10) & \\
N11-069 & 04 56 20.80 & $-$66 27 03.3 & 14.56         & $-$0.19 & B1 III            &279\p~(12) & P1037; UVES spectrum \\
N11-070 & 04 57 04.61 & $-$66 24 22.3 & 14.59         & $-$0.13 & B3 III            & 274\p~(8)   & P3239 \\
N11-071 & 04 57 24.19 & $-$66 26 00.0 & 14.61         & $-$0.19 & O8: V             & Binary (SB2)    & \\
N11-072 & 04 55 51.63 & $-$66 21 57.5 & 14.61         & $-$0.27 & B0.2 V            &296\p~(13) & \\
N11-073 & 04 57 11.26 & $-$66 26 53.8 & 14.63         & $-$0.06 & B0.5 (Be-Fe)      & 288:~(6)   & H$\alpha =$ broad em.\\
N11-074 & 04 56 19.49 & $-$66 27 37.9 & 14.63         & $-$0.01 & B0.5 (Be-Fe)      & 288:~(5)   & P1028; H$\alpha =$ broad em.\\
N11-075 & 04 57 22.05 & $-$66 19 20.7 & 14.67         & $-$0.18 & B2 III            & Binary?   & Gaps in spectra due to bad pixels \\
N11-076 & 04 56 33.58 & $-$66 23 28.0 & 14.67         & \pp0.33 & B0.2 Ia           & 331\p~(9)  & H$\alpha =$ P Cyg., variable $v_r$?\\
N11-077 & 04 56 01.36 & $-$66 20 51.2 & 14.68         & $-$0.14 & B2 III            & Binary (SB1)    & \\
N11-078 & 04 56 04.94 & $-$66 31 24.8 & 14.69         & $-$0.19 & B2 (Be-Fe)        & Binary (SB1)    & H$\alpha =$ broad em.\\
N11-079 & 04 56 47.22 & $-$66 24 42.1 &\a14.71$^\ast$ & $-$0.12 & B0.2 V            & 305\p~(8)  & P3128 \\
N11-080 & 04 56 54.71 & $-$66 24 54.3 &\a14.71$^\ast$ & $-$0.15 & O7: V $+$ O9:     & Binary (SB2)    & P3173 \\
N11-081 & 04 55 17.87 & $-$66 31 49.7 & 14.72         & \pp0.09 & B0: n (Be-Fe)     & $-$    &H$\gamma =$twin em.$+$str. nebular; {\it HeII \lam4686: $v_r = 286$} \\
N11-082 & 04 57 32.49 & $-$66 29 49.1 & 14.72         & $-$0.16 & B1-2 $+$early-B   & Binary (SB2)    & \\
N11-083 & 04 57 00.24 & $-$66 32 16.3 & 14.73         & $-$0.20 & B0.5 V            &287\p~(12) & variable $v_r$?\\
N11-084 & 04 56 09.20 & $-$66 27 08.5 & 14.75         & $-$0.21 & B0.5 V            & 306:~(8)   & \\
N11-085 & 04 56 39.34 & $-$66 28 59.0 & 14.75         & $-$0.16 & B0.5 V            &293:~(10) & variable $v_r$?\\
N11-086 & 04 56 43.69 & $-$66 28 36.0 & 14.75         & $-$0.21 & B1 V              & 284\p~(7)  & \\
N11-087 & 04 56 39.18 & $-$66 24 50.0 &\a14.76$^\ast$ & $-$0.10 & O9.5 Vn           & 309\p~(2)  & P3042 \\
N11-088 & 04 56 32.93 & $-$66 28 52.1 & 14.78         & $-$0.21 & B1 III            & 281:~(6)   & P1160 \\
N11-089 & 04 55 28.33 & $-$66 29 41.6 & 14.81         & $-$0.03 & B2 III            & Binary (SB1)   & \\
N11-090 & 04 57 32.51 & $-$66 25 56.8 & 14.83         & $-$0.08 & B2e               & 292:~(5)   & H$\alpha =$ broad em.\\
N11-091 & 04 57 01.32 & $-$66 20 13.0 & 14.85         & $-$0.10 & O9 V             & Binary     & \\
N11-092$\dagger$ & 04 55 54.19 & $-$66 25 01.5 & 14.87         & \pp0.11 & O7 V              & Binary    & \\
N11-093 & 04 55 29.95 & $-$66 30 37.5 & 14.87         & $-$0.14 & B2.5 III          &293\p~(10) & \\
N11-094 & 04 57 17.08 & $-$66 30 22.5 & 14.87         & $-$0.18 & B1 III            & Binary (SB1)    & \\
N11-095 & 04 57 37.92 & $-$66 24 59.7 & 14.89         & $-$0.18 & B1 Vn             & 295:~(5)   & \\
N11-096 & 04 56 54.08 & $-$66 21 56.3 & 14.89         & $-$0.19 & B1.5 III          & Binary (SB2)   & \\
N11-097 & 04 55 17.66 & $-$66 27 35.8 & 14.90         & \pp0.05 & B3 II             &294\p~(12) & \\
N11-098 & 04 56 58.97 & $-$66 34 46.3 & 14.93         & $-$0.18 & B2 III            & Binary  (SB1)    & \\
N11-099 & 04 57 46.69 & $-$66 28 27.8 & 14.93         & $-$0.11 & B0.2 V            & Binary  (SB1)   & strong nebular contamination \\
N11-100 & 04 57 21.21 & $-$66 25 01.2 & 14.94         & $-$0.20 & B0.5 V            &345\p~(12) & P3270; UVES spectrum \\
N11-101 & 04 56 38.12 & $-$66 23 54.2 &\a14.95$^\ast$ & $-$0.14 & B0.2 V            &302\p~(10) & P3033; UVES spectrum \\
N11-102 & 04 55 53.84 & $-$66 34 29.4 & 14.95         & $-$0.16 & B0.2 V            & 296:~(5)    & \\
N11-103 & 04 56 43.96 & $-$66 28 14.5 & 14.95         & $-$0.14 & B1-2 $+$early-B   & Binary     & \\
N11-104 & 04 55 55.23 & $-$66 21 57.2 & 14.96         & $-$0.19 & B1.5 V            & 263:~(9)   & \\
N11-105 & 04 56 46.37 & $-$66 27 16.7 &\a14.97$^\ast$ & $-$0.25 & B1 V              & 295\p~(8)  & P1378; variable $v_r$?\\
N11-106 & 04 56 42.08 & $-$66 31 19.0 & 14.99         & $-$0.26 & B0 V              &290\p~(13) & \\
N11-107 & 04 55 32.90 & $-$66 32 31.3 & 15.00         & $-$0.14 & B1-2 $+$early-B   & Binary (SB2)    & \\
N11-108 & 04 57 09.31 & $-$66 22 11.8 & 15.04         & $-$0.18 & O9.5 V          &306\p~(15) & \\
N11-109 & 04 55 49.10 & $-$66 27 39.1 & 15.07         & \pp0.10 & B0.5 Ib           &300\p~(15) & \\
N11-110 & 04 57 37.11 & $-$66 23 44.7 & 15.08         & $-$0.05 & B1 III            &298\p~(14) & \\
N11-111 & 04 56 05.86 & $-$66 21 54.8 & 15.11         & $-$0.24 & B1.5 III          &285\p~(10) & \\
N11-112 & 04 56 58.50 & $-$66 32 43.6 & 15.12         & $-$0.11 & B1 Vn             & Binary    & \\
N11-113 & 04 56 46.72 & $-$66 29 11.4 & 15.13         & $-$0.26 & B0.5 III          & Binary (SB2)     & Early B-type companion \\
N11-114 & 04 57 37.69 & $-$66 24 48.5 & 15.15         & $-$0.21 & B0 Vn             & 296\p~(6)   & variable $v_r$?\\
N11-115 & 04 57 43.26 & $-$66 24 38.0 & 15.15         & $-$0.14 & B1 III            &299\p~(10) & \\
N11-116 & 04 55 54.80 & $-$66 27 41.8 & 15.16         & $-$0.07 & B2 III            & 295\p~(8)  & \\
N11-117 & 04 57 44.46 & $-$66 23 59.2 & 15.21         & $-$0.19 & B1 Vn             & 291:~(6)   & variable $v_r$?\\
N11-118 & 04 56 44.30 & $-$66 23 04.8 & 15.21         & $-$0.17 & B1.5 V            & Binary (SB1)    & \\
N11-119 & 04 55 49.45 & $-$66 26 03.5 & 15.22         & $-$0.21 & B1.5 V            & Binary (SB2)    & \\
N11-120 & 04 57 15.22 & $-$66 22 29.5 & 15.24         & $-$0.12 & B0.2 Vn           & Binary     & \\
N11-121 & 04 56 44.52 & $-$66 30 03.3 & 15.24         & $-$0.19 & B1 Vn             & 299:~(4)    & \\
N11-122 & 04 56 28.76 & $-$66 28 47.1 & 15.27         & $-$0.27 & O9.5 V            &289\p~(10) & \\
N11-123 & 04 56 50.50 & $-$66 28 23.1 & 15.29         & $-$0.25 & O9.5 V            &295\p~(11) & \\
N11-124 & 04 56 18.63 & $-$66 28 37.2 & 15.34         & $-$0.21 & B0.5 V            & Binary (SB1)    & \\
\end{longtable}
\end{center}
}

\newpage

{\scriptsize
\begin{center}
\begin{longtable}{lccccclll}
\caption[]{NGC\,2004: Observational parameters of target stars.  
The cross-references in the final column are to \citet[][Sk]{sk69}, \citet[][R74]{r74},
\citet[][Brey]{brey}, \citet{w87} and \citet[][MHF]{mhf}.  All spectral types are those of the authors, 
excepting NGC\,2004-057 (Brey 45), taken from \citet{ssm96}.  Photometry for the first eight stars
comes from \citet{ard72} and \citet{bal93}, as described in the erratum.
\label{2004}} \\
\hline
ID & $\alpha$(2000) & $\delta$(2000) & r$_{\rm d}$ & $V$ & $B-V$ & Sp. Type & $v_{\rm r}$  & Comments \\
\hline 
\endfirsthead
\caption[]{\it{continued}} \\
\hline
ID & $\alpha$(2000) & $\delta$(2000) & r$_{\rm d}$ & $V$ & $B-V$ & Sp. Type & $v_{\rm r}$  & Comments \\
\hline 
\endhead
\hline 
\multicolumn{9}{r}{\it{continued on next page}} \\
\endfoot
\hline
\endlastfoot
\hline 
NGC\,2004-001 & 05 30 07.07 & $-$67 15 43.3 &  3.54 & 11.46 & \pp0.05  & A1 Ia     & 314\p~(6)  & Sk$-$67$^\circ$143; H$\alpha => v_r$ variable or var. em.\\
NGC\,2004-002 & 05 31 12.82 & $-$67 15 08.0 &  3.79 & 11.60 & \pp0.09  & A3: Iab   & 295\p~(2)  & Sk$-$67$^\circ$155; FLAMES-UVES target \\
NGC\,2004-003 & 05 30 40.40 & $-$67 16 09.0 &  1.09 & 12.09 & $-$0.06  & B5 Ia     & 309\p~(12)  & R74-C01; binary?\\
NGC\,2004-004 & 05 31 27.90 & $-$67 24 43.9 &  8.79 & 11.95 & \pp0.00  & B9 Ia     & 322\p~(8)   & Sk$-$67$^\circ$157 \\
NGC\,2004-005 & 05 29 42.61 & $-$67 20 47.5 &  6.60 & 11.93 & \pp0.04  & B8 Ia     &313\p~(10)  & Sk$-$67$^\circ$137 \\
NGC\,2004-006 & 05 30 01.22 & $-$67 14 36.9 &  4.59 & 12.01 & \pp0.07  & A2 Iab    & 313\p~(3)  & Sk$-$67$^\circ$141 \\
NGC\,2004-007 & 05 32 00.76 & $-$67 20 22.6 &  8.39 & 12.04 & $-$0.03  & B8 Ia     &310\p~(10) & Sk$-$67$^\circ$171 \\
NGC\,2004-008 & 05 30 40.10 & $-$67 16 37.9 &  0.61 & 12.43 & $-$0.03  & B9 Ia     &305\p~(10)  & R74-B01 \\
NGC\,2004-009 & 05 31 52.98 & $-$67 12 15.4 &  8.61 & 12.47 & \pp0.01  & A1 Iab    & 308\p~(4)    & \\
NGC\,2004-010 & 05 29 21.72 & $-$67 20 11.0 &  8.13 & 12.60 & $-$0.15  & B2.5 Iab  &296\p~(12) & Sk$-$67$^\circ$133 \\
NGC\,2004-011 & 05 31 03.75 & $-$67 21 20.1 &  4.69 & 12.64 & $-$0.13  & B1.5 Ia   &309\p~(12)  & Sk$-$67$^\circ$154 \\
NGC\,2004-012 & 05 30 37.48 & $-$67 16 53.7 &  0.43 & 13.39 & $-$0.20  & B1.5 Iab  &305\p~(12)  & R74-B09 \\
NGC\,2004-013 & 05 30 46.55 & $-$67 19 39.8 &  2.50 & 13.40 & $-$0.17  & B2 II     &289:~(10)   & R74-D13; H$\alpha => v_r$ variable or var. em.\\
NGC\,2004-014 & 05 30 44.47 & $-$67 21 01.5 &  3.81 & 13.43 & $-$0.12  & B3 Ib     &306\p~(12)   & \\
NGC\,2004-015 & 05 29 11.46 & $-$67 15 24.4 &  8.76 & 13.48 & $-$0.20  & B1.5 II   & Binary (SB1)     & \\
NGC\,2004-016 & 05 30 40.11 & $-$67 18 59.1 &  1.75 & 13.54 & $-$0.05  & B9 Ib     & 296\p~(5)   & R74-D18 \\
NGC\,2004-017 & 05 30 46.59 & $-$67 14 21.7 &  2.94 & 13.55 & \pp0.56  & G2        & $-$         & FLAMES-UVES target; foreground \\
NGC\,2004-018 & 05 31 02.69 & $-$67 20 49.8 &  4.20 & 13.58 & \pp0.57  & G8:       & $-$         & FLAMES-UVES target; foreground \\
NGC\,2004-019 & 05 30 44.62 & $-$67 18 37.9 &  1.46 & 13.60 & $-$0.20  & O9.5 IIIn &318:~(12)   & R74-D16 \\
NGC\,2004-020 & 05 30 53.80 & $-$67 15 48.8 &  1.94 & 13.61 & $-$0.13  & B1.5 II   & Binary (SB1)     & \\
NGC\,2004-021 & 05 30 42.01 & $-$67 21 41.4 &  4.46 & 13.67 & $-$0.14  & B1.5 Ib   &311\p~(12) & \\
NGC\,2004-022 & 05 30 47.37 & $-$67 17 23.4 &  0.71 & 13.77 & $-$0.17  & B1.5 Ib   &300\p~(12) & R74-B30 \\
NGC\,2004-023 & 05 30 58.01 & $-$67 18 14.8 &  1.99 & 13.91 & $-$0.08  & B2 (Be-Fe) & 311:~(6)   & H$\alpha =$ str. em.\\
NGC\,2004-024 & 05 29 53.99 & $-$67 17 22.2 &  4.46 & 14.07 & $-$0.15  & B1.5 IIIn & 356\p~(7)  & \\
NGC\,2004-025 & 05 30 00.73 & $-$67 21 56.4 &  6.05 & 14.17 & $-$0.16  & B2 (Be-Fe) & 305\p~(8)  & H$\alpha =$ str. em.\\
NGC\,2004-026 & 05 30 36.35 & $-$67 17 42.9 &  0.60 & 14.18 & $-$0.17  & B2 II     & Binary (SB1)     & R74-B15; H$\alpha =$ shell star\\
NGC\,2004-027 & 05 29 34.53 & $-$67 11 56.1 &  8.26 & 14.18 & $-$0.11  & B0e       & 282\p~(5)  & H$\alpha$ \& He6678 variable $v_r$ or var. em.?\\
NGC\,2004-028 & 05 31 32.59 & $-$67 16 40.3 &  5.09 & 14.26 & $-$0.19  & B2 II     & 311:~(6)   & \\
NGC\,2004-029 & 05 31 00.11 & $-$67 14 59.2 &  2.96 & 14.27 & $-$0.20  & B1.5e     &322\p~(10) & H$\alpha =$ str. em.\\
NGC\,2004-030 & 05 30 11.03 & $-$67 22 57.1 &  6.37 & 14.28 & $-$0.23  & B0.2 Ib   & Binary (SB1)    & \\
NGC\,2004-031 & 05 30 38.77 & $-$67 20 23.9 &  3.16 & 14.29 & $-$0.16  & B2 II     & Binary (SB1)    & \\
NGC\,2004-032 & 05 30 37.30 & $-$67 10 34.0 &  6.68 & 14.30 & $-$0.14  & B2 II     & 305:~(7)   & \\
NGC\,2004-033 & 05 29 32.69 & $-$67 17 05.0 &  6.52 & 14.31 & $-$0.18  & B1.5e     & 309\p~(6)   & H$\alpha =$ twin em., var \\
NGC\,2004-034 & 05 30 28.11 & $-$67 15 16.7 &  2.28 & 14.40 & $-$0.11  & B1.5e     & 310\p~(6)  & H$\alpha =$ twin em. \\
NGC\,2004-035 & 05 30 37.16 & $-$67 15 44.3 &  1.53 & 14.40 & $-$0.02  & B1: (Be-Fe) & 312:~(4) & H$\alpha =$ twin em. \\
NGC\,2004-036 & 05 29 20.73 & $-$67 17 54.8 &  7.70 & 14.43 & $-$0.22  & B1.5 III  &308\p~(12) & \\
NGC\,2004-037 & 05 30 26.38 & $-$67 09 00.8 &  8.33 & 14.45 & $-$0.10  & B2e       & 270:~(4)   & H$\alpha =$twim em, variable $v_r$?\\
NGC\,2004-038 & 05 31 22.14 & $-$67 17 39.1 &  4.07 & 14.46 & $-$0.24  & B0.7 V    & 290:~(6)   & \\
NGC\,2004-039 & 05 30 27.22 & $-$67 13 27.6 &  3.98 & 14.47 & $-$0.11  & B1.5e     & 310:~(6)  & H$\alpha =$ wk twin em \\
NGC\,2004-040 & 05 30 07.27 & $-$67 14 23.3 &  4.27 & 14.49 & \pp0.07  & A7 II     & 292\p~(3)  & \\
NGC\,2004-041 & 05 30 32.64 & $-$67 15 25.9 &  1.95 & 14.52 & $-$0.14  & B2.5 III  & Binary (SB1)    & MHF98013 \\
NGC\,2004-042 & 05 31 11.98 & $-$67 23 14.3 &  6.74 & 14.53 & $-$0.18  & B2.5 III  &294\p~(10) & \\
NGC\,2004-043 & 05 30 38.74 & $-$67 15 13.2 &  2.02 & 14.58 & $-$0.16  & B1.5 III  &333\p~(12)  & \\
NGC\,2004-044 & 05 29 33.98 & $-$67 18 06.9 &  6.45 & 14.60 & $-$0.22  & B1.5:     & Binary (SB2)    & Similar temperature secondary; MHF83937 \\
NGC\,2004-045 & 05 30 50.19 & $-$67 14 47.6 &  2.63 & 14.60 & $-$0.18  & B2 III    & Binary (SB1)    & \\
NGC\,2004-046 & 05 30 42.96 & $-$67 16 43.5 &  0.58 & 14.64 & $-$0.19  & B1.5 III  &312\p~(12) & R74-B50 \\
NGC\,2004-047 & 05 30 13.54 & $-$67 14 27.3 &  3.79 & 14.70 & $-$0.21  & B2 III    & Binary (SB1)    & MHF103207 \\
NGC\,2004-048 & 05 31 47.04 & $-$67 17 54.6 &  6.49 & 14.70 & $-$0.20  & B2.5e     & 307:~(7)  & \\
NGC\,2004-049 & 05 30 06.19 & $-$67 14 32.9 &  4.24 & 14.71 & $-$0.24  & O2-3 III(f$^\ast$) $+$ OB & Binary (SB2)  & \\
NGC\,2004-050 & 05 30 43.07 & $-$67 17 43.9 &  0.57 & 14.71 & $-$0.19  & B2.5 III  & Binary (SB1)    & R74-B24 \\
NGC\,2004-051 & 05 30 53.44 & $-$67 17 10.6 &  1.28 & 14.73 & $-$0.19  & B2 III    &308\p~(12) & R74-C16 \\
NGC\,2004-052 & 05 32 06.28 & $-$67 19 49.2 &  8.70 & 14.75 & $-$0.23  & B2 III    &291:~(10)  & \\
NGC\,2004-053 & 05 30 43.79 & $-$67 15 22.6 &  1.89 & 14.75 & $-$0.21  & B0.2 Ve   &303\p~(14) & H$\alpha =$ em.\\
NGC\,2004-054 & 05 31 46.38 & $-$67 13 28.9 &  7.41 & 14.76 & $-$0.20  & B2 III    & Binary (SB1) & \\
NGC\,2004-055 & 05 30 14.55 & $-$67 15 27.2 &  3.05 & 14.76 & $-$0.17  & B2.5 III  & 309:~(8)   & H$\alpha =$ complex, in-filled from em.\\
NGC\,2004-056 & 05 30 35.92 & $-$67 11 08.5 &  6.11 & 14.84 & $-$0.04  & B1.5e     & 300\p~(4) & H$\alpha =$ twin em. \\
NGC\,2004-057 & 05 31 34.37 & $-$67 16 29.4 &  5.28 & 14.85 & $-$0.51  & WN4b      & $-$       & Sk$-$67$^\circ$160; Brey 45\\
NGC\,2004-058 & 05 30 05.13 & $-$67 18 45.8 &  3.71 & 14.86 & $-$0.20  & O9.5 V (N str) &303:~(10) & \\
NGC\,2004-059 & 05 30 45.92 & $-$67 18 24.4 &  1.29 & 14.86 & $-$0.20  & B2 III    & Binary (SB1)    & R74-D17 \\
NGC\,2004-060 & 05 30 32.94 & $-$67 11 22.5 &  5.91 & 14.86 & $-$0.14  & B2 III    & 295\p~(7)  & \\
NGC\,2004-061 & 05 29 15.96 & $-$67 19 45.4 &  8.51 & 14.88 & $-$0.25  & B2 III    &312\p~(12) & \\
NGC\,2004-062 & 05 31 10.97 & $-$67 22 51.4 &  6.36 & 14.90 & $-$0.23  & B0.2 V    & 319:~(8)   & \\
NGC\,2004-063 & 05 30 34.49 & $-$67 17 34.9 &  0.65 & 14.93 & $-$0.15  & B2 III    & 311\p~(8)  & R74-C08 \\
NGC\,2004-064 & 05 31 19.11 & $-$67 16 54.7 &  3.77 & 14.96 & $-$0.22  & B0.7-B1 III    &310\p~(12)  & \\
NGC\,2004-065 & 05 30 13.10 & $-$67 21 06.5 &  4.67 & 14.96 & $-$0.19  & B2.5 III  & 306:~(7)    & \\
NGC\,2004-066 & 05 30 27.78 & $-$67 17 22.8 &  1.21 & 14.96 & $-$0.16  & B1.5 Vn   & 308:~(6)  & \\
NGC\,2004-067 & 05 30 34.64 & $-$67 14 45.6 &  2.54 & 14.96 & $-$0.08  & B1.5e     & 284:~(5)    & H$\alpha =$ twin; MHF101350 \\
NGC\,2004-068 & 05 30 23.00 & $-$67 08 53.6 &  8.51 & 14.97 & $-$0.08  & B2.5 III  &281\p~(10) & \\
NGC\,2004-069 & 05 30 12.15 & $-$67 10 19.7 &  7.42 & 14.99 & $-$0.24  & B0.7 V    & 332\p~(6)  & \\
NGC\,2004-070 & 05 31 03.61 & $-$67 16 07.7 &  2.52 & 15.01 & $-$0.22  & B0.7-B1 III    &310\p~(12) & \\
NGC\,2004-071 & 05 31 31.31 & $-$67 11 33.3 &  7.53 & 15.05 & $-$0.23  & B1.5 III  & 298\p~(9)   & \\
NGC\,2004-072 & 05 29 50.50 & $-$67 15 10.4 &  5.22 & 15.08 & $-$0.22  & B1.5 V    & Binary (SB1)   & \\
NGC\,2004-073 & 05 30 55.73 & $-$67 18 48.6 &  2.17 & 15.08 & $-$0.20  & B2 III    &304\p~(12) & R74-D12 \\
NGC\,2004-074 & 05 30 29.03 & $-$67 22 52.8 &  5.74 & 15.08 & $-$0.19  & B0.7-B1 V      & Binary (SB1)    & \\
NGC\,2004-075 & 05 30 53.42 & $-$67 18 03.7 &  1.52 & 15.08 & $-$0.18  & B2 III    &289:~(11)  & R74-D10 \\
NGC\,2004-076 & 05 30 20.65 & $-$67 08 41.2 &  8.76 & 15.08 & $-$0.08  & B2.5 III  & Binary (SB1)     & \\
NGC\,2004-077 & 05 29 16.94 & $-$67 20 33.8 &  8.70 & 15.09 & $-$0.29  & B0.5 V    & 296:~(6)    & \\
NGC\,2004-078 & 05 31 31.33 & $-$67 15 05.9 &  5.38 & 15.09 & $-$0.20  & B2 III    & Binary (SB1)    & \\
NGC\,2004-079 & 05 30 40.75 & $-$67 11 43.9 &  5.51 & 15.09 & $-$0.16  & B2 III    & Binary (SB1)    & \\
NGC\,2004-080 & 05 29 58.83 & $-$67 16 17.8 &  4.10 & 15.11 & $-$0.21  & B2.5 III  &313\p~(10)  & \\
NGC\,2004-081 & 05 29 52.86 & $-$67 18 14.6 &  4.68 & 15.11 & $-$0.20  & B1 V      & 298\p~(8)   & \\
NGC\,2004-082 & 05 30 20.37 & $-$67 20 55.4 &  4.15 & 15.11 & $-$0.19  & B1.5 V    & 305:~(7)   & \\
NGC\,2004-083 & 05 30 28.43 & $-$67 17 00.6 &  1.16 & 15.11 & $-$0.11  & B1.5: e   & Binary (SB1)    & R74-D22; H$\alpha =$ twin \\
NGC\,2004-084 & 05 31 08.01 & $-$67 20 00.3 &  3.85 & 15.14 & $-$0.21  & B1.5 III  &306\p~(12) & \\
NGC\,2004-085 & 05 29 56.32 & $-$67 17 55.5 &  4.29 & 15.14 & $-$0.19  & B2.5 III  &305:~(10)   & \\
NGC\,2004-086 & 05 30 58.02 & $-$67 14 04.8 &  3.60 & 15.15 & $-$0.20  & B2 III    &312\p~(10) & \\
NGC\,2004-087 & 05 31 01.27 & $-$67 20 08.7 &  3.55 & 15.15 & $-$0.20  & B1.5 V    &316\p~(12) & \\
NGC\,2004-088 & 05 30 57.15 & $-$67 15 14.3 &  2.58 & 15.15 & $-$0.18  & B2.5 III  & 307:~(5)    & No $\lambda$4124 data \\
NGC\,2004-089 & 05 30 48.35 & $-$67 21 58.6 &  4.80 & 15.16 & $-$0.14  & B2.5e     & 304:~(9)     & H$\alpha =$ twin \\
NGC\,2004-090 & 05 29 22.31 & $-$67 16 40.1 &  7.54 & 15.17 & $-$0.28  & O9.5 III  &310\p~(16) & \\
NGC\,2004-091 & 05 29 34.32 & $-$67 15 01.9 &  6.73 & 15.17 & $-$0.22  & B1.5 III  &310\p~(12)  & \\
NGC\,2004-092 & 05 30 24.95 & $-$67 12 46.8 &  4.70 & 15.18 & $-$0.11  & B2e       & 304:~(5)  & H$\alpha =$ twin \\
NGC\,2004-093 & 05 29 20.33 & $-$67 19 04.3 &  7.92 & 15.19 & $-$0.21  & B3 III    & 287:~(9)   & \\
NGC\,2004-094 & 05 31 09.27 & $-$67 15 22.3 &  3.37 & 15.20 & $-$0.19  & B2.5 III  & Binary     & \\
NGC\,2004-095 & 05 30 59.22 & $-$67 15 31.7 &  2.51 & 15.20 & $-$0.19  & B1.5 V    & 287\p~(8)   & \\
NGC\,2004-096 & 05 30 29.95 & $-$67 15 53.5 &  1.67 & 15.21 & $-$0.09  & B1.5e     & 317:~(4)   & H$\alpha =$ twin \\
NGC\,2004-097 & 05 30 13.21 & $-$67 08 40.4 &  8.95 & 15.24 & $-$0.20  & B2 III    &298\p~(10) & variable $v_r$?\\
NGC\,2004-098 & 05 31 00.30 & $-$67 19 05.9 &  2.69 & 15.24 & $-$0.18  & B2 III    & 287:~(7)   & \\
NGC\,2004-099 & 05 31 16.52 & $-$67 20 55.7 &  5.09 & 15.26 & $-$0.21  & B2 III    & 292:~(7)   & \\
NGC\,2004-100 & 05 29 29.89 & $-$67 18 22.1 &  6.88 & 15.29 & $-$0.18  & B1 Vn     & 298:~(6)   & \\
NGC\,2004-101 & 05 30 48.23 & $-$67 17 13.0 &  0.78 & 15.30 & $-$0.18  & B2 III    & 309:~(10)   & R74-B28 \\
NGC\,2004-102 & 05 31 00.23 & $-$67 16 04.0 &  2.26 & 15.30 & $-$0.18  & B2 III    & Binary    & \\
NGC\,2004-103 & 05 30 20.63 & $-$67 13 40.7 &  4.03 & 15.30 & $-$0.12  & B2 III    &303\p~(11) & \\
NGC\,2004-104 & 05 29 38.76 & $-$67 12 07.6 &  7.83 & 15.31 & $-$0.19  & B1.5 V    & 301:~(6)     & \\
NGC\,2004-105 & 05 30 10.20 & $-$67 18 42.3 &  3.25 & 15.31 & $-$0.19  & B1.5 V    & 297:~(7)   & \\
NGC\,2004-106 & 05 30 29.49 & $-$67 16 46.0 &  1.14 & 15.31 & $-$0.14  & B2 III    &307\p~(11) & R74-D23; Walker 4 \\
NGC\,2004-107 & 05 31 45.13 & $-$67 23 01.8 &  8.53 & 15.32 & $-$0.25  & B0.5 V    &Binary (SB1)  & \\
NGC\,2004-108 & 05 30 18.15 & $-$67 12 58.6 &  4.76 & 15.32 & $-$0.21  & B2.5 III  &296\p~(10)  & \\
NGC\,2004-109 & 05 29 11.05 & $-$67 18 59.2 &  8.78 & 15.32 & $-$0.20  & B2.5 III  &298:~(11)   & MHF79301 \\
NGC\,2004-110 & 05 30 50.48 & $-$67 10 38.4 &  6.67 & 15.32 & $-$0.19  & B2 III    & 296:~(7)     & \\
NGC\,2004-111 & 05 32 05.74 & $-$67 15 58.1 &  8.35 & 15.39 & $-$0.21  & B2.5 III  &310\p~(10)&  \\
NGC\,2004-112 & 05 32 04.73 & $-$67 19 03.7 &  8.36 & 15.39 & $-$0.19  & B2 III    & 307\p~(8)  & \\
NGC\,2004-113 & 05 30 53.86 & $-$67 16 54.4 &  1.36 & 15.39 & $-$0.17  & B2.5 IIIn & 293:~(7)   & R74-D08 \\
NGC\,2004-114 & 05 30 47.87 & $-$67 24 45.4 &  7.55 & 15.43 & $-$0.18  & B2 III    &298\p~(11)  & \\
NGC\,2004-115 & 05 31 04.64 & $-$67 18 35.1 &  2.72 & 15.44 & $-$0.19  & B2e       & Binary (SB1)    & H$\alpha =$ twin$+$central abs\\
NGC\,2004-116 & 05 30 23.58 & $-$67 10 41.7 &  6.74 & 15.44 & $-$0.12  & B2 III    &309\p~(13) & H$\alpha => v_r$ variable or var. em.\\
NGC\,2004-117 & 05 30 29.14 & $-$67 09 51.2 &  7.46 & 15.47 & $-$0.14  & B2 III    &305\p~(10) & \\
NGC\,2004-118 & 05 31 17.28 & $-$67 18 42.5 &  3.87 & 15.52 & $-$0.21  & B1.5 V    & Binary (SB1)    & \\
NGC\,2004-119 & 05 31 14.65 & $-$67 11 59.8 &  6.21 & 15.53 & $-$0.20  & B2 III    &309\p~(14) & \\
\end{longtable}
\end{center}
}

\newpage
{\small
\begin{center}
\begin{longtable}{llll}
\caption[]{
Comparison of current classifications with published spectral types.
Sources of classifications are W77 \citep{w77}; FB80 \citep{fb80};
CJF85 \citep{cjf85}; NMC \citep{nmc}; G87 \citep{gar87}; MPG
\citep{mpg}; F91 \citep{f91}; P92 \citep{p92}; L93 \citep{l93}; M95
\citep{m95}; G96 \citep{grb96}; LG96 \citep[classifications from
Lennon, given by][]{grb96}; HM00 \citep[][who adopt identifications
from \citealt{woo}]{hm00}; W00 \citep{wal00}; L03 \citep{l03}; EH04
\citep{eh04}.
\label{cftypes}} \\
\hline
ID & Alias & FLAMES & Published \\
\hline 
\endfirsthead
\caption[]{\it{continued}} \\
\hline
ID & Alias & FLAMES & Published \\
\hline 
\endhead
\hline 
\multicolumn{4}{r}{\it{continued on next page}} \\
\endfoot
\hline 
\endlastfoot
\hline
NGC\,346-001 &  AzV 232/Sk 80   & O7 Iaf$+$       & O7 Iaf$+$ [W77]; O7 If [MPG\,789] \\
NGC\,346-004 &  AzV 191         & Be (B1:)        & B extr [G87] \\ 
NGC\,346-007 &  MPG\,324        & O4 V((f$+$))    & O4-5 V [NMC]; O4 V((f)) [MPG]; O4 ((f)) [W00] \\
NGC\,346-008 &  AzV 224         & B1e             & B1 III [G87] \\
NGC\,346-009 &  MPG\,845        & B0e             & O9.5 V [MPG] \\
NGC\,346-010 &  AzV 226         & O7 IIIn((f))    & O7 III [G87] \\
NGC\,346-012 &  AzV 202         & B1 Ib           & B1 III [G87] \\
NGC\,346-014 &  2dFS\#1425      & A0 II           & A0 (Ib) [EH04] \\
NGC\,346-015 &  AzV 217         & B1 V $+$?       & B1 III [M95]; B1-3 (II) [EH04: 2dFS\#1357] \\
NGC\,346-016 &  2dFS\#5100      & B0.5 Vn $+$?    & B0-3 (III) [EH04] \\
NGC\,346-020 &  2dFS\#1259      & B1 V$+$early-B  & B0-3 (III) [EH04] \\
NGC\,346-021 &  2dFS\#5099      & B1 III          & B1-3 (III) [EH04] \\
NGC\,346-022 &  MPG\,682        & O9 V            & O8 V [MPG] \\
NGC\,346-023 &  MPG\,178        & Be (B0.2:)      & O8-8.5 V:: [MPG]; B0: (IV) [EH04: 2dFS\#5097] \\ 
NGC\,346-025 &  MPG\,848        & O9 V            & O8.5 V [MPG] \\
NGC\,346-026 &  MPG\,12         & B0 IV (Nstr)    & O9.5 V [MPG]; O9.5-B0 V (N str) [W00]; O9.5 III [EH04: 2dFS\#1299] \\
NGC\,346-028 &  MPG\,113        & OC6 Vz          & O6 V [MPG]; OC6 Vz [W00, the source of the type adopted here] \\
NGC\,346-029 &  MPG\,637        & B0 V $+$?       & B0 V [MPG] \\
NGC\,346-030 &  2dFS\#5098      & B0 V $+$?       & B0.5 (V) [EH04] \\
NGC\,346-034 &  MPG\,467        & O8.5 V          & O8 V$+$neb [MPG] \\
NGC\,346-035 &  2dFS\#1418      & B1 V $+$?       & B1-5 (II) [EH04] \\
NGC\,346-036 &  MPG\,729        & B0.5 Ve         & B0$+$neb [MPG] \\
NGC\,346-039 &  2dFS\#1262      & B0.7 V          & B1-3 (III) [EH04] \\
NGC\,346-043 &  MPG\,11         & B0 V            & B0 V [MPG] \\
NGC\,346-047 &  2dFS\#1189      & B2.5 III        & B1-5 (III) [EH04] \\
NGC\,346-050 &  MPG\,299        & O8 Vn           & O9 V [MPG] \\
NGC\,346-051 &  MPG\,523        & O7 Vz           & O7 V$+$neb [MPG] \\
NGC\,346-056 &  MPG\,310        & B0 V            & O9.5 V [MPG] \\
NGC\,346-061 &  2dFS\#1277      & Be (B1-2)       & B1-5 (II)e [EH04] \\
NGC\,346-063 &  2dFS\#1413      & A0 II           & A0 (II) [EH04] \\
NGC\,346-075 &  2dFS\#1389      & B1 V $+$?       & B1-3 (IV) [EH04] \\
NGC\,346-077 &  MPG\,238        & O9 V            & B0: [MPG] \\
NGC\,346-084 &  2dFS\#1296      & B1 V            & B0-5 (IV) [EH04] \\
\hline
NGC\,330-002 &  R74-A02         & B3 Ib           & B5 I [FB80]; B6 I [CJF85]; B4 Iab/b [L93]; B5 I [G96]; B4 Ib [L03] \\ 
NGC\,330-003 &  2dFS\#1183      & B2 Ib           & B2.5 (Ib) [EH04] \\
NGC\,330-004 &  R74-B37         & B2.5 Ib         & B5 I [FB80]; B5 I [CJF85]; B3 Ib [L93]; B2.5 (Ib) [EH04: 2dFS\#5090] \\
NGC\,330-012 &  Arp 211         & A0 Ib           & A0 I [CJF85] \\
NGC\,330-013 &  AzV 186         & O8 III ((f))    & O7 III [G87]; O8 III((f)) [EH04: 2dFS\#1230]] \\
NGC\,330-017 &  2dFS\#1171      & B2 II           & B1-3 (II) [EH04] \\
NGC\,330-018 &  R74-B30         & B3 II           & B6 I [CJF85]; B2 II [L93] \\
NGC\,330-020 &  2dFS\#1232      & B3 II           & B3 (II) [EH04] \\
NGC\,330-021 &  2dFS\#1242      & B0.2 III        & B0.5 (IV) [EH04] \\
NGC\,330-022 &  2dFS\#1062      & B3 II           & B3 (II) [EH04] \\
NGC\,330-024 &  2dFS\#1034      & B5 Ib           & B5 (II) [EH04] \\
NGC\,330-025 &  2dFS\#1224      & B1.5e           & B2 (II) [EH04] \\
NGC\,330-028 &  2dFS\#1195      & B1 V            & B2 (III) [EH04] \\
NGC\,330-029 &  2dFS\#5093      & B0.2 Ve         & B0-3 (III)e [EH04] \\
NGC\,330-031 &  2dFS\#5088      & B0.5 Ve         & B0.5 (IV) [EH04] \\
NGC\,330-033 &  2dFS\#5094      & B1.5 V          & B1-5 (II) [EH04] \\
NGC\,330-036 &  R74-B32         & B2 II           & B2 III [L03] \\
NGC\,330-037 &  2dFS\#1058      & A2 II           & A0 (II) [EH04] \\
NGC\,330-038 &  2dFS\#1206      & B1 V            & B1-2 (III) [EH04] \\
NGC\,330-039 &  2dFS\#1109      & B0 V            & B0.5 (V) [EH04] \\
NGC\,330-040 &  2dFS\#1041      & B2 III          & B1-2 (III) [EH04] \\
NGC\,330-041 &  2dFS\#1241      & B0 V            & B0 (V) [EH04] \\
NGC\,330-045 &  2dFS\#5087      & B3 III          & B1-5 (III) [EH04] \\
NGC\,330-046 &  2dFS\#1293      & O9.5 V          & B0 (V) [EH04] \\
NGC\,330-047 &  2dFS\#1190      & B1 V            & B1-3 (III) [EH04] \\
NGC\,330-051 &  2dFS\#1276      & B1.5 V          & B1-5 (III) [EH04] \\
NGC\,330-052 &  2dFS\#1152      & O8.5 Vn         & O8 V [EH04] \\
NGC\,330-061 &  2dFS\#1118      & A0 II           & A0 (II) [EH04] \\
NGC\,330-073 &  2dFS\#1087      & B8 Ib           & B8 (II) [EH04] \\
NGC\,330-104 &  2dFS\#1037      & B0: V           & B0-5 (IV) [EH04] \\
NGC\,330-118 &  2dFS\#1013      & B1-2            & B1-2 (V) [EH04] \\
NGC\,330-121 &  R74-B38         & A5 II           & A1 I [FB80]\\
NGC\,330-123 &  R74-B18         & O9.5 V          & B0 Ve [L93]; O9 III/Ve [LG96] \\
NGC\,330-124 &  R74-B28         & B0.2 V          & B0 Ve [L93]; B0.2 IIIe [LG96] \\
NGC\,330-125 &  R74-B13         & B2 III          & B2 III/IVe [L93]; B2 III/IVe [LG96] \\
\hline
N11-001 &  Sk$-$66$^\circ$36/P3252 & B2 Ia                & B2 I $+$ neb [F91]; B2 II (Hwk) [P92] \\
N11-002 &  Sk$-$66$^\circ$27/P1062 & B3 Ia                & B2.5-3 Ia [F91]; B4 Ia [P92] \\
N11-003 &  P3157                   & B1 Ia                & BC1 Ia (Nwk) [P92] \\
N11-010 &  P1310                   & O9.5 III $+$ B1-2:   & B0 V [P92] \\
N11-013 &  P3223                   & O8 V                 & O8.5 IV [P92] \\
N11-015 &  Sk$-$66$^\circ$37/P3271 & B0.7 Ib              & B1 II (Hwk) [P92] \\
N11-016 &  Sk$-$66$^\circ$26/P1036 & B1 Ib                & B1.5 Ia (Nwk) [P92]; B1 III [M95] \\
N11-018 &  P3053                   & O6 II(f$^\ast$)      & O5.5 I-III(f) [P92] \\
N11-022 &  Sk$-$66$^\circ$20       & O6.5 II(f)           & O6 III [M95] \\
N11-023 &  P1014                   & B0.7 Ib              & B0.5 II (blend?) [P92] \\
N11-028 &  P3264                   & O6-8 V               & O3-6 V (ZAMS in N11A) [P92] \\
N11-031 &  P3061                   & ON2 III(f$^\ast$)    & O3 III(f$^\ast$) [P92]; O2 III(f$^\ast$) [W02]; ON2 III(f$^\ast$) [W04, adopted here] \\
N11-032 &  P3168                   & O7 II(f)             & O7 II(f) [P92] \\
N11-033 &  P1005                   & B0 IIIn              & B0 III [P92] \\
N11-034 &  P1332                   & B0.5 III             & B0.7 II [P92] \\
N11-037 &  P1052                   & B0 III               & B0.2 III [P92] \\
N11-038 &  P3100                   & O5 II(f$+$)          & O6.5 V((f)): [P92] \\
N11-042 &  P1017                   & B0 III               & B0 III [P92] \\
N11-048 &  P3204                   & O6.5 V((f))          & O6-7 V (ZAMS) [P92] \\
N11-049 &  P1110                   & O7.5 V               & O7.5 V [P92] \\
N11-050 &  P3224                   & O4-5 $+$ O7:         & O6 III (blend?) [P92] \\
N11-059 &  P1125                   & O9 V                 & O8.5 V [P92] \\
N11-060 &  P3058                   & O3 V((f$^\ast$))     & O3 V((f$^\ast$)) [P92]; O3 V((f$^\ast$)) [W02, adopted here] \\
N11-063 &  Wo597                   & O9: Vn               & O9 V [HM00] \\
N11-065 &  P1027                   & O6.5 V((f))          & O6.5 V((f)) [P92] \\
N11-069 &  P1037                   & B1 III               & B1.5 II [P92] \\
N11-070 &  P3239                   & B3 III               & B2 V [P92] \\
N11-074 &  P1028                   & B0.5e                & O8-9: III: [P92] \\
N11-079 &  P3128                   & B0.2 V               & O9.5 IV [P92] \\
N11-080 &  P3173                   & O7: V $+$ O9:        & O4-6 V [P92] \\
N11-087 &  P3042                   & O9.5 Vn              & O9.5: III: [P92] \\
N11-088 &  P1160                   & B1 III               & B1 V [P92] \\
N11-099 &  Wo622                   & B0.2 V               & O9.7 III/B0.2 V [HM00] \\
N11-100 &  P3270                   & B0.5 V               & B1 V [P92] \\
N11-101 &  P3033                   & B0.2 V               & B0.2 IV [P92] \\
N11-105 &  P1378                   & B1 V                 & B1 V [P92] \\
\hline
NGC\,2004-001 & Sk$-$67$^\circ$143      & A1 Ia                & B7 Ia$+$ [F91] \\
\hline
\end{longtable}
\end{center}
}

\newpage

\begin{center}
\begin{figure}
\caption{FLAMES targets in the NGC\,346 field.  Due to crowding in the core of the cluster
346-007, 034, 079, 086, 111, and 115 are not labelled.  Each of these is included in the
\citet{mpg} study and the reader is referred to their finding charts for these stars.}\label{fchart_346}
\end{figure}
\end{center}


\begin{center}
\begin{figure}
\caption{FLAMES targets in the NGC\,330 field.  The five additional UVES targets
(nos. 121-125) are not included, the reader is directed to the finding charts by
\citet{r74}.}\label{fchart_330}
\end{figure}
\end{center}


\begin{center}
\begin{figure}
\caption{FLAMES targets in the N11 field.  The nine 
additional UVES targets are not included, the reader is directed to the finding
charts by \citet{p92}.\label{fchart_n11}}
\end{figure}
\end{center}


\begin{center}
\begin{figure}
\caption{FLAMES targets in the NGC\,2004 field.}\label{fchart_2004}
\end{figure}
\end{center}


\begin{center}
\begin{figure}
\caption{O-type (open blue circles) and B-type (gold) FLAMES \& UVES targets in the N11 field}
\label{n11_mess}
\end{figure}
\end{center}

\newpage

\appendix
\section{Detailed record of observations}\label{mjd}
Following the detailed discussion of numerous binaries in this paper,
for completeness we also include the Modified Julian Dates (MJD) of
each of our observations in Tables \ref{346_obs}, \ref{330_obs}, 
\ref{lh9_obs}, and \ref{2004_obs}.

\begin{table}[h]
\caption[]{Modified Julian Dates (MJD) of the NGC\,346 FLAMES observations. 
The Giraffe wavelength settings (e.g. HR02) and central wavelengths (\lam$_c$)
are given.  The exposure time for each observation was 2275s, excepting HR06/\#08, 
for which it was 2500s.}
\label{346_obs}
\begin{center}
\begin{tabular}{lccc}
\hline\hline
Giraffe & \lam$_c$ & \# & MJD \\
setting & & & \\
\hline
HR02 & 3958 & 01 &  52954.135894 \\
HR02 & 3958 & 02 &  52954.162910 \\
HR02 & 3958 & 03 &  52954.189858 \\
HR02 & 3958 & 04 &  52955.089567 \\
HR02 & 3958 & 05 &  52955.116573 \\
HR02 & 3958 & 06 &  52955.143518 \\
\\
HR03 & 4124 & 01 &  52981.045452\\
HR03 & 4124 & 02 &  52981.072478 \\
HR03 & 4124 & 03 &  52981.099430 \\
HR03 & 4124 & 04 &  52981.132751 \\
HR03 & 4124 & 05 &  52981.159698 \\
HR03 & 4124 & 06 &  52981.186653 \\
\\
HR04 & 4297 & 01 &  52989.130391 \\
HR04 & 4297 & 02 &  52989.157348 \\
HR04 & 4297 & 03 &  52989.184302 \\
HR04 & 4297 & 04 &  53005.048810 \\
HR04 & 4297 & 05 &  53005.075776 \\
HR04 & 4297 & 06 &  53005.102733 \\
\\
HR05 & 4471 & 01 &  52978.115490 \\
HR05 & 4471 & 02 &  52978.142443 \\
HR05 & 4471 & 03 &  52978.169401 \\
HR05 & 4471 & 04 &  53006.047328 \\
HR05 & 4471 & 05 &  53006.074344 \\
HR05 & 4471 & 06 &  53006.101307 \\
\\
HR06 & 4656 & 01 &  52926.086233 \\
HR06 & 4656 & 02 &  52926.138196 \\
HR06 & 4656 & 03 &  52926.165224 \\
HR06 & 4656 & 04 &  52926.203441 \\
HR06 & 4656 & 05 &  52926.230394 \\
HR06 & 4656 & 06 &  52988.111039 \\
HR06 & 4656 & 07 &  52988.155055 \\
HR06 & 4656 & 08 &  52988.183316 \\
\\
HR14 & 6515 & 01 &  52922.133208 \\
HR14 & 6515 & 02 &  52922.177677 \\
HR14 & 6515 & 03 &  52922.204633 \\
HR14 & 6515 & 04 &  52922.240118 \\
HR14 & 6515 & 05 &  52922.267069 \\
HR14 & 6515 & 06 &  52922.308471 \\
HR14 & 6515 & 07 &  52925.152684 \\
HR14 & 6515 & 08 &  52925.179646 \\
HR14 & 6515 & 09 &  52925.206601 \\
\hline
\end{tabular}
\end{center}
\end{table}

\begin{table}
\caption[]{Modified Julian Dates (MJD) of the NGC\,330 FLAMES observations. 
The Giraffe wavelength settings (e.g. HR02) and central wavelengths (\lam$_c$)
are given.  The exposure time for each observation was 2275s, excepting HR02/\#06
(1743s); HR05/\#05 (2221s); HR04/\#06 (1900s). }
\label{330_obs}
\begin{center}
\begin{tabular}{lccc}
\hline\hline
Giraffe & \lam$_c$ & \# & MJD \\
setting & & & \\
\hline
HR02 & 3958 & 01 &  52831.335190\\
HR02 & 3958 & 02 &  52831.362917 \\
HR02 & 3958 & 03 &  52831.389880 \\
HR02 & 3958 & 04 &  52833.343405 \\
HR02 & 3958 & 05 &  52833.380020 \\
HR02 & 3958 & 06 &  52833.407722 \\
HR02 & 3958 & 07 &  52834.272660 \\
HR02 & 3958 & 08 &  52834.310219 \\
HR02 & 3958 & 09 &  52834.338165 \\
\\
HR03 & 4124 & 01 &  52832.306332 \\
HR03 & 4124 & 02 &  52832.334079 \\
HR03 & 4124 & 03 &  52832.361042 \\
HR03 & 4124 & 04 &  52832.389991 \\
HR03 & 4124 & 05 &  52832.417687 \\
HR03 & 4124 & 06 &  52833.431610 \\
HR03 & 4124 & 07 &  53571.338966 \\
HR03 & 4124 & 08 &  53571.365957 \\
HR03 & 4124 & 09 &  53571.392904 \\
HR03 & 4124 & 10 &  53575.350926 \\
HR03 & 4124 & 11 &  53575.377924 \\
HR03 & 4124 & 12 &  53575.404861 \\
\\
HR04 & 4297 & 01 &  52835.365846 \\
HR04 & 4297 & 02 &  52835.392841 \\
HR04 & 4297 & 03 &  52835.419862 \\
HR04 & 4297 & 04 &  52839.383749 \\
HR04 & 4297 & 05 &  52839.410778 \\
HR04 & 4297 & 06 &  52839.435562 \\
\\
HR05 & 4471 & 01 &  52836.311151 \\
HR05 & 4471 & 02 &  52836.338693 \\
HR05 & 4471 & 03 &  52836.365649 \\
HR05 & 4471 & 04 &  52836.406499 \\
HR05 & 4471 & 05 &  52836.433197 \\
HR05 & 4471 & 06 &  52837.228526 \\
HR05 & 4471 & 07 &  52837.255493 \\
HR05 & 4471 & 08 &  52837.282458 \\
HR05 & 4471 & 09 &  52837.409609 \\
HR05 & 4471 & 10 &  52839.296100 \\
HR05 & 4471 & 11 &  52839.323059 \\
HR05 & 4471 & 12 &  52839.350034 \\
\\
HR06 & 4656 & 01 &  52834.365133 \\
HR06 & 4656 & 02 &  52834.400818 \\
HR06 & 4656 & 03 &  52834.428742 \\
HR06 & 4656 & 04 &  52837.320664 \\
HR06 & 4656 & 05 &  52837.347739 \\
HR06 & 4656 & 06 &  52837.375036 \\
\\
HR14 & 6515 & 01 &  52829.311413 \\
HR14 & 6515 & 02 &  52829.339117 \\
HR14 & 6515 & 03 &  52829.366074 \\
HR14 & 6515 & 04 &  52830.302236 \\
HR14 & 6515 & 05 &  52830.329997 \\
HR14 & 6515 & 06 &  52830.356985 \\
\hline
\end{tabular}
\end{center}
\end{table}

\begin{table}
\caption[]{Modified Julian Dates (MJD) of the N11 FLAMES observations. 
The Giraffe wavelength settings (e.g. HR02) and central wavelengths (\lam$_c$)
are given.  The exposure time for each observation was 2275s.} 
\label{lh9_obs}
\begin{center}
\begin{tabular}{lccc}
\hline\hline
Giraffe & \lam$_c$ & \# & MJD \\
setting & & & \\
\hline
HR02 & 3958 & 01 & 52928.313905 \\
HR02 & 3958 & 02 & 52928.340935 \\
HR02 & 3958 & 03 & 52928.367887 \\
HR02 & 3958 & 04 & 52978.209107 \\
HR02 & 3958 & 05 & 52978.236069 \\
HR02 & 3958 & 06 & 52978.263031 \\
\\
HR03 & 4124 & 01 & 52978.298954 \\
HR03 & 4124 & 02 & 52978.325913 \\
HR03 & 4124 & 03 & 52978.352876 \\
HR03 & 4124 & 04 & 52979.120168 \\
HR03 & 4124 & 05 & 52979.147117 \\
HR03 & 4124 & 06 & 52979.174080 \\
\\
HR04 & 4297 & 01 & 52979.209306 \\
HR04 & 4297 & 02 & 52979.236270 \\
HR04 & 4297 & 03 & 52979.263219 \\
HR04 & 4297 & 04 & 52979.297524 \\
HR04 & 4297 & 05 & 52979.324479 \\
HR04 & 4297 & 06 & 52979.351427 \\
\\
HR05 & 4471 & 01 & 52980.104396 \\
HR05 & 4471 & 02 & 52980.131349 \\
HR05 & 4471 & 03 & 52980.158315 \\
HR05 & 4471 & 04 & 52980.193917 \\
HR05 & 4471 & 05 & 52980.220880 \\
HR05 & 4471 & 06 & 52980.247825 \\
\\
HR06 & 4656 & 01 & 52980.282525 \\
HR06 & 4656 & 02 & 52980.309565 \\
HR06 & 4656 & 03 & 52980.336517 \\
HR06 & 4656 & 04 & 52981.222102 \\
HR06 & 4656 & 05 & 52981.249059 \\
HR06 & 4656 & 06 & 52981.276019 \\
\\
HR14 & 6515 & 01 & 52924.309781 \\
HR14 & 6515 & 02 & 52924.336804 \\
HR14 & 6515 & 03 & 52924.363751 \\
HR14 & 6515 & 04 & 52955.178138 \\
HR14 & 6515 & 05 & 52955.205160 \\
HR14 & 6515 & 06 & 52955.232125 \\
\hline
\end{tabular}
\end{center}
\end{table}

\begin{table}
\caption[]{Modified Julian Dates (MJD) of the NGC\,2004 FLAMES observations. 
The Giraffe wavelength settings (e.g. HR02) and central wavelengths (\lam$_c$)
are given.  The exposure time for each observation was 2275s, excepting 
HR06/\#06 (1468s).} 
\label{2004_obs}
\begin{center}
\begin{tabular}{lccc}
\hline\hline
Giraffe & \lam$_c$ & \# & MJD \\
setting & & & \\
\hline
HR02 & 3958 & 01 & 52982.201147 \\
HR02 & 3958 & 02 & 52982.228098 \\
HR02 & 3958 & 03 & 52982.255044 \\
HR02 & 3958 & 04 & 52988.219081 \\
HR02 & 3958 & 05 & 52988.250594 \\
HR02 & 3958 & 06 & 52988.277546 \\
\\
HR03 & 4124 & 01 & 53005.136738 \\
HR03 & 4124 & 02 & 53005.163753 \\
HR03 & 4124 & 03 & 53005.190706 \\
HR03 & 4124 & 04 & 53005.221334 \\
HR03 & 4124 & 05 & 53005.248354 \\
HR03 & 4124 & 06 & 53005.275313 \\
\\
HR04 & 4297 & 01 & 53006.135301 \\
HR04 & 4297 & 02 & 53006.162257 \\
HR04 & 4297 & 03 & 53006.189220 \\
HR04 & 4297 & 04 & 53008.047982 \\
HR04 & 4297 & 05 & 53008.075016 \\
HR04 & 4297 & 06 & 53008.101980 \\
\\
HR05 & 4471 & 01 & 53008.163059 \\
HR05 & 4471 & 02 & 53008.190005 \\
HR05 & 4471 & 03 & 53008.216952 \\
HR05 & 4471 & 04 & 53009.161824 \\
HR05 & 4471 & 05 & 53009.188777 \\
HR05 & 4471 & 06 & 53009.215739 \\
\\
HR06 & 4656 & 01 & 53012.095573 \\
HR06 & 4656 & 02 & 53012.122595 \\
HR06 & 4656 & 03 & 53012.149545 \\
HR06 & 4656 & 04 & 53012.183447 \\
HR06 & 4656 & 05 & 53012.210458 \\
HR06 & 4656 & 06 & 53012.232744 \\
\\
HR14 & 6515 & 01 & 52955.266266 \\
HR14 & 6515 & 02 & 52955.293277 \\
HR14 & 6515 & 03 & 52955.320221 \\
HR14 & 6515 & 04 & 52989.259276 \\
HR14 & 6515 & 05 & 52989.286232 \\
HR14 & 6515 & 06 & 52989.313190 \\
\hline
\end{tabular}
\end{center}
\end{table}



\clearpage

\begin{center}
\vspace*{0.2in}   
{\huge{\it Erratum}}\\
\smallskip
\end{center}   

\vspace{0.1in}
Incorrect photometry was given for a total of ten bright stars in
Tables 6 and 7.  These targets were saturated
in the Wide Field Imager (WFI) frames and should have been replaced
with published values as given below in Table~\ref{errphot}.  Values for the two
stars in N11 are taken from \citet{p92}.  NGC\,2004-003 and
NGC\,2004-008 are from \citet{bal93}, with the remaining six from
\citet{ard72}.  The only consequence of these changes for the
published version is the position of these stars in the
Hertzsprung-Russell diagrams in Figure~12.  These were used for a
qualitative discussion of the populations in each FLAMES field --
because of the difference in reddenings the two apparently massive
stars in N11 (with log(T$_{\rm eff}$) $\sim$ 4.2 and 4.3) will have
lower luminosities, therefore corresponding to lower-mass evolutionary
tracks.

\begin{table}[h]
\caption[]{Replacement photometry for bright stars in Tables 6 and 7 of the published
version. These values have been corrected in Tables 6 and 7 of the replacement astro-ph 
copy.\label{errphot}}
\begin{center}
\begin{tabular}{lcccc}
\hline\hline
ID & $\alpha$ (2000) & $\delta$ (2000) & $V$ & $B - V$ \\
\hline
N11-001 & 04 57 08.85 & $-$66 23 25.1 & 11.35 & $\phantom{-}$0.06 \\
N11-002 & 04 56 23.51 & $-$66 29 51.7 & 11.90 & $\phantom{-}$0.37 \\
\hline
NGC\,2004-001 & 05 30 07.07 & $-$67 15 43.3 & 11.46 & $\phantom{-}$0.05 \\ 
NGC\,2004-002 & 05 31 12.82 & $-$67 15 08.0 & 11.60 & $\phantom{-}$0.09 \\
NGC\,2004-003 & 05 30 40.40 & $-$67 16 09.0 & 12.09 & $-$0.06 \\
NGC\,2004-004 & 05 31 27.90 & $-$67 24 43.9 & 11.95 & $\phantom{-}$0.00 \\
NGC\,2004-005 & 05 29 42.61 & $-$67 20 47.5 & 11.93 & $\phantom{-}$0.04 \\
NGC\,2004-006 & 05 30 01.22 & $-$67 14 36.9 & 12.01 & $\phantom{-}$0.07 \\
NGC\,2004-007 & 05 32 00.76 & $-$67 20 22.6 & 12.04 & $-$0.03 \\
NGC\,2004-008 & 05 30 40.10 & $-$67 16 37.9 & 12.43 & $-$0.03 \\
\hline
\end{tabular}
\end{center}
\end{table}

We have also noticed two typographical errors in the published
version.  The classification for N11-020 should be given in Sections
5.5 and 8.2, and Table 6 as O5~I(n)fp.  Secondly, in the heading of Table 7, the
number in the parentheses after $\alpha$ should, of course, read
`(2000)'.

\end{document}